\documentclass{aa}  
\usepackage{graphicx}
\usepackage{txfonts}
\usepackage{placeins}

\usepackage{xcolor}
\usepackage[pdfpagelabels=false]{hyperref}
\hypersetup{colorlinks=true,linkcolor=blue,citecolor=blue,filecolor=blue,urlcolor=blue,}
\newcommand{\kms}{\hbox{$\mathrm{km \, s^{-1}}$}}
\newcommand{\kmskpc}{\hbox{$\mathrm{km \, s^{-1} \, kpc^{-1}}$}}

\begin{document}
\title{Linking dynamics and chemistry in barred galaxies through action-space mapping}

\author{Tetsuro Asano \inst{1,2,3,4}\corrauth{asano@fqa.ub.edu}
	\and
	Victor P. Debattista \inst{5}
	\and
	Michiko S. Fujii \inst{4}
}
\authorrunning{T. Asano et al.}

\institute{Departament de F\'isica Qu\`antica i Astrof\'isica (FQA), Universitat de Barcelona (UB), c. Mart\'i i Franqu\`es, 1, 08028 Barcelona, Spain
	\and
	Institut de Ci\`encies del Cosmos (ICCUB), Universitat de Barcelona (UB), c. Mart\'i i Franqu\`es, 1, 08028 Barcelona, Spain
	\and
	Institut d’Estudis Espacials de Catalunya (IEEC), c. Gran Capit\`a, 2-4, 08034 Barcelona, Spain
	\and
	Department of Astronomy, Graduate School of Science, The University of Tokyo, 7-3-1 Hongo, Bunkyo-ku, Tokyo 113-0033, Japan
	\and
	Jeremiah Horrocks Institute, University of Lancashire, Preston, PR1 2HE, UK
}

\date{Received MM DD, YYYY; accepted mm dd, yyyy}

\abstract
{}
{We aim to extend the action-based framework, proposed by a previous study, for assigning chemical information to collisionless $N$-body simulations of barred galaxies, enabling more realistic comparisons with observations.
}
{We developed a new method to map chemistry of star-forming $N$-body+hydrodynamical simulations (donor models) onto pure $N$-body simulations (target models) using a matching algorithm. We applied the method to two barred $N$-body galaxies, one Milky Way (MW)-like with a classical bulge and another with a strongly buckled bar, combined with two star-forming donor models exhibiting bimodal and unimodal chemical tracks in the [Fe/H]–[$\alpha/$Fe] plane.
}
{Our models reproduce key chemo-dynamical trends observed in the MW and external galaxies. Metal-rich populations are more strongly associated with the X-shaped bulge morphology, leading to vertically pinched metallicity maps consistent with observations and chemo-hydrodynamical simulations. The [Fe/H]–[O/Fe] distributions show latitude-dependent bimodality, and the mock magnitude distributions of red clump stars reveal a stronger X-shape for metal-rich populations. In the model with a classical bulge, metal-poor stars exhibit a more spherical morphology on the longitude-latitude plane, while metal-rich stars show a boxy distribution, consistent with observations of the MW bulge.
}
{
The proposed action-based chemical assignment provides a computationally efficient and flexible approach to link the dynamical and chemical evolution of barred galaxies. It enables realistic chemo-dynamical modelling of the galactic bulges.
}


\maketitle

\nolinenumbers

\section{Introduction}
Boxy-, peanut-, X-shaped (BPX) bulges are commonly observed in disc galaxies.
In the nearby Universe, roughly half of edge-on galaxies exhibit BPX structures \citep{2000A&AS..145..405L, 2014MNRAS.444L..80L}, although the detailed fraction depends on the stellar mass \citep{2017MNRAS.468.2058E, 2022MNRAS.512.1371M}.
The Milky Way (MW) also hosts a BPX bulge, as evidenced by the bimodal distribution in the distances of red clump stars in the bulge regions \citep{2010ApJ...724.1491M, 2010ApJ...721L..28N, 2015MNRAS.447.1535N, 2011AJ....142...76S, 2012ApJ...756...22N, 2013MNRAS.435.1874W, 2015A&A...583L...5G}.
Many numerical simulations have demonstrated that BPX bulges can form either through the buckling instability of bars \citep[e.g.][]{1991Natur.352..411R, 1994ApJ...425..551M, 2004ApJ...604L..93D, 2006ApJ...645..209D, 2017MNRAS.469.1587D, 2005ApJ...626..159B, 2017A&A...606A..47F, 2020MNRAS.492.2241C, 2020RAA....20..159S, 2024A&A...683A.196G} or through the secular evolution due to vertical resonances \citep[e.g.][]{1981A&A....96..164C, 1990A&A...233...82C, 1991A&A...252...75P, 2014MNRAS.437.1284Q, 2020MNRAS.495.3175S, 2022MNRAS.513.2850B, 2024MNRAS.527.2919A, 2025MNRAS.537.1475M, 2025MNRAS.542..464D}. 
Since buckling is a short-lived process, direct signatures of it are detected in only a small number of galaxies \citep{2016ApJ...825L..30E}.
Most edge-on galaxies with BPX bulges, including the MW, appear nearly symmetric with respect to their mid-planes \citep[e.g.][]{2023MNRAS.518.2300C}, making it difficult to determine whether they were shaped by bar buckling long ago or by vertical resonances.
\citet{2024MNRAS.527.2919A} found roughly half of the BPX galaxies in the TNG50 cosmological simulation \citep{2019MNRAS.490.3234N, 2019MNRAS.490.3196P} exhibit clear buckling signatures, while the other half do not and their BPX structures are likely formed through vertical resonances or weak buckling.

Spectroscopic surveys, such as BRAVA \citep{2007ApJ...658L..29R, 2012AJ....143...57K}, ARGOS \citep{2013MNRAS.428.3660F}, GIBS \citep{2014A&A...562A..66Z}, APOGEE \citep{2017AJ....154...94M}, PIGS \citep{2020MNRAS.491L..11A}, and \textit{Gaia}-ESO \citep{2022A&A...666A.120G}, provided detailed insights into the chemical and kinematic properties of the MW bulge, offering important clues to its present structure, formation, and evolutionary history.
One of the key observational features is that the X-shape is more prominent in metal-rich populations than in metal-poor ones \citep{2012ApJ...756...22N, 2012A&A...546A..57U, 2014A&A...569A.103R, 2021A&A...647A..34L}.
The process of `kinematic fractionation'---the bar-induced separation of stellar populations according to their kinematics---offers a mechanism for this metallicity dependence.
\citet{2017MNRAS.469.1587D} demonstrated that populations which were radially cooler prior to bar formation evolve into a stronger X-shaped structure than radially hotter populations.
In this way, the initial correlation between kinematics and metallicity becomes imprinted on the metallicity dependence of the bulge morphology after the bar forms. 

\citet{2020MNRAS.498.3334D} further extended this idea and examined BPX bulge formation and evolution in action space using  $N$-body simulations.
They showed that disc particles contribute differently to the BPX structure depending on their initial actions: populations with low radial, low vertical, and high azimuthal actions form the strongest BPX feature.
Since pure $N$-body simulations offer significant computational advantages for exploring dynamical processes, being able to paint them with realistic metallicities extends their utility for chemodynamical studies.
They therefore proposed a practical method of assigning metallicities to $N$-body models based on initial actions.
As discussed in Section 5 of their paper, their method using all three actions can reproduce the bulge metallicity distribution more realistically than previous methods relying on single actions, spatial positions, or membership in the thin or thick disc \citep[e.g.][]{2011MNRAS.416L..60B, 2013ApJ...766L...3M, 2016PASA...33...27D, 2024ApJ...976..232C}.
This action-based modelling offers a pragmatic alternative to $N$-body+hydrodynamical simulations with star formation and chemical evolution, which are computationally expensive for achieving  high resolution and broad coverage of parameter space.

In this study, we extend the action-based metallicity assignment framework. We present a method to map chemical information from star-forming $N$-body+hydrodynamical simulations onto pure $N$-body models using a matching algorithm called the Hungarian algorithm \citep{kuhn1955hungarian}.
We apply this method to two $N$-body models of barred galaxies and two star-forming models.
Both $N$-body models develop BPX bulges, but with differing properties: one is identical to the model used in \citet{2020MNRAS.498.3334D}, with a strongly buckled bar, while the other is a MW-like model with a classical bulge component and features a BPX bulge formed without buckling.
The star-forming models also display contrasting chemical behaviour: one exhibits a bimodal distribution in the [Fe/H]--[$\alpha$/Fe] plane, while the other follows a single sequence.
We confirm consistency with the results of \citet{2020MNRAS.498.3334D}, and also discuss both the similarities and differences between the two $N$-body models.

The structure of this paper is as follows.
In Section~\ref{sec:models_methods}, we describe the simulations used in this study and our chemical assignment method.
Section~\ref{sec:results} presents basic results, including face-on and edge-on metallicity maps, bulge metallicity distributions, and kinematic properties.
In Section~\ref{sec:comparison_with_GIBS}, we discuss the implications of our results for the MW bulge comparing with GIBS observations.
Finally, we summarise our findings in Section~\ref{sec:summary}.

\section{Models and methods}\label{sec:models_methods}
\subsection{$N$-body models}
We use two $N$-body models of disc galaxies.
The first one is MWa from \citet{2019MNRAS.482.1983F}. It is a MW-like $N$-body model which consists of a dark matter (DM) halo, a stellar disc, and a classical bulge. The DM halo follows the Navarro-Frenk-White \citep[NFW;][]{1997ApJ...490..493N} profile:
\begin{equation}
	\rho(r) = \frac{\rho_{\mathrm{h},0}}{(r/a_{\mathrm{h}})(1+r/a_{\mathrm{h}})^2}.
\end{equation}
The scale radius, $a_{\mathrm{h}}$, and the characteristic velocity, $\sigma_{\mathrm{h}} = \sqrt{4\pi G a_{\mathrm{h}}^2 \rho_{\mathrm{h},0}}$, are 10~kpc and 420~\kms, respectively.
The  rotation parameter, which represents the fraction of the DM particles rotating in the prograde direction with respect to the disc, is 0.8.
The stellar disc follows the radially exponential and vertically isothermal profile:
\begin{equation}
	\rho(R,z) = \rho_{\mathrm{d},0} \exp \left(- \frac{R}{R_{\mathrm{d}}} \right) \mathrm{sech}^2 \left( \frac{z}{z_{\mathrm{d}}} \right).
	\label{eq:disc_density}
\end{equation}
The scale radius, $R_{\mathrm{d}}$, and the scale height, $z_{\mathrm{d}}$, are set to 2.3~kpc and 0.2~kpc, respectively.
The total mass of the disc is $3.61\times10^{10} M_{\sun}$. The radial velocity dispersion exponentially decreases, and the central velocity dispersion and the dispersion scale radius are 94~\kms and 2.3~kpc, respectively.
The classical bulge follows the Hernquist profile \citep{1990ApJ...356..359H}:
\begin{equation}
	\rho(r) = \frac{\rho_{\mathrm{b},0}}{(r/a_{\mathrm{b}})(1+r/a_{\mathrm{b}})^3}.
\end{equation}
The scale radius and the characteristic velocity are $a_{\mathrm{b}}=0.75$~kpc and $\sigma_{\mathrm{b}}=\sqrt{4\pi G a_{\mathrm{b}}^2 \rho_{\mathrm{b},0}}=330$~\kms, respectively. The classical bulge does not  spin initially.
The total number of the particles is 5.1 billion, and the particle mass for each component is $178 M_{\sun}$.
\citet{2019MNRAS.482.1983F}  generated the initial condition with \texttt{Galactics} \citep{1995MNRAS.277.1341K, 2005ApJ...631..838W, 2008ApJ...679.1239W} and performed the simulation with the \texttt{Bonsai} code \citep{2012JCoPh.231.2825B, 2014hpcn.conf...54B}.
The simulation starts at $t=0$, and the bar forms at $t\sim3$~Gyr. The bar slows down from $t\sim3$~Gyr to 7~Gyr due to the angular momentum transfer from the bar to the DM halo.
The non-zero halo spin in the initial condition prevents the bar from slowing down too much and becoming too long.
The bar length and the pattern speed in the final snapshot ($t=10$~Gyr) are $\sim4$ kpc and $\sim45\,\kmskpc$, respectively.
The model gradually develops a BPX bulge without strong buckling events.

The second $N$-body model is model 2 from \citet{2020MNRAS.498.3334D}. They also generated the initial condition with \texttt{Galactics}, but the version is different from the one used in \citet{2019MNRAS.482.1983F}.
The model consists of a DM halo and a stellar disc. The DM halo follows the NFW profile with a smooth truncation:
\begin{align}
	\rho(r) &= \frac{2^{2-\gamma}\rho_{\mathrm{h,0}}}{(r/a_{\mathrm{h}})^{\gamma}(1+r/a_{\mathrm{h}})^{3-\gamma}} \times \frac{1}{2} \mathrm{erfc} \left( \frac{r-r_{\mathrm{h}}}{\delta r_{\mathrm{h}}} \right).
\end{align}	
The parameters are set as follows: $\gamma=0.873$, $a_{\mathrm{h}}=16.7$~kpc, $\sigma_{\mathrm{h}}=400$~\kms,  $r_{\mathrm{h}}=100$~kpc, and $\delta r_{\mathrm{h}}=25$~kpc. 
The disc follows the radially exponential and vertically isothermal profile as in Eq.~\ref{eq:disc_density} with  $R_{\mathrm{d}}=2.5$~kpc and $z_{\mathrm{d}}=0.3$~kpc. The total mass of the disc is $5.2 \times 10^{10} M_{\sun}$.
The central velocity dispersion and the scale radius are 128~\kms and 2.5~kpc, respectively.
The model was evolved for 5 Gyr with \texttt{pkdgrav} \citep{2001PhDT........21S}. A bar forms at $t\sim1$~Gyr and buckles at $t\sim3$~Gyr, producing a prominent BPX bulge. 

While the two galaxy models are both barred, they are several important differences.
The MWa has a classical bulge, whereas model 2 does not. The bar of model 2 strongly buckles, whereas the bar of MWa does not. The bar of model 2 is longer and thicker than that of MWa. MWa is more similar to the MW than model 2 in terms of the disc and bulge properties. More details of the models can be found in \citet{2019MNRAS.482.1983F} and \citet{2020MNRAS.498.3334D}.

In this study, we label the $N$-body particles with actions before the bar formation $\mathbf{J}_0 = (J_{\phi,0}, J_{R,0}, J_{z,0})$.
We compute the actions using \texttt{Agama} \citep{2019MNRAS.482.1525V}.
We first construct axisymmetric potential models from the initial snapshots of the $N$-body simulations, performing a multipole expansion for the DM halos and the classical bulge (in model MWa), and a cylindrical harmonic  expansion for the stellar discs.
In these model potentials, we then compute the actions of the particles with the St\"ackel fudge method \citep{2012MNRAS.426.1324B}.

In Appendix~\ref{sec:kinematic_fractionation}, we show the face-on and edge-on maps colour-coded by mean $\mathbf{J}_0$ and density distributions of different $\mathbf{J}_0$ populations in MWa.
We confirm that particles with different initial actions contribute differently to the BPX structure, consistent with model 2 (see Figures 4, 7, and 8 of \citealt{2020MNRAS.498.3334D}).

\subsection{Star-forming models}
We briefly describe two $N$-body$+$hydrodynamical simulations with star formation.
We do not directly use their results but map their chemistry to the $N$-body models.

The first model is the clumpy model of \citet{2019MNRAS.484.3476C}, which was also studied by \citet{2020MNRAS.492.4716B}, \citet{2020ApJ...891L..30A},  \citet{2023ApJ...946..118D}, and \citet{2023ApJ...953..128G}.
The initial conditions consists of a DM halo and a gas corona with  NFW profiles.
The virial radius and the virial mass of the DM halo are $r_{200}\simeq 200$~kpc and $10^{12} M_{\sun}$, respectively.
The gas corona contains 10\% of the total mass within the virial radius and has a spin of $\lambda=0.065$ \citep{2001ApJ...555..240B}.
Gas settles into a disc, cooling via metal line cooling \citep{2010MNRAS.407.1581S}.
Stars form from the gas when the density exceeds a threshold of $1$~cm$^{-3}$ and the temperature falls below 15\,000~K.
The feedback from supernovae, following the blastwave implementation of  \citet{2006MNRAS.373.1074S}, injects 10\% of the $10^{51}$~erg per supernova into the interstellar medium as thermal energy.
Feedback from asymptotic giant branch stars is also accounted for.
Gas diffusion uses  \citet{2010MNRAS.407.1581S}'s method.
The model is evolved for 10~Gyr with the $N$-body + smoothed particle hydrodynamics code \texttt{Gasoline2} \citep{2004NewA....9..137W, 2017MNRAS.471.2357W}.
The model forms clumps  within the first 4~Gyr.

The second model is described in \citet{2025MNRAS.537.1620D} and is referred to as M1\_c\_b in \citet{2021MNRAS.503.1418F}.
The initial conditions are the same as the clumpy model but with five times more particles.
It was evolved with the same star formation, cooling and diffusion prescriptions, using \texttt{Gasoline2}.
However, it features stronger supernova feedback than the clumpy model; 40\% of the $10^{51}$~erg per supernova is injected into the interstellar medium as thermal energy.
Clumps also form in this model but at a lower rate than in the clumpy model due to the stronger feedback.

\begin{figure}
	\begin{center}
		\includegraphics[width=0.95\hsize]{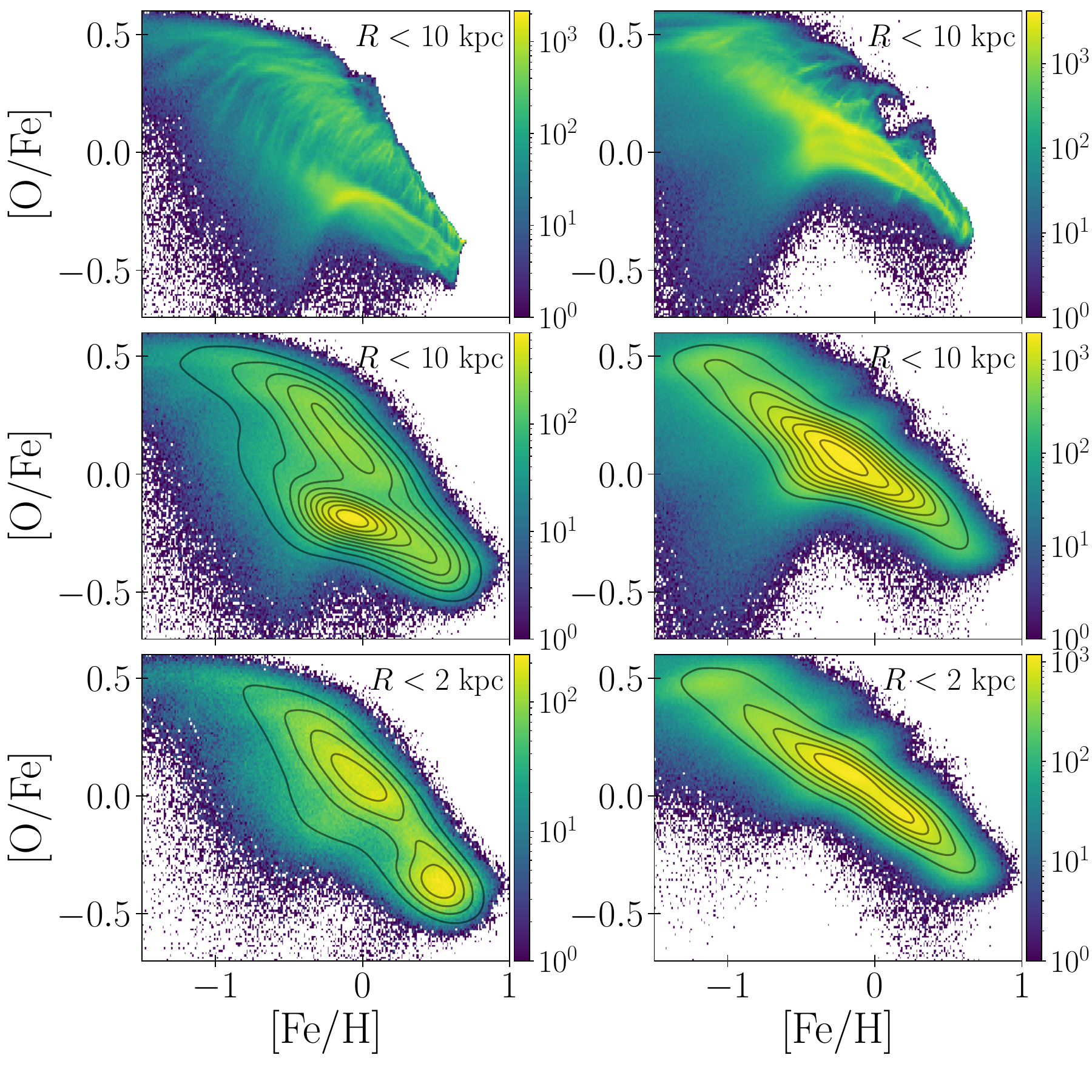}
		\caption{[Fe/H] vs [O/Fe] distribution of the clumpy model (\textit{left}) and M1\_c\_b (\textit{right}). 
			\textit{Top:} Original distributions of all star particles within 10~kpc of the galactic centre.
			\textit{Middle:} Same maps but smoothed with a Gaussian kernel to represent observational uncertainties.
			\textit{Bottom:} Smoothed chemical distributions of stars within 2~kpc of the galactic centre.
		}\label{fig:feh_ofe_donors}
	\end{center}
\end{figure}

Figure~\ref{fig:feh_ofe_donors} shows the [Fe/H] versus [O/Fe] distributions of stellar particles in the clumpy model (left) and M1\_c\_b (right).
The top panels show the original distributions for all stellar particles within 10~kpc of the galactic centre, whereas the middle panels show the same distributions smoothed with a Gaussian kernel with $\sigma_{\mathrm{[Fe/H]}} = 0.05$~dex and $\sigma_{\mathrm{[O/Fe]}} = 0.03$~dex, which correspond to typical observational errors in the APOGEE data \citep{2014ApJ...796...38N}.
Numerous fine streaks are seen in the unsmoothed maps, which trace the chemical evolution of individual clumps \citep{2019MNRAS.484.3476C, 2023ApJ...953..128G}.
Such structures are smoothed out in the middle panels, and the clumpy model and M1\_c\_b exhibit two-track and single-track distributions, respectively.
As discussed in previous studies \citep[e.g.][]{2019MNRAS.484.3476C, 2023ApJ...946..118D}, clumpy star formation produces two tracks (high-$\alpha$ and low-$\alpha$), similar to the chemical distribution of solar neighbourhood stars.
In contrast, clump formation is suppressed in M1\_c\_b due to the strong feedback, resulting in a single-track distribution.
The bottom panels show the same distributions, but restricted to within 2~kpc of the galactic centre. The clumpy model exhibits two peaks along a single track, whereas M1\_c\_b exhibits a single peak along a single track. The distribution in the clumpy model is qualitatively consistent with that of the MW bulge (see Section~\ref{sec:feh_ofe_distribution} and \citealt{2023ApJ...946..118D} for detailed discussions).

We use snapshots at $t = 6$~Gyr for both models, at which time neither model has a strong bar, and compute actions, at this time, in the same manner as for the $N$-body models. The actions are used to assign chemistry to the $N$-body particles using the method described in Section~\ref{sec:chem_assign}. In effect we are assuming that the hydrodynamical models at $t=6$~Gyr produce galaxies which are more or less equivalent to the pure $N$-body models at their $t=0$.

\subsection{Mapping chemistry to $N$-body particles}\label{sec:chem_assign}
\begin{figure}[ht!]
	\begin{center}
		\includegraphics[width=0.8\hsize]{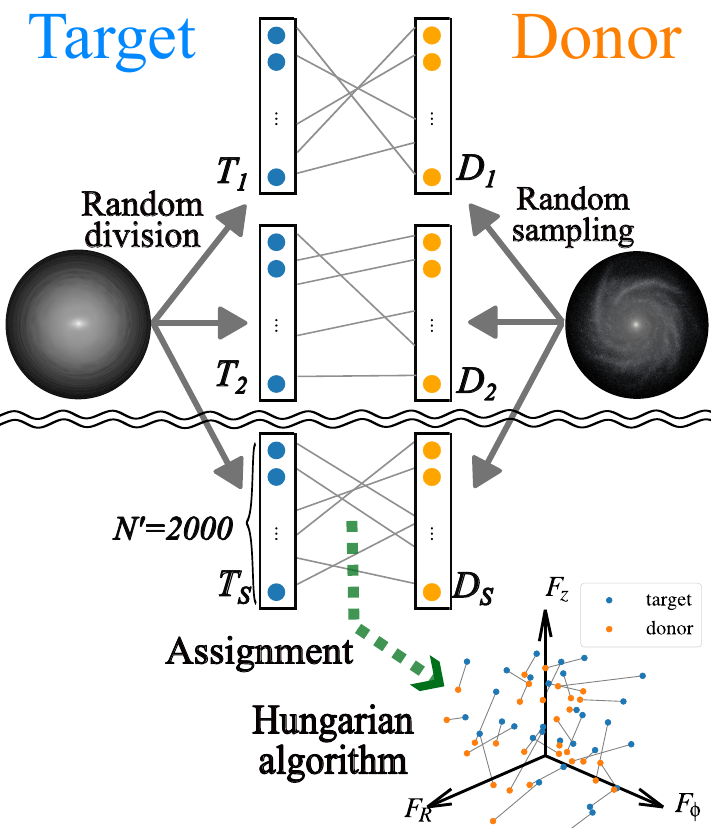}
		\caption{Schematic illustration of the assignment method. The target model is divided into subgroups: $T_1, T_2, \ldots T_S$. $D_i$ represents a set of $N' = 2000$ particles randomly selected from the donor model. For each subgroup $i$ ($i = 1, 2, \ldots S$), the assignment problem between $T_i$ and $D_i$ is solved using the Hungarian algorithm, based on the distance in CDF space.}\label{fig:illustration_hungarian}
	\end{center}
\end{figure}

\citet{2020MNRAS.498.3334D} proposed a method to map metallicity to $N$-body simulations from hydrodynamical simulations based on actions.
In real galaxies and star-forming simulations, metallicity of stars is correlated with actions \citep[e.g.][]{2024MNRAS.527.1915B}.
Azimuthal actions, vertical actions, and radial actions roughly correspond to orbital radii, vertical heights, and ages of stars, respectively.
Furthermore, stars contribute differently to the BPX structure depending on their initial actions (kinematic fractionation; \citealt{2017MNRAS.469.1587D}), therefore their method using three actions successfully reproduced the following observational trends:
(1) the dependence of the density bimodality on metallicity observed in the MW bulge \citep{2012ApJ...756...22N, 2012A&A...546A..57U, 2014A&A...569A.103R},
(2) the pinched metallicity distribution in the edge-on view of self-consistent chemo-hydrodynamical simulations \citep{2017MNRAS.469.1587D, 2019MNRAS.485.5073D, 2017MNRAS.467L..46A} and real galaxies \citep{2017MNRAS.466L..93G},
(3) the vertical metallicity gradient in the MW bulge \citep{2011A&A...534A...3G, 2013A&A...552A.110G, 2013MNRAS.432.2092N, 2017A&A...599A..12Z}.

In this study, we propose a method to assign chemical information ([Fe/H], [O/Fe]) to particles in $N$-body models (hereafter target models) based on actions, modifying \citet{2020MNRAS.498.3334D}'s method.
The chemical information is mapped from  star-forming $N$-body+hydrodynamical models (hereafter donor models).
The action-space distributions in target and donor models are generally very different, therefore we do not use the actions directly. Instead, we use the cumulative distribution functions (CDFs), $F_{\phi}$, $F_R$, and $F_z$, for $|J_{\phi}|$, $J_R$, and $J_z$. An action triplet $(J_{\phi}, J_R, J_z)$ is mapped to a point in  $[0,1] \times [0,1] \times [0,1] $ through the CDFs in individual models. 
\citet{2020MNRAS.498.3334D} first computed the averages and the standard deviations of the metallicity in bins in the  CDF space $F_{\phi} \otimes F_R \otimes F_z$ for the donor model. They then matched the binned CDFs of the target model to the donor model and assigned the metallicity from the Gaussian distribution of the metallicity in the corresponding bin.

We use a similar method but without binning the CDFs.
The following constraints are imposed on the assignment process.
First, chemistry of each target particle is mapped from a single donor particle.
Second, the assigned donor particle must be located close to the target particle in CDF space.
Third, the donor particles are selected equally; no donor particle should be assigned to more target particles than others.
This third condition is crucial to ensuring that the chemical distributions in the donor and target models remain consistent.
We can formalise this  problem as an assignment (or matching) problem as follows. Let $d_{ij}$ be the Euclidean distance between the target particle $i$ and the donor particle $j$ in CDF space. We define the cost function $C$ as 
\begin{align}
	C = \sum_{i=1}^{N} \sum_{j=1}^{N} d_{ij} x_{ij},
\end{align}
where $x_{ij}$ takes the value 1 if the donor particle $j$ is assigned to the target particle $i$, and 0 otherwise. For simplicity, we assume that the target model and the donor model have the same number of particles.  From the constraint of the perfect matching, $x_{ij}$ must satisfy the following conditions: 
\begin{align}
	\sum_{i}^{N} x_{ij} = 1 \quad \forall j \in \{1,\ldots,N \},\; 
	\sum_{j}^{N} x_{ij} = 1 \quad \forall i \in \{1,\ldots,N \}.
\end{align}
Our goal is to find the assignment $x_{ij}$ that minimises the cost function $C$. This problem is known as the linear sum assignment problem, which, in theory,  can be solved by the Hungarian algorithm \citep{kuhn1955hungarian}.
However, it is impractical to apply this algorithm directly to our task due to the high computational cost. The computational complexity of the Hungarian algorithm scales as $\mathcal{O}(N^3)$  \citep{tomizawa1971some, edmonds1972theoretical}, and even for $N=5000$, the computation of the algorithm takes $\sim 25$ seconds on a 3.3 GHz single CPU core.
If we applied the original Hungarian algorithm to our matching task, the computation does not finish within a reasonable amount of time, as our donor model has $N \gtrsim 10^6$ particles.

Even though the direct application of the Hungarian algorithm is impractical, we can still use it by dividing the target and donor models into subgroups provided each subgroup is an unbiased subsample of the full sample.
We can obtain  an approximate solution of the assignment problem by the following procedure. 
\begin{enumerate}
	\item Divide the target model  into $S$ random subgroups. Here, each group has $N'=2000$ particles\footnote{We also tested the method increasing $N'$ up to 5000, but the final results did not change except for statistical fluctuations.}, and $S=N/N'$. We refer to the $i$-th subgroup as $T_i$ ($i=1,2\ldots, S$).
	\item Select a subgroup $T_i$.
	\item Randomly sample $N'$ particles from the donor model. We refer to the group of the selected particles as $D_i$.
	\item Compute the distance matrix $d_{ij}$ between the particles in $T_i$ and $D_i$.
	\item Solve the assignment problem between $T_i$ and $D_i$ with the Hungarian algorithm.
	\item Repeat steps 2--5 for $i=1,2,\ldots,S$.
	\item Combine the results of the assignment for each subgroup and
copy the chemistry from the assigned donor particles to the target particles.
\end{enumerate}
Fig.~\ref{fig:illustration_hungarian} schematically illustrates this method. 
In step 4, we use \texttt{optimize.linear\_sum\_assignment}, an implementation of the Hungarian algorithm, from \texttt{scipy} package \citep{2020NatMe..17..261V}.
The computation takes $\sim 6$ hours for $N=2\times10^8$ particles on a 14-core 3.3 GHz CPU.

In this paper, we have two target models (MWa and model 2) and two donor models (clumpy and M1\_c\_b), resulting in four target-donor pairs.
We represent each pair as (\textit{target name}, \textit{donor name}).

\section{Results}\label{sec:results}
\subsection{Face-on and edge-on maps colour-coded by chemistry}
\begin{figure*}[ht!]
	\begin{center}
		\includegraphics[width=\hsize]{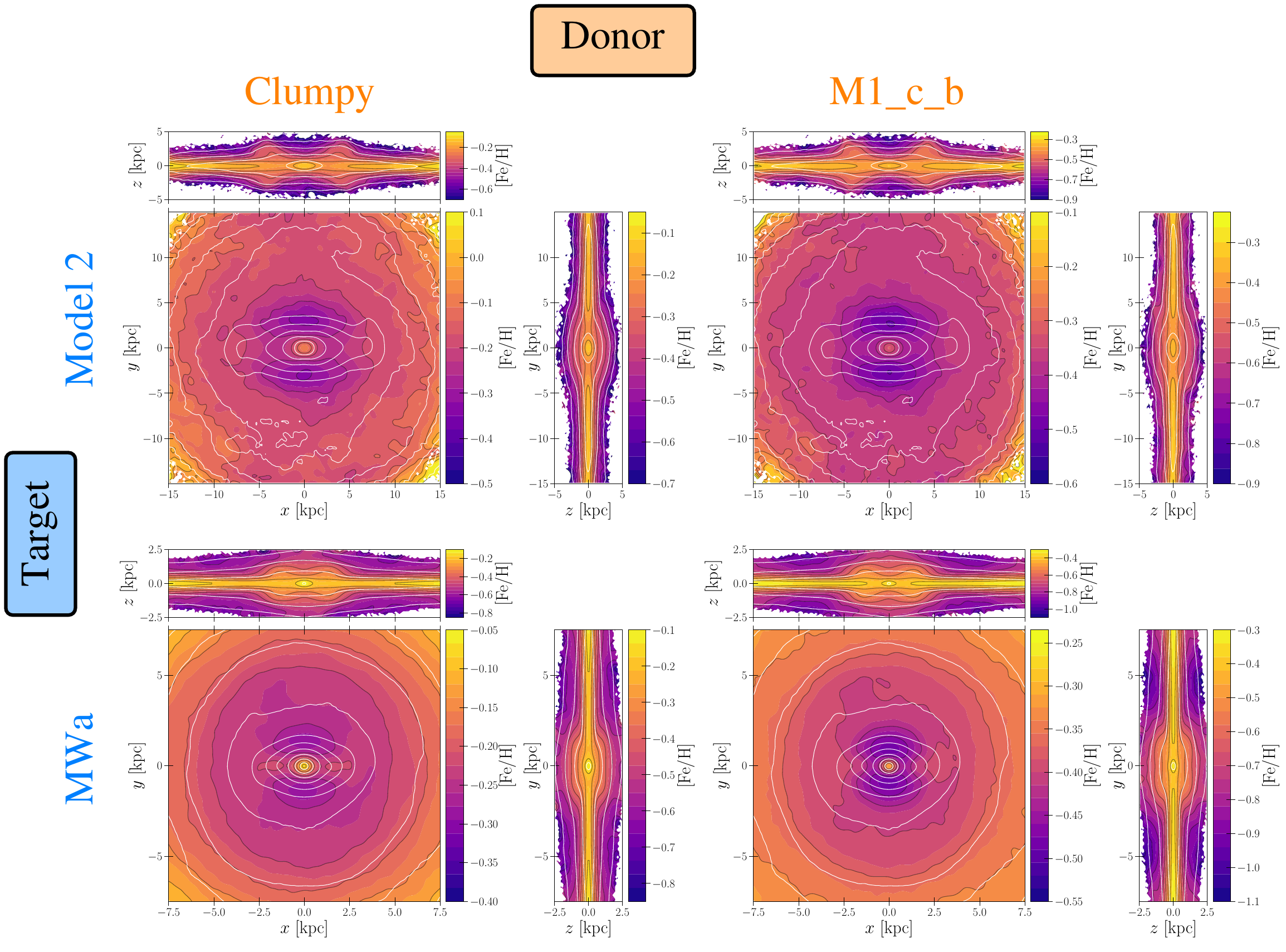}
		\caption{Face-on and edge-on maps  colour-coded by mean [Fe/H].  White and black lines represent the contours for the surface density and the metallicity, respectively. The four panels correspond to the following target–donor pairs: \textit{top left} (model 2, clumpy), \textit{top right} (model 2, M1\_c\_b), \textit{bottom left} (MWa, clumpy), and \textit{bottom right} (MWa, M1\_c\_b). } \label{fig:face_on_edge_on_feh}
	\end{center}
\end{figure*}

Fig.~\ref{fig:face_on_edge_on_feh} shows the face-on and edge-on maps of the target models colour-coded by mean [Fe/H] for the four target-donor pairs.
To obtain smooth distributions, the maps of model 2 and MWa were convolved with Gaussian filters of $0.2\times0.2$ kpc$^2$ and $0.4\times 0.4$ kpc$^2$, respectively.
The metallicity increases radially towards the outer galaxies and vertically towards the mid-plane.
There are compact metal-rich regions at the centres of the galaxies.

In the side-on views, the metallicity distributions within the bulge regions appear more pinched compared to the density distribution.
Outside the bulge regions, the metallicity contours are flat.
The metallicity assignment method  of \citet{2020MNRAS.498.3334D} also reproduced the pinching in the metallicity distribution.
Their model shows pinching in both side-on and end-on views, although the pinching is less prominent in the end-on view.
On the other hand, our model shows pinching only in the side-on views.
Our results are more consistent with  self-consistent $N$-body+hydrodynamical simulations with star formation \citep{2017MNRAS.469.1587D, 2019MNRAS.485.5073D}, in which the metallicity distributions are pinched in the side-on view but not in the end-on view.

The face-on maps show metallicity hollows on the bar's minor axis.
This feature is not observed in real galaxies but is not entirely unexpected as discussed in \citet{2020MNRAS.498.3334D}.
These regions correspond to star formation deserts \citep{2016MNRAS.457..917J, 2018MNRAS.474.3101J} in real galaxies.
Face-on metallicity maps of the MW disc reconstructed from the orbit superposition method \citep{2025A&A...695A.220K} also show low-metallicity regions on the bar's minor axis (see figure 15 of \citealt{2025A&A...700A..90K}) .

The difference of donor models little affects the face-on and edge-on metallicity maps except for the metallicity ranges, 
but we find more qualitative differences between the two target models in the edge-on maps.
In the side-on views, model 2 exhibits strongly pinched  X-shapes, whereas  MWa exhibits weakly pinched peanut-shapes. 
In the end-on views, the contours of MWa are more boxy than those of model 2.
This difference might be related to the presence of a classical bulge component in MWa.

\begin{figure*}
	\begin{center}
		\includegraphics[width=\hsize]{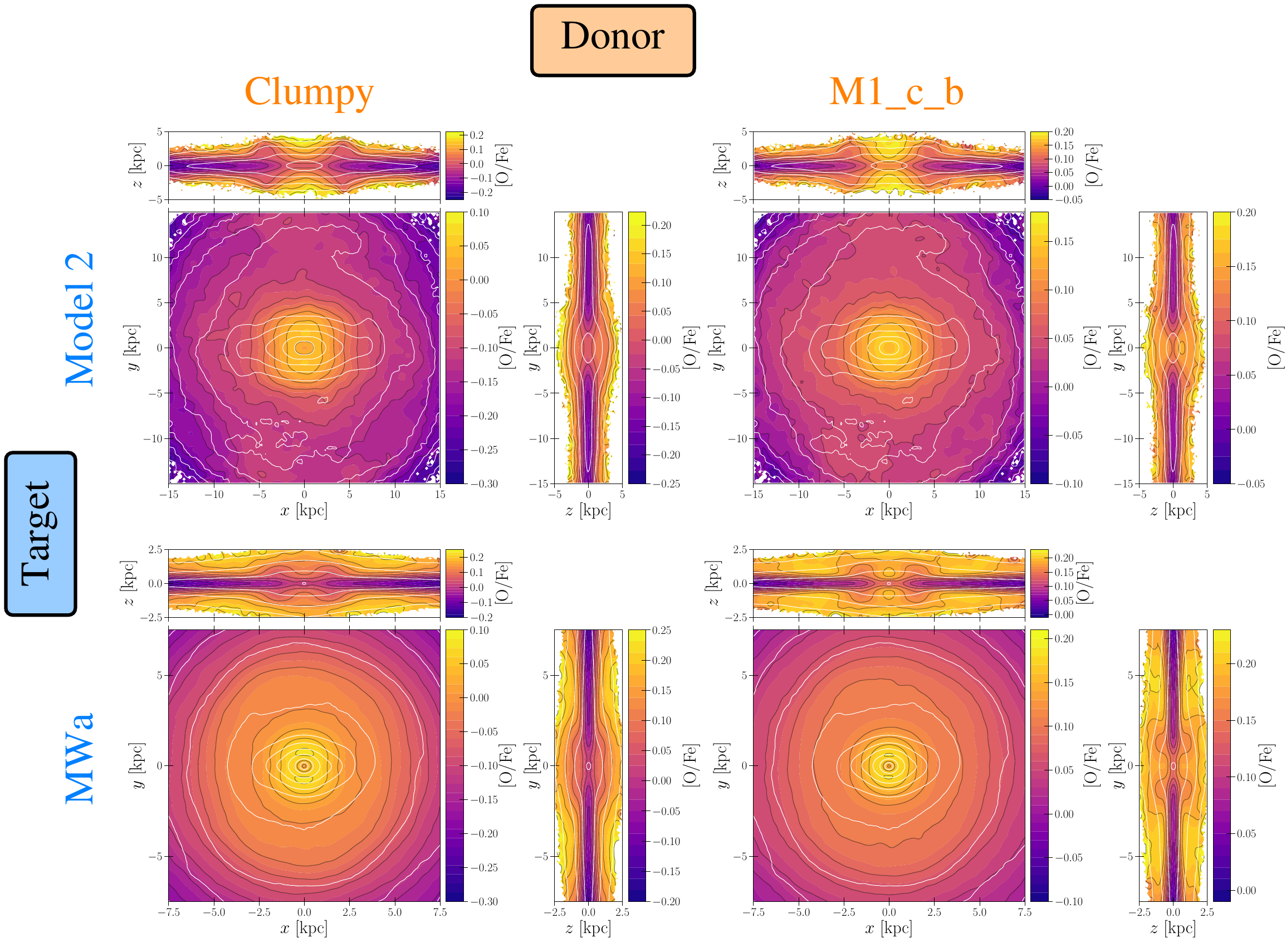}
		\caption{Same as Fig.~\ref{fig:face_on_edge_on_feh} but colour-coded by mean [O/Fe]. }\label{fig:face_on_edge_on_chem_ofe}
	\end{center}
\end{figure*}
Fig.~\ref{fig:face_on_edge_on_chem_ofe} shows the face-on and edge-on maps colour-coded by mean [O/Fe].
[O/Fe] decreases radially towards the outer galaxy and vertically towards the mid-plane. 
This opposite trend of [O/Fe] compared to [Fe/H] reflects the anti-correlation between [O/Fe] and [Fe/H] seen in Fig.~\ref{fig:feh_ofe_donors}.
In the face-on maps, the contours of [O/Fe] are elongated along the bar’s minor axis.
Similar to the [Fe/H] distribution, the [O/Fe] distribution is more pinched than the density distribution within the bulge regions.
For (model 2, clumpy) and (model 2, M1\_c\_b), the pinching is stronger in the side-on view.
For (MWa, clumpy), the [O/Fe] distribution shows a pinched peanut shape and a boxy shape in the side-on and end-on views, respectively.
For (MWa, M1\_c\_b),  we see the pinching clearly at low latitudes ($|z|\lesssim 1$~kpc). 
In the side-on view, high-[O/Fe] holes are observed at $(x, z) \sim (0, \pm1)$~kpc.
The contours appears boxy in both the side-on and end-on views at high latitudes.

To summarise, side-on maps of the galaxies, and the X- or peanut-shapes consist of high-[Fe/H] and low-[O/Fe] populations.
Such pinched metallicity distributions are seen in hydrodynamical simulations \citep{2017MNRAS.469.1587D, 2019MNRAS.485.5073D, 2017MNRAS.467L..46A} and in external galaxies \citep{2017MNRAS.466L..93G}.
Similarly, in the MW bulge, metal-rich stars exhibit a pronounced X-shaped density distribution \citep{2012ApJ...756...22N, 2013MNRAS.432.2092N, 2021A&A...653A.143W}.

\subsection{[Fe/H] versus [O/Fe] distribution}\label{sec:feh_ofe_distribution}
\begin{figure}
	\begin{center}
		\includegraphics[width=0.9\hsize]{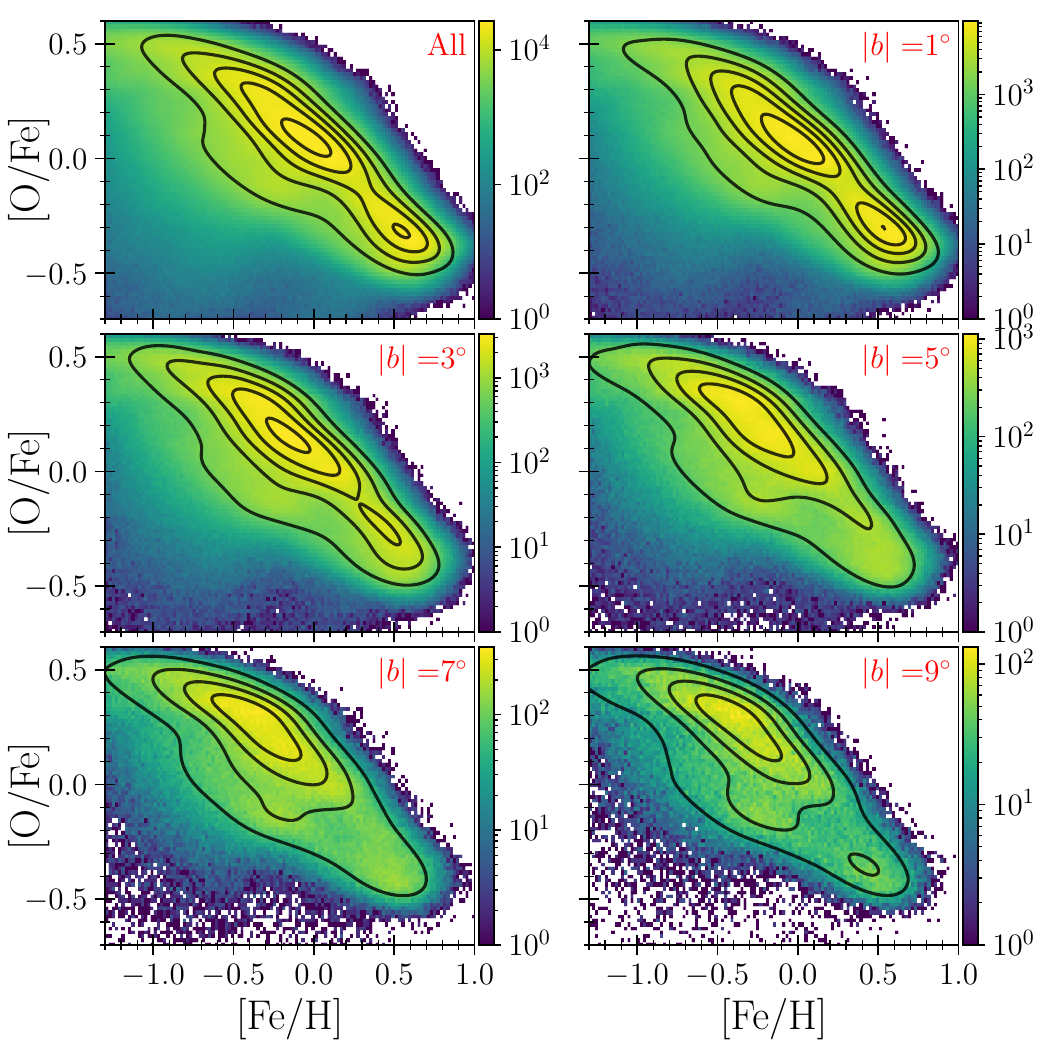}
		\caption{[Fe/H]--[O/Fe] distribution of the (MWa, clumpy). \textit{Top left panel} shows the distribution for all of the particles in the bulge region. The other panels show the distribution at five different latitude slices of $\pm 0.5^{\circ}$ centred at $|b|$ indicated in the top right corner of each panel. Contours indicate the density on a liner scale.}\label{fig:feh_ofe_MWa_run733}
	\end{center}
\end{figure}
\begin{figure}
	\begin{center}
		\includegraphics[width=0.85\hsize]{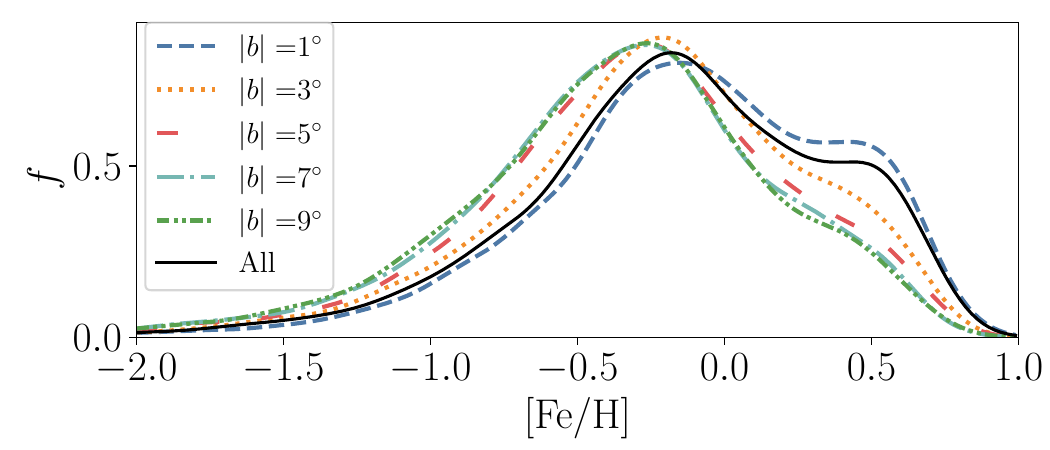}
		\includegraphics[width=0.85\hsize]{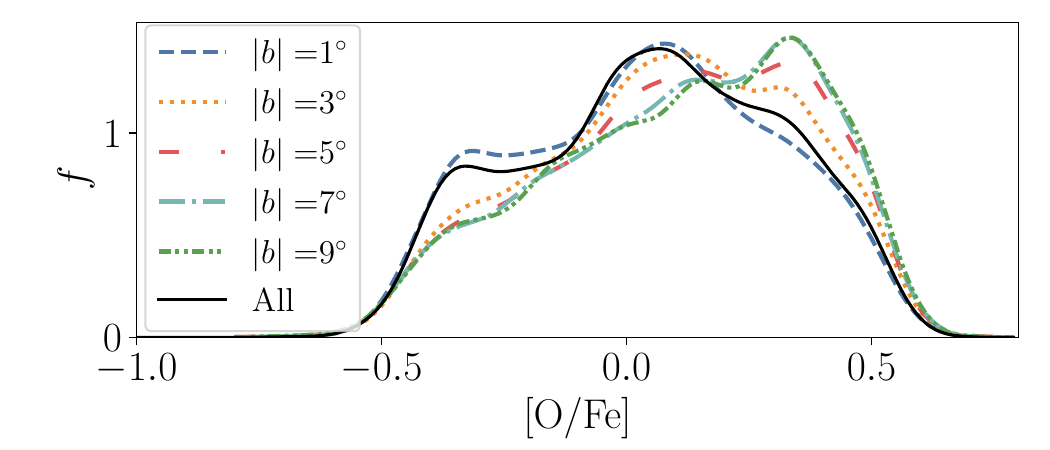}
		\caption{\textit{Top:} [Fe/H] distribution in the bulge region of (MWa, clumpy) for different latitude slices of $\pm 0.5^\circ$. The distribution is normalised at each latitude. \textit{Bottom:} Same as the top panel but for [O/Fe].}\label{fig:feh_ofe_1Dhist_MWa_run733}
	\end{center}
\end{figure}
We plot the [Fe/H]--[O/Fe] space distribution for the MWa's bulge.
We define the bulge region as $7<R_s/\mathrm{kpc}<9$ and $|l|<7^{\circ}$, where $R_s$ is the distance from the Sun and $l$ is the Galactic longitude. We assume that the distance of the `Sun' from the galactic centre is 8~kpc and that the relative angle between the Sun and the  major axis of the bar is $25^{\circ}$.
The top left panel of Fig.~\ref{fig:feh_ofe_MWa_run733} shows the distribution for all particles in the bulge region.
The chemistry is mapped from the clumpy donor model.
The  distribution  is smoothed with a Gaussian filter as in Fig.~\ref{fig:feh_ofe_donors}, to take into account observational errors.
We generated the contours with a Gaussian kernel density estimate (KDE) for 10\,000 randomly sampled particles using the \texttt{scipy} package \citep{2020NatMe..17..261V}.
The distribution follows a single track and has two peaks at [Fe/H] $\sim -0.1$ and $0.5$.
This feature is qualitatively consistent with observations of the MW bulge \citep[e.g.][]{2019A&A...626A..16R, 2020MNRAS.497.3557L, 2021A&A...656A.156Q}.

The other panels of Fig.~\ref{fig:feh_ofe_MWa_run733} show the [Fe/H]--[O/Fe] maps of the bulge at five Galactic latitude slices of $\pm 0.5^{\circ}$, centred at $|b|=1^{\circ}$, $3^{\circ}$, $5^{\circ}$, $7^{\circ}$, and $9^{\circ}$. 
The double-peak feature is clearly identified only at low latitudes.
The high-[Fe/H] peak becomes faint, and the low-[Fe/h] peak moves to lower [Fe/H] values along the track (hence higher [O/Fe]) as $|b|$ increases.
In addition, the low-$\alpha$ track becomes prominent at higher latitudes.

Figure~\ref{fig:feh_ofe_1Dhist_MWa_run733}  presents the 1D distributions of [Fe/H] and [O/Fe] derived using a Gaussian KDE.
The solid line shows the distribution for all particles within the bulge region ($7 < R_s/\mathrm{kpc} < 9$ and $|l| < 7^\circ$), while the other lines correspond to different latitude bins.
The [Fe/H] distribution at $|b| = 1^\circ$, as well as the combined distribution for all latitudes, exhibits a bimodal structure with a prominent peak at [Fe/H] $\sim -0.1$ and a secondary peak at $\sim0.5$.
As $|b|$ increases, the high-[Fe/H] peak gradually weakens.
In contrast, the low-[Fe/H] peak shifts towards lower [Fe/H] with increasing latitude, reaching [Fe/H] $\sim -0.3$ at $|b| = 9^\circ$.
The [O/Fe] distribution shows more complex latitude-dependent behaviour than the [Fe/H] distribution.
At $|b| = 1^\circ$, it displays two peaks at [O/Fe] $\sim -0.3$ and 0.
As illustrated in the [Fe/H]--[O/Fe] maps in Fig.~\ref{fig:feh_ofe_MWa_run733}, three effects drive the variation of the [O/Fe] distribution with latitude.
First, the low-[O/Fe] peak, which corresponds to the high-[Fe/H] peak in the top panel, decreases with increasing $|b|$.
Second, the peak of the high-$\alpha$ sequence shifts towards higher [O/Fe] values as $|b|$ increases.
Third, the low-$\alpha$ sequence becomes progressively more prominent.
As a result, at $|b| \ge 3^\circ$, the distribution develops two peaks at [O/Fe] $\sim 0.1$ and $\sim0.35$.
The low-$\alpha$ sequence primarily contributes to the peak at [O/Fe] $\sim 0.1$, whereas the contribution of the high-$\alpha$ sequence to the peak at [O/Fe] $\sim 0.35$ increases with latitude.
At $|b| = 7^\circ$ and $9^\circ$, the distribution becomes effectively single-peaked at [O/Fe] $\sim 0.35$, with a gradual decline towards lower [O/Fe] values.
In the combined distributions, three peaks are present due to these three effects.

\begin{figure}
	\begin{center}
		\includegraphics[width=0.9\hsize]{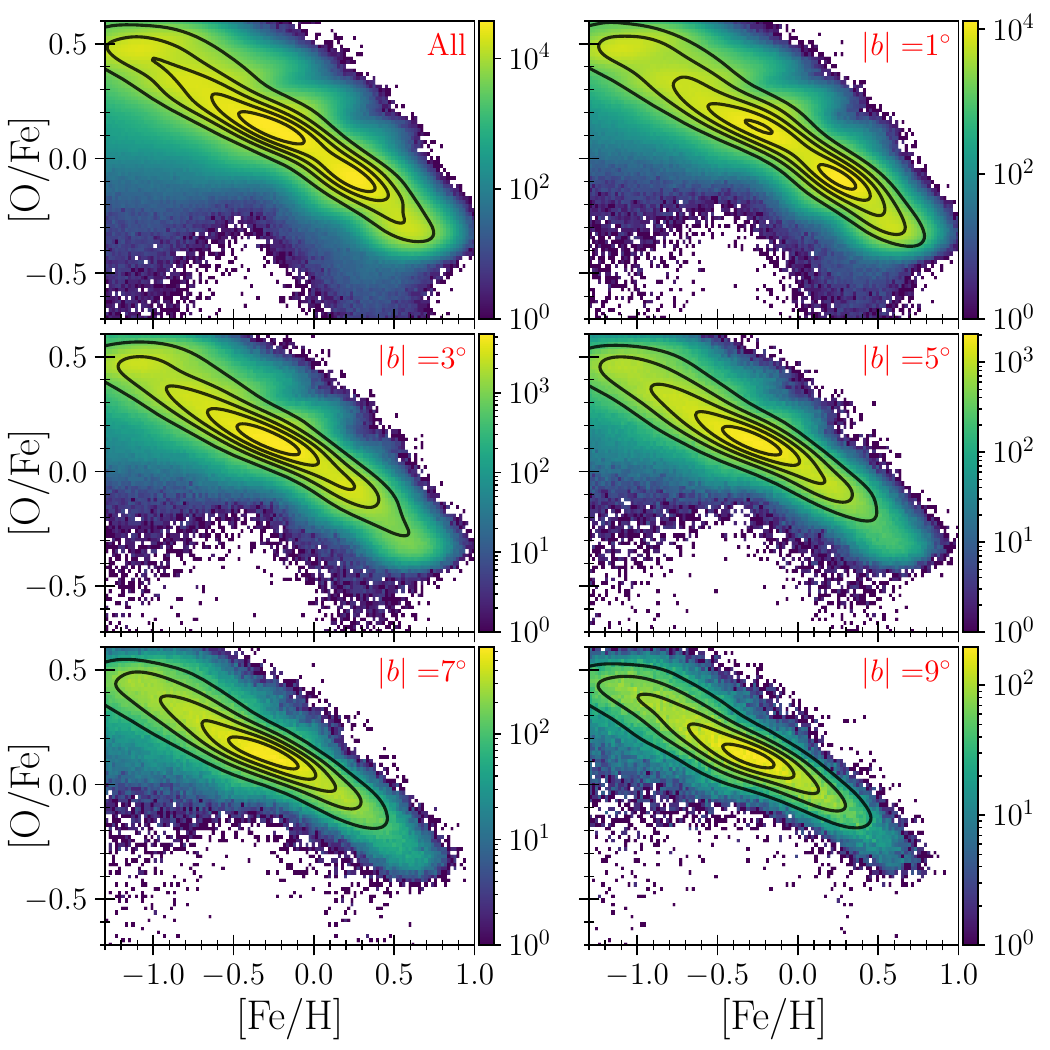}
		\caption{Same as Fig.~\ref{fig:feh_ofe_MWa_run733} but for (MWa, M1\_c\_b).}\label{fig:feh_ofe_MWa_run739}
	\end{center}
\end{figure}
\begin{figure}
	\begin{center}
		\includegraphics[width=0.85\hsize]{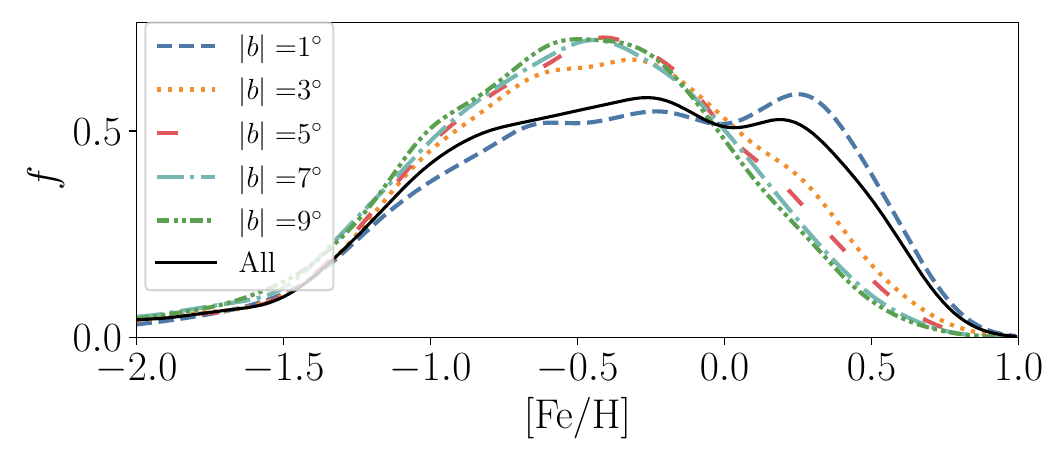}
		\includegraphics[width=0.85\hsize]{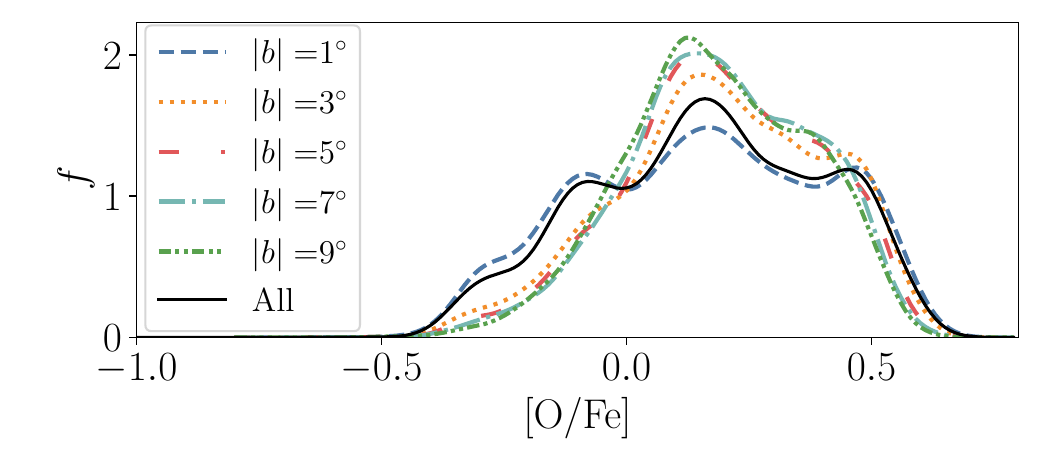}
		\caption{Same as Fig.~\ref{fig:feh_ofe_1Dhist_MWa_run733} but for (MWa, M1\_c\_b).}\label{fig:feh_ofe_1Dhist_MWa_run739}
	\end{center}
\end{figure}
Figure~\ref{fig:feh_ofe_MWa_run739} shows the [Fe/H]--[O/Fe] maps for (MWa, M1\_c\_b).
The [Fe/H]--[O/Fe] distribution of the entire bulge region (top left panel of fig~\ref{fig:feh_ofe_MWa_run739}) follows a single track with a peak at [Fe/H] $\sim-0.4$.
This contrasts with the  chemical distributions in the MW bulge and (MWa, clumpy), which show two distinct peaks.
At $|b|=1^{\circ}$, the [Fe/H]--[O/Fe] distribution exhibits two peaks, but the bimodality is less prominent compared with (MWa, clumpy).
These arise from individual clumps that are not completely smoothed out, even after Gaussian filtering. The track in the [Fe/H]–[O/Fe] plane becomes shorter as $|b|$ increases, but its overall shape does not change significantly.
The one-dimensional [Fe/H] and [O/Fe] distributions are shown in Fig.~\ref{fig:feh_ofe_1Dhist_MWa_run739}. 
At $|b| = 1^\circ$ and in the combined distribution across all latitudes, the [Fe/H] distribution shows weak bimodality.
At higher latitudes, the [Fe/H] distributions show a single peak and little variation with $|b|$.
In the [O/Fe] distribution, several small peaks are seen at $|b| = 1^\circ$ and in the combined distribution, but only a single peak is observed at higher latitudes.

The clear bimodality in the [Fe/H]--[O/Fe] distribution and its variation with latitude seen in (MWa, clumpy) are qualitatively consistent with observations of the MW bulge \citep[e.g.][]{2014A&A...569A.103R, 2019A&A...626A..16R, 2016PASA...33...22N, 2021A&A...656A.156Q}.
The same trend is found when model 2 is used as the target model (see Appendix~\ref{sec:supplementary_figures}).
In the following sections, we mainly focus on the clumpy donor model, as it reproduces the observed chemical properties of the MW bulge better than M1\_c\_b.

\subsection{X-shape bulge of the MW}

The X-shaped structure of the MW bulge is identified as a bimodal distribution in the distances of red clump stars \citep{2010ApJ...724.1491M, 2010ApJ...721L..28N, 2015MNRAS.447.1535N, 2011AJ....142...76S,  2013MNRAS.435.1874W, 2015A&A...583L...5G}.
The bimodality is stronger in metal-rich samples than in metal-poor samples \citep{2012ApJ...756...22N, 2012A&A...546A..57U, 2014A&A...569A.103R}.
The metallicity or age dependence of the bulge morphology was observed by other stellar tracers \citep{2022MNRAS.517.6060S, 2022MNRAS.509.4532S}.
We investigate how the apparent magnitude (or distance) distribution of bulge stars varies with metallicity in (model 2, clumpy) and (MWa, clumpy).
We measure the distribution of red clump stars in the bulge region of model 2 and MWa, assuming that all particles have an absolute magnitude of $M_K=-1.61$~mag. 
We rescale the length in model 2  by a factor of $5/9$ since the bar of model 2 is larger than that of the MW.
We convolve the distribution with a Gaussian kernel with a standard deviation of 0.17~mag, taking into account the intrinsic dispersion of the red clump magnitude \citep{2012ApJ...744L...8G}.

\begin{figure}
	\begin{center}
			\includegraphics[width=0.9\hsize]{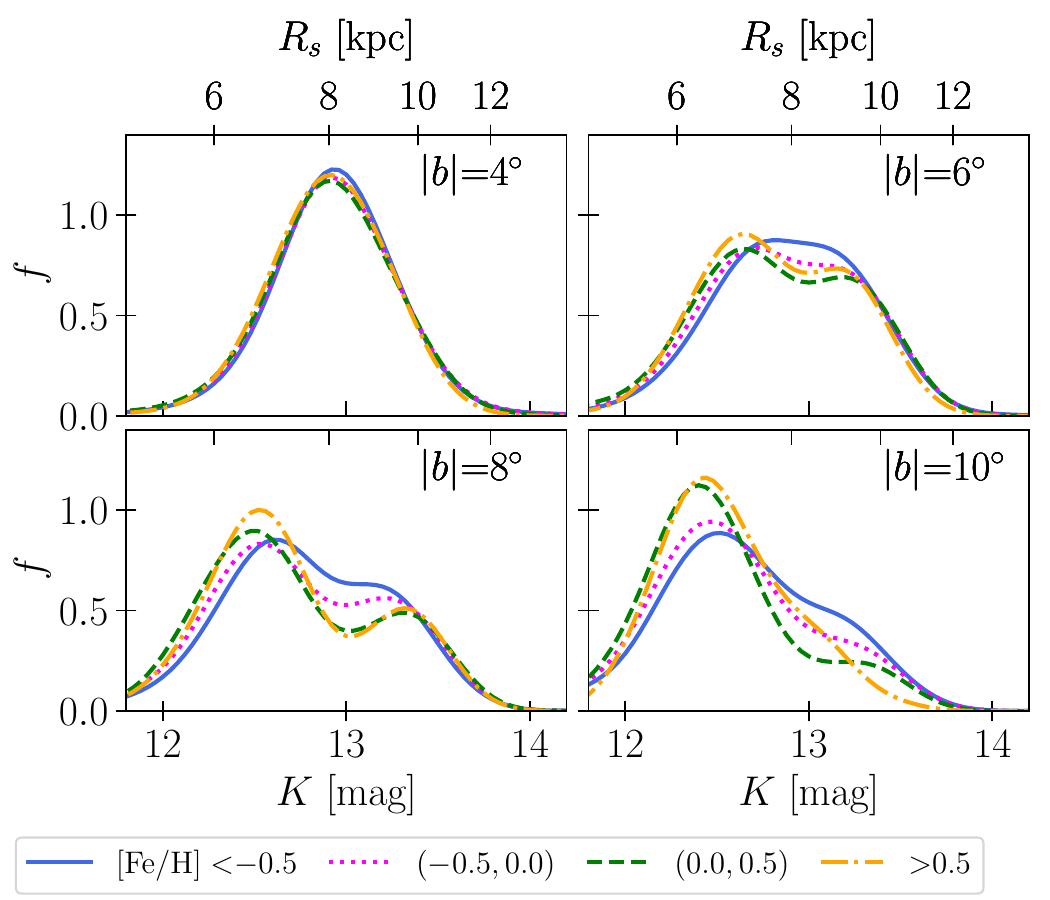}
			\caption{$K$-band magnitude distribution of red clump stars in the bulge region of (model 2, clumpy) at four different latitudes slices of $\pm 0.5^{\circ}$ centred at $|b|=4^{\circ}$ (\textit{top left}), $6^{\circ}$ (\textit{top right}), $8^{\circ}$ (\textit{bottom left}), and $10^{\circ}$ (\textit{bottom right}). The distribution is shown for different metallicity bins: [Fe/H] $<-0.5$ (blue solid), $-0.5<$ [Fe/H] $<0$ (pink dotted), $0<$ [Fe/H] $<0.5$ (green dashed), and [Fe/H] $>0.5$ (orange dash-dotted). The distribution is normalised at each latitude and metallicity bin.
            }\label{fig:Kmag_run741_run733}
		\end{center}
\end{figure}

Figure~\ref{fig:Kmag_run741_run733} shows the $K$-band magnitude distribution of red clump stars in the bulge region for (model 2, clumpy).
The distributions are normalised within each latitude.
At $|b|=4^{\circ}$, the distribution is symmetric and single-peaked across all metallicity bins.
At $|b|=6^{\circ}$, the distribution is weakly bimodal for [Fe/H] $>-0.5$, with the bimodality being more pronounced in the metal-rich population compared to the metal-poor population.
For the lowest metallicity bin ([Fe/H] $<-0.5$), the bimodality is not clearly seen, but the distribution is more broadened compared to that at $|b|=4^{\circ}$.
At $|b|=8^{\circ}$, the bimodality is more distinct than at $|b|=6^{\circ}$, and the peak on the bright side is higher than that on the faint side for all metallicity bins.
This is because the line of sight crosses lower altitude (i.e. higher density regions) of the X-shape on the near (bright) side than on the far (faint) side.
The bimodality and the asymmetry are more pronounced for metal-richer populations.
Finally, at $|b|=10^{\circ}$, the asymmetry becomes more evident, while the bimodality is less clear.
The distribution is more skewed to the bright side for metal-richer populations.

\begin{figure}
	\begin{center}
			\includegraphics[width=0.9\hsize]{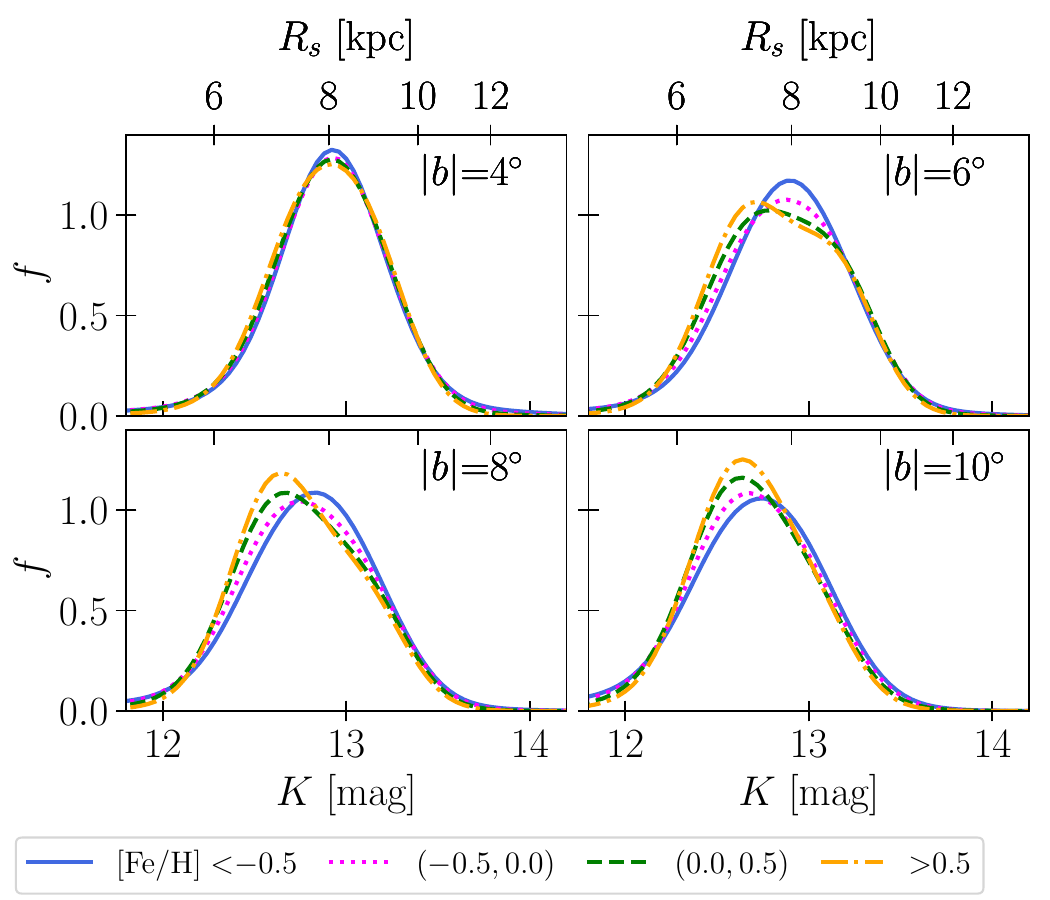}
			\caption{Same as Fig.~\ref{fig:Kmag_run741_run733} but for (MWa, clumpy).}\label{fig:Kmag_MWa_run733}
		\end{center}
\end{figure}

Figure~\ref{fig:Kmag_MWa_run733} shows the same plot for (MWa, clumpy).
In contrast to model 2, MWa does not display a clear bimodality in any metallicity bin at any latitude.
In this case, the Gaussian convolution smooths out the intrinsic bimodality in the distance distribution.
For comparison, the unconvolved distribution is presented in Fig.~\ref{fig:Kmag_MWa_run733_unconvolved}, which reveals a bimodality among the metal-rich populations.
The absence of such bimodality in the mock $K$-band magnitude distribution suggests that the MWa model has a weaker X-shaped structure than the actual MW.
One of the reasons for the weak X-shape in MWa is the presence of a classical bulge component, which exhibits a more spherical morphology and is located within the gap between the two arms of the X-shape.
The MWa model hosts a relatively massive classical bulge (15\% of the disc mass), and its influence on the total density distribution is not negligible.

Although a distinct bimodality is absent, the shape of the MWa distribution still varies with both latitude and metallicity, showing trends similar to those in model 2.
At $|b| = 4^\circ$, the distribution remains symmetric and single-peaked across all metallicity bins.
At $|b| = 6^\circ$, it becomes slightly broader and flatter for the high-metallicity bins ($0<$ [Fe/H] $<0.5$ and [Fe/H] $>0.5$) near the central region ($K \sim 12.9$~mag).
As $|b|$ increases further, the distributions become more skewed towards the bright side, with the asymmetry being more pronounced in the metal-rich populations.

\subsection{Bulge kinematics}
\begin{figure*}
	\begin{center}
		\includegraphics[width=0.9\hsize]{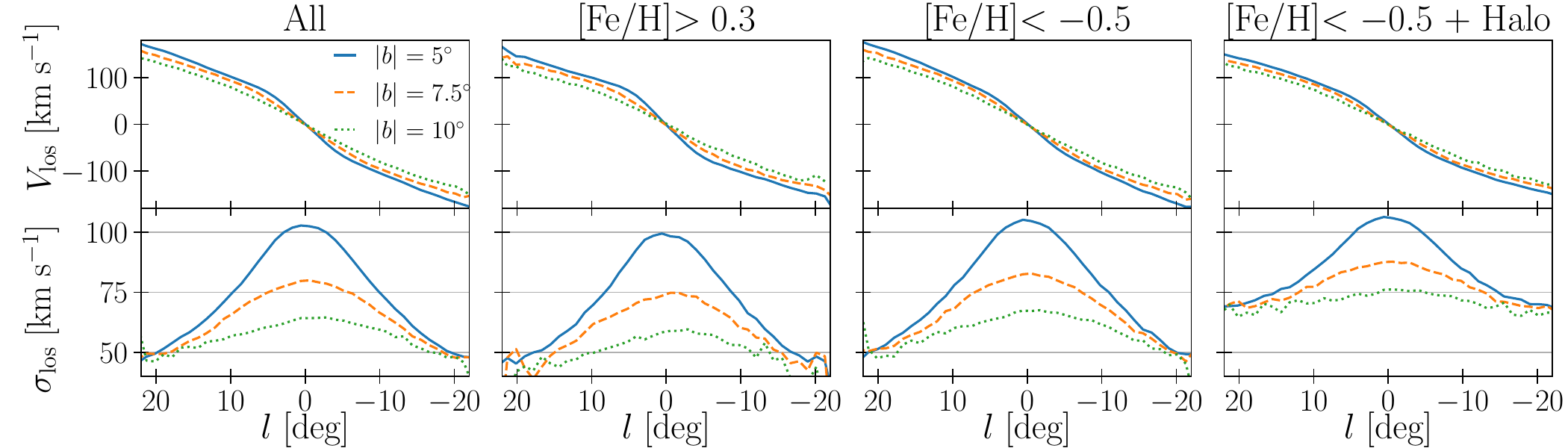}
		\caption{Mean line-of-sight velocity (\textit{top row}) and velocity dispersion (\textit{bottom row}) as functions of Galactic longitude in MWa. Solid, dashed and dotted lines correspond to $|b|=5^{\circ}$, $7.5^{\circ}$, and $10^{\circ}$, respectively. 
		The first, second, and third columns show the kinematics for all particles, metal-rich particles ($[\mathrm{Fe/H}]>0.3$), and metal-poor particles ($\mathrm{[Fe/H]}<-0.5$), respectively. The fourth column shows the kinematics for the metal-poor particles  additionally including a hot population corresponding to the stellar halo (see text for details).
		}\label{fig:bulge_kinematics_MWa_run733}
	\end{center}
\end{figure*}

The top and bottom panels in the first column of Fig.~\ref{fig:bulge_kinematics_MWa_run733} show the mean line-of-sight velocity and the velocity dispersion, respectively in the bulge region ($R<3$~kpc) of (MWa, clumpy) (see Appendix~\ref{sec:supplementary_figures} for the same plot for model 2).
As shown in \citet{2019MNRAS.482.1983F}, they fit the observational data of the BRAVA survey \citep{2007ApJ...658L..29R, 2012AJ....143...57K} well.
The line-of-sight velocity does not differ much across latitudes, which indicates cylindrical rotation of the bulge.
Such cylindrical rotation of BPX bulges is observed in the MW \citep[e.g.][]{2008ApJ...688.1060H, 2012AJ....143...57K, 2013MNRAS.432.2092N, 2016ApJ...819....2N, 2014A&A...562A..66Z, 2021A&A...653A.143W}, external galaxies \citep[e.g.][]{2006MNRAS.369..529F, 2016A&A...591A...7G} and  $N$-body simulations \citep[e.g.][]{1990A&A...233...82C, 2002MNRAS.330...35A, 2013MNRAS.430.2039S}.
The velocity dispersion is peaked at $l=0^{\circ}$ and decreases as $|l|$ increases. The peak is higher at lower latitudes, and the profile is flatter at higher latitudes. 

The second and third columns in Fig.~\ref{fig:bulge_kinematics_MWa_run733} show the bulge kinematics for the metal-rich ($[\mathrm{Fe/H}]>0.3$) and metal-poor ($\mathrm{[Fe/H]}<-0.5$) populations, respectively.
The upper and lower boundaries of the two metallicity bins roughly correspond to the 20th and 80th percentiles of the metallicity distribution in the bulge region, respectively.
There is no significant difference in the mean line-of-sight velocity between the metal-rich and metal-poor populations.
This is consistent with observations of the MW bulge \citep{2013MNRAS.432.2092N}.

The velocity dispersion of the metal-poor population is slightly ($\sim5\text{--}10\;\mathrm{km \, s^{-1}}$) higher than that of the metal-rich population, but the overall profile is similar for both populations.
In the MW bulge, however, the velocity dispersion profile varies strongly with metallicity.
In the ARGOS data \citep{2013MNRAS.432.2092N}, the metal-poor population exhibits a higher velocity dispersion and a flatter profile than the metal-rich population, which is inconsistent with our model.
One possible reason of this discrepancy is that our model does not include a metal-poor population originating from the stellar halo.
To assess its impact on the bulge kinematics, following \citet{2017MNRAS.469.1587D}, we introduced a hot population corresponding to component D of \citet{2013MNRAS.430..836N} into the model.
The fourth column of Fig.~\ref{fig:bulge_kinematics_MWa_run733} shows the bulge kinematics for the metal-poor population with 15\% contamination from this hot component, whose rotation speed is 50\% of that of the metal-rich population and whose velocity dispersion is $120\;\mathrm{km \,s^{-1}}$.
The velocity dispersion increases, particularly at higher longitudes, and the profile becomes flatter across all latitudes. This trend is now qualitatively more consistent with the observations.

However, the relatively weak central peak in the velocity dispersion of the metal-rich population still suggests a limitation of our method. The classical bulge is not the cause of this issue because model 2 also presents the same behaviour (see Fig.~\ref{fig:bulge_kinematics_run741_run733}). 
Further investigation and improvements to the modelling are needed to understand and resolve this discrepancy.

\section{Comparison with GIBS observation}\label{sec:comparison_with_GIBS}
\begin{figure*}
    \centering
    \includegraphics[width=0.95\hsize]{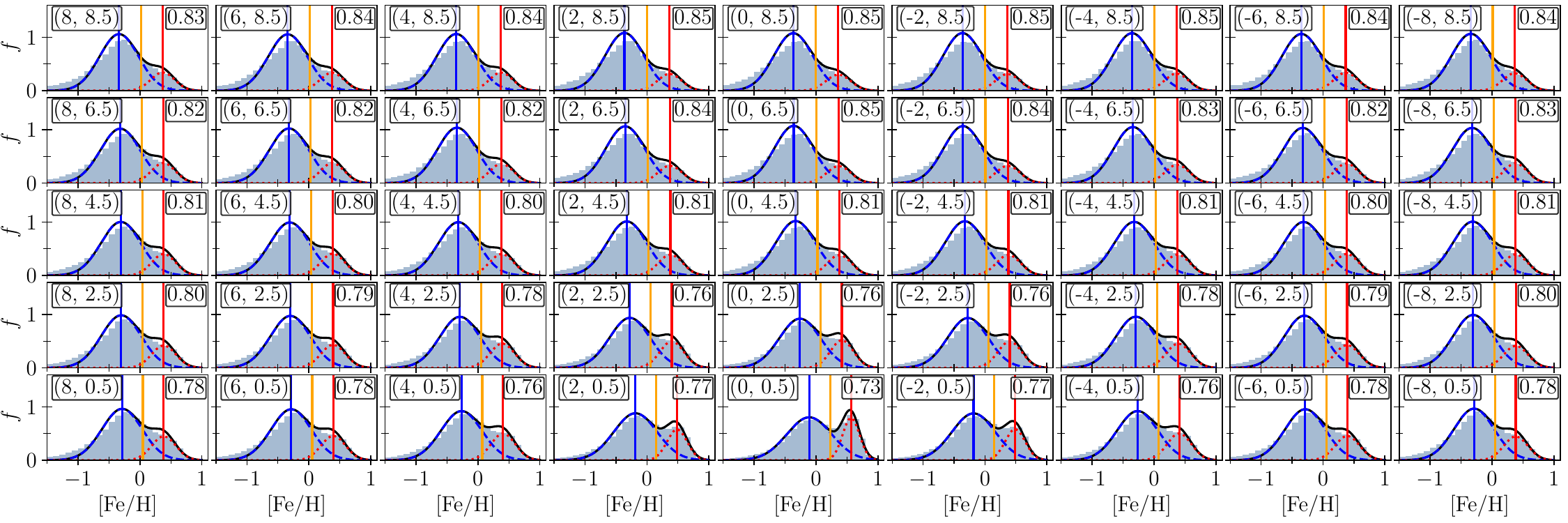}
    \caption{Metallicity distribution binned by $l$ and $|b|$ for (MWa, clumpy). The numbers in the top left corner of each panel indicate the values of $l$ and $|b|$. The blue-grey histogram represents the normalised metallicity distribution within each $l$-$|b|$ bin. The blue dashed, red dotted and black solid curves show the distributions of the metal-poor, metal-rich populations and their combined distribution as estimated by the GMM, respectively.  Blue, red and orange vertical lines mark the peaks of the metal-poor, metal-rich Gaussian components and the midpoint between them, respectively. The number in the top right corner  indicates the fraction of the metal-poor population.}\label{fig:feh_hist_lb_MWa_run733}
\end{figure*}

In this section, we compare the metallicity distribution in the bulge region of our models with that of the MW from the GIBS survey \citep{2014A&A...562A..66Z, 2017A&A...599A..12Z, 2015A&A...584A..46G}.
Fig.~\ref{fig:feh_hist_lb_MWa_run733} shows the metallicity distribution at $9 \times 5$ points in the $l$-$|b|$ plane for (MWa, clumpy).
In each panel, the blue-grey histogram represents the normalised metallicity distribution for particles within $7 < R_s/\mathrm{kpc} < 9$ and within a $1^\circ \times 1^\circ$ bin centred at the $(l, |b|)$ values noted in the top left corner of the panel.
We fit the metallicity distribution using a two-component Gaussian mixture model (GMM), as done in \citet{2017A&A...599A..12Z}.
The black curve represents the fitted distribution, while the blue dashed and red dotted curves represent the metal-poor and metal-rich Gaussian components, respectively.
The number in the top right corner of each panel indicates the fraction of the metal-poor component.

The bimodality is most pronounced in the low-$|l|$ and low-$|b|$ bins.
As $|l|$ and $|b|$ increase, the metal-poor fraction rises, and the bimodality weakens.
The blue and red vertical lines indicate the peaks of the metal-poor and metal-rich Gaussian components, respectively.
The orange vertical line marks the midpoint between these peaks and roughly corresponds to the minimum of the bimodal distribution.
The locations of the metal-poor and metal-rich peaks are almost the same in all panels, except for those with $|l| \lesssim 2^\circ$ and $|b| \lesssim 2.5^\circ$.
In the central region, both peaks shift towards the metal-rich side.

\begin{figure}
    \centering
    \includegraphics[width=0.95\hsize]{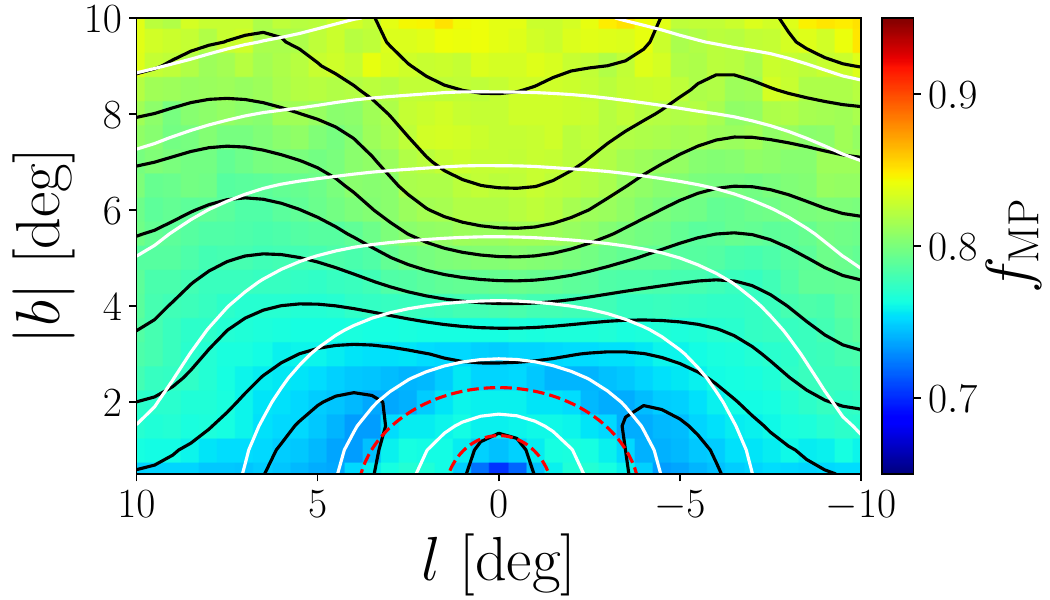}
    \caption{Metal-poor fraction as a function of $l$ and $|b|$ for (MWa, clumpy). The colour represents the fraction of the metal-poor population estimated by the GMM. White and black lines indicate the contours of density and metal-poor fraction, respectively.}\label{fig:feh_mp_lb_MWa_run733}
    \label{fig:enter-label}
\end{figure}
Fig.~\ref{fig:feh_mp_lb_MWa_run733} shows the metal-poor fraction ($f_\mathrm{MP}$) as a function of $l$ and $|b|$.
The contours of the metal-poor fraction (black lines) were obtained by smoothing the $f_\mathrm{MP}$ map with a Gaussian kernel with $\sigma_l = 1^\circ$ and $\sigma_{|b|} = 1^\circ$.
The distribution of  $f_{\mathrm{MP}}$ is strongly pinched compared with the density distribution (white contours).
This pinching resembles what is observed in the edge-on maps in Fig.~\ref{fig:face_on_edge_on_feh}. 
Both figures indicate that the metal-rich (metal-poor) population contributes more (less) to the X-shaped structure.
In Fig.~\ref{fig:feh_mp_lb_MWa_run733}, we see a weak asymmetry in the metal-poor fraction about $l = 0^\circ$ (the metal-poor fraction is higher at $l < 0^\circ$ than at $l > 0^\circ$ at $|b| \gtrsim 5^\circ$) due to projection effects.

\begin{figure*}
    \centering
    \includegraphics[width=0.95\hsize]{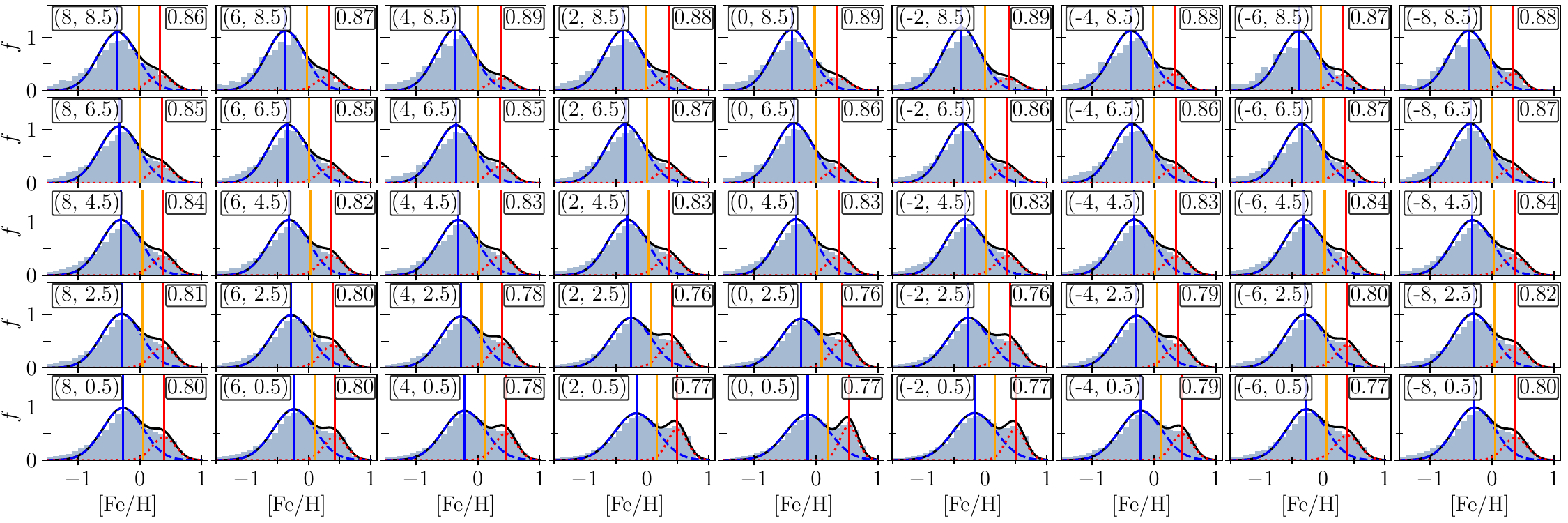}
    \caption{Same as Fig.~\ref{fig:feh_hist_lb_MWa_run733} but for (model 2, clumpy).}\label{fig:feh_hist_lb_run741_run733}
\end{figure*}
\begin{figure}
    \centering
    \includegraphics[width=0.95\hsize]{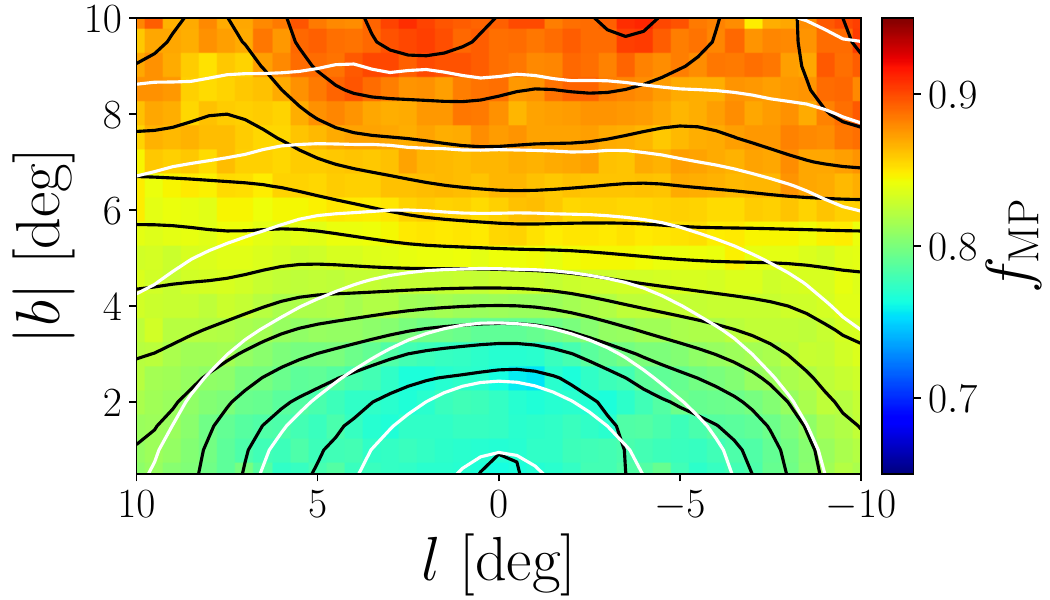}
    \caption{Same as Fig.~\ref{fig:feh_mp_lb_MWa_run733} but for (model 2, clumpy).}\label{fig:feh_mp_lb_run741_run733}
\end{figure}

Figs.~\ref{fig:feh_hist_lb_run741_run733} and \ref{fig:feh_mp_lb_run741_run733} are analogous to Figs.~\ref{fig:feh_hist_lb_MWa_run733} and \ref{fig:feh_mp_lb_MWa_run733}, but here we show the results for (model 2, clumpy).
In this model, we adopted a bin size of $2^\circ\times2^\circ$ to ensure a sufficient number of particles in each bin and rescaled the length by a factor of $5/9$ to match the bar length of MW.
Overall trends are similar to those in (MWa, clumpy): the metal-poor fraction rises with increasing $|l|$ and $|b|$, and the peaks of the two Gaussian components shift towards higher [Fe/H] only in the central region.
In the $f_\mathrm{MP}$ map (Fig.~\ref{fig:feh_mp_lb_run741_run733}), the contours of $f_\mathrm{MP}$ appear more boxy than the density contours and show pinching at high latitudes.

General trends of the metallicity distributions of the two models are qualitatively consistent with those derived from the GIBS survey data: the metallicity distribution of the bulge is bimodal, and the relative fraction of the metal-poor component is higher in the outer bulge than in the inner bulge.
However, detailed $|b|$- (and $l$-) dependence on the metal-poor fraction differs among the models and the GIBS data in the central region.
In the data, the metal-poor fraction decreases from high to intermediate latitudes, reaching a minimum at $|b|\sim3.5^\circ$, and then increases again towards the Galactic plane.
In (model 2, clumpy), the metal-poor fraction decreases almost monotonically towards the centre. 
On the other hand, (MWa, clumpy) shows non-monotonic dependence on $|b|$ and $l$.
Its metal-poor fraction decreases from the outer bulge inward, exhibits a weak upturn at $|b| \lesssim 2^\circ$ in the region highlighted by the red dashed lines in Fig.~\ref{fig:feh_mp_lb_MWa_run733}, and then decreases again.\footnote{\citet{2017A&A...599A..12Z} presented the latitude dependence of the metal-poor fraction by co-adding data along the longitude  (see their Figure 7).
	We also checked the longitude-averaged metal-poor fraction in our models, but we did not observe a clear re-increase towards the mid-plane.
}
Possible causes of these discrepancies include the presence of a classical bulge, differences in BPX strength, or different BPX structure formation mechanisms, but the exact reason is not clear at this stage and requires further investigation.

\begin{figure}
		\centering
		\includegraphics[width=0.8\linewidth]{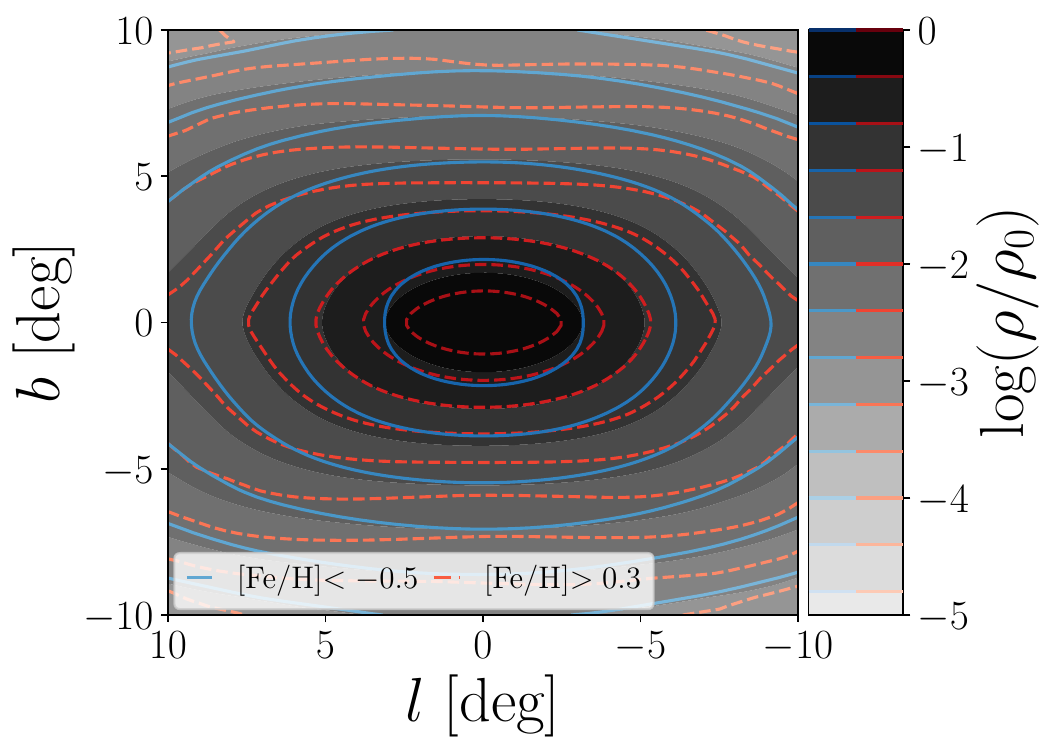}
		\includegraphics[width=0.49\linewidth]{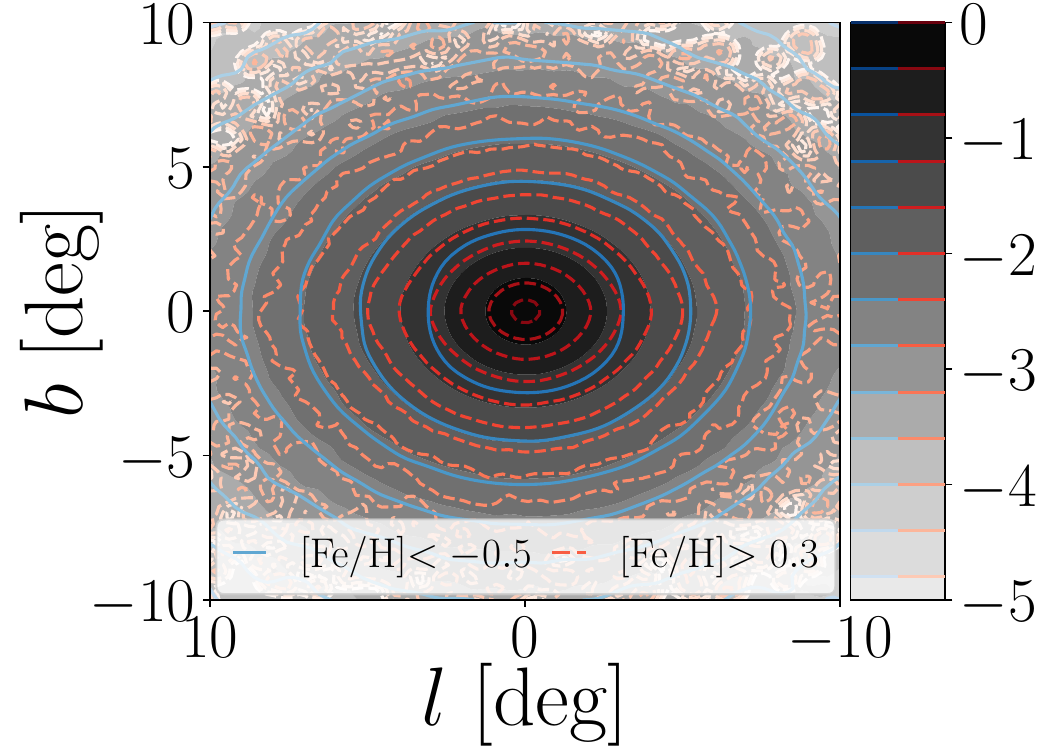}
		\includegraphics[width=0.49\linewidth]{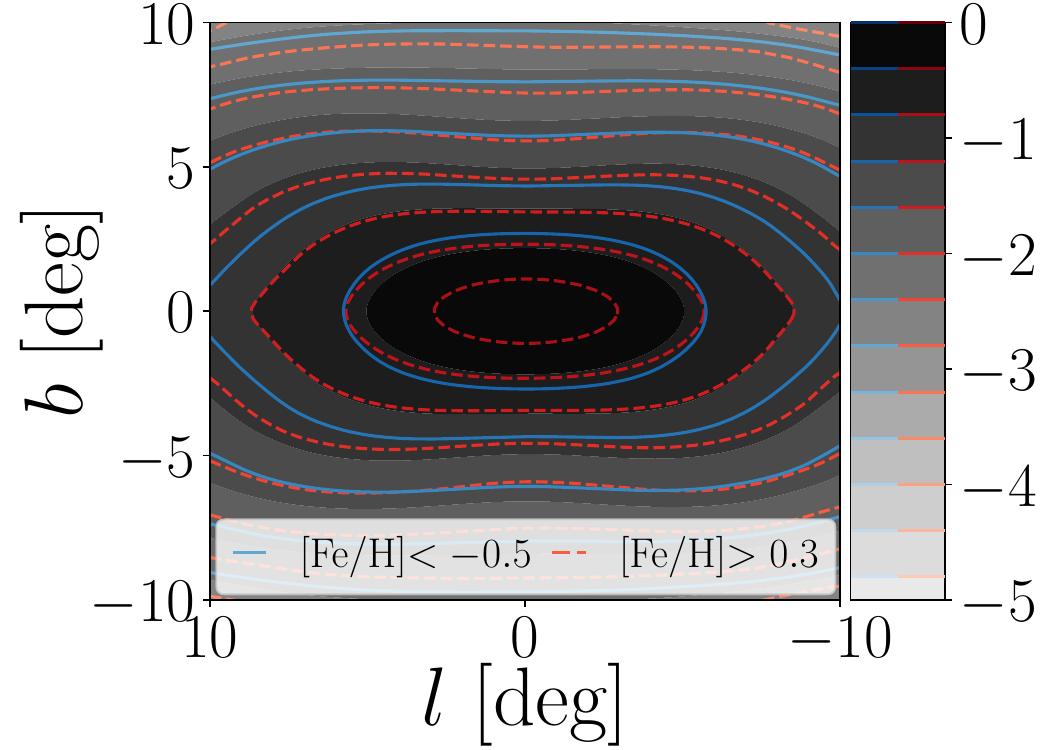}
		\caption{Density map of the bulge region of (MWa, clumpy). The top, bottom left, and bottom right panels show the density distributions of all stellar particles (classical bulge and disc), classical bulge particles, and disc particles, respectively. The blue solid line and red dashed line indicate the density contours of metal-poor ($\mathrm{[Fe/H]}< -0.5$) and metal-rich ($\mathrm{[Fe/H]} > 0.3$) populations, respectively. The background greyscale map shows the density distribution without binning by [Fe/H]. Densities are displayed on a logarithmic scale and normalised by the central density. The contours follow the same scaling and normalisation.}\label{fig:lb_dens_MWa_run733}
\end{figure}

\begin{figure}
		\centering
		\includegraphics[width=0.8\linewidth]{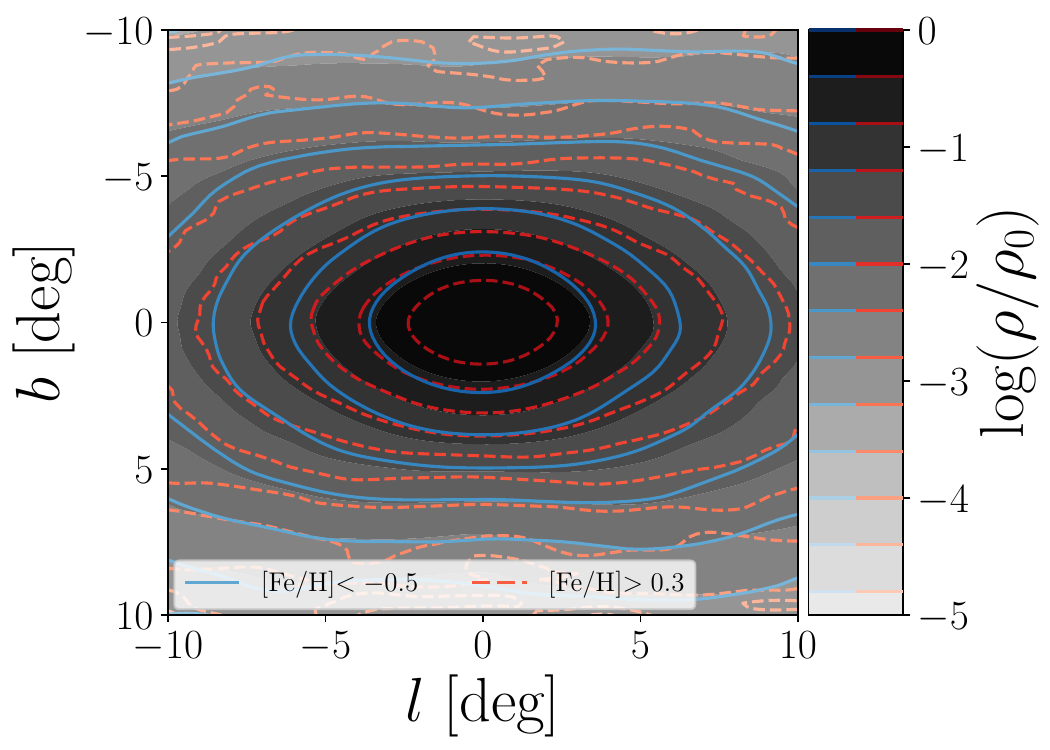}
		\caption{Same as Fig.~\ref{fig:lb_dens_MWa_run733} but for (model 2, clumpy). The length is rescaled by a factor of $5/9$ to match the bar length of the MW.}\label{fig:lb_dens_run741_run733}
\end{figure}
\citet{2017A&A...599A..12Z} found that the metal-poor and metal-rich populations have different spatial distributions in the MW bulge: the metal-poor population show a spheroidal distribution in the $l$-$b$ plane, while the metal-rich population exhibits a boxy distribution.
We divide the particles of (MWa, clumpy) into the metal-poor ([Fe/H] $< -0.5$) and metal-rich ([Fe/H] $> 0.3$) populations.
The top panel of Fig.~\ref{fig:lb_dens_MWa_run733} shows the density distribution of the two populations (blue contour: metal-poor, red contour: metal-rich, and background: total) in the $l$-$b$ plane for the particle within $7.5< R_s/\mathrm{kpc} < 8.5$.
The density is normalised by the central density of the total population and displayed on a logarithmic scale.
Consistent with the real observations, the distribution of the metal-poor population (blue contour) is more spheroidal, and the metal-rich population (red contour) is elongated along the longitude direction.
Especially, at high latitudes ($|b| \gtrsim 4^\circ$), the metal-poor population shows round isodensity contours, while the metal-rich population exhibits straighter, more horizontal ones.
Figure~\ref{fig:lb_dens_run741_run733} shows the same density map for (model 2, clumpy).
Compared with (MWa, clumpy), the metallicity dependence of the density distribution is weaker, although the metal-rich population is slightly more elongated horizontally than the metal-poor population.
In the MWa model, the spheroidal shape of the metal-poor population is likely a consequence of the presence of a classical bulge component.
The lower panels of Fig.~\ref{fig:lb_dens_MWa_run733} isolate the classical bulge and disc particles.
When plotting the two components separately, the metallicity dependence becomes much weaker.
The classical bulge is spheroidal regardless of [Fe/H], while the disc shows an elongated shape in both metallicity bins.  

However, it is too early to conclude, from these results only, whether the presence of a classical bulge is necessary to reproduce the observed chemo-morphological features of the MW bulge.
Some numerical studies \citep[e.g.][]{2017A&A...606A..47F, 2018A&A...616A.180F, 2019A&A...628A..11D} suggested that the thick disc contributes the metallicity dependence of the bulge morphology and kinematics.
Since our target models do not include a geometrically distinct thick disc component (although the clumpy donor contains high-$\alpha$ population corresponding to the thick disc), the influence of the thick disc cannot be fully explored in this study.
Moreover, we note the tension with the failure to produce a bimodality in the magnitude distribution when the classical bulge is included, which model 2 without a classical bulge does not suffer from. The fact that both models fail in different ways suggests that more target models need to be explored than the two considered here.
Our action-based chemical mapping method offers a strong advantage in this context.
We can systematically explore a wide range of target models with different parameters (e.g. classical bulge mass fraction, inclusion of a thick disc component) at orders of magnitude lower computational cost than full $N$-body+hydrodynamical simulations.
We expect to gain deeper insights into its chemo-dynamical structure and to better constrain its formation and evolution by exploring a broad parameter space.

\section{Summary}\label{sec:summary}
We present a new method for assigning chemical abundances to high-resolution pure $N$-body simulations (target models) by transferring information from lower-resolution star-forming hydrodynamical simulations (donor models).
Building on previous action-based approaches \citep{2020MNRAS.498.3334D}, our method introduces a particle-level mapping using the Hungarian algorithm \citep{kuhn1955hungarian}, enabling one-to-one correspondence between donor and target particles.

We applied it to two targets: model 2 from \citet{2020MNRAS.498.3334D}, which features a strongly buckled bar but no classical bulge, and MWa from \citet{2019MNRAS.482.1983F}, which includes a classical bulge and a weaker BPX feature.
Two star-forming hydrodynamical simulations were used as donors: the clumpy model \citep{2019MNRAS.484.3476C}, which produces a bimodal [Fe/H]--[$\alpha$/Fe] distribution, and M1\_c\_b \citep{2021MNRAS.503.1418F}, which yields a single chemical track due to stronger supernova feedback.
The main results are as follows:

\begin{itemize}
	\item The models successfully reproduce key structural trends of the bulges seen in the MW and external barred galaxies.
		The side-on [Fe/H] and [O/Fe] maps show a pinched structure because metal-rich, low-$\alpha$ populations form stronger peanut- or X-shaped structures than metal-poor, high-$\alpha$ populations.

	\item The [Fe/H]–[O/Fe] distributions in the models using M1\_c\_b as a donor exhibit a single track, while those using the clumpy model show a clear latitude-dependent bimodality, in qualitative agreement with MW observations.

	\item In the mock $K$-band magnitude distributions of bulge red clump stars, (model 2, clumpy) shows a clear bimodality that varies with latitude and metallicity, consistent with MW observations. On the other hand, (MWa, clumpy) does not exhibit a distinct bimodality due to its weaker X-shaped structure.

	\item The models show cylindrical rotation regardless of metallicity, consistent with MW observations. The velocity dispersion is slightly higher for the metal-poor population than for the metal-rich population, but the metallicity dependence of the velocity dispersion is weaker than that of the MW bulge.
	When taking into account of the contamination from the halo stars to the metal-poor population, the velocity dispersion increases and its profile as a function of longitude becomes flatter, which is more qualitatively consistent with the observations.
    In the metal-rich population, the velocity dispersion is not strongly peaked, suggesting a limitation of our method.

	\item The spatial variation of the metallicity distribution derived from GMM matches the main features of the GIBS survey: the metal-poor fraction increases toward larger $|l|$ and $|b|$. $l$--$b$ maps of the metal-poor fraction show a pinched structure similar to the mean metallicity maps.

	\item In (MWa, clumpy), the metal-poor and metal-rich populations exhibit spheroidal and boxy density distributions, respectively, consistent with GIBS results. This trend is less pronounced in (model 2, clumpy) even after rescaling the length to match the bar length of the MW.
		In (MWa, clumpy), the classical bulge mainly contributes to the spheroidal morphology of the metal-poor population.
\end{itemize}

Our models generally reproduce the chemo-dynamical properties of BPX bulges expected from the observations and full $N$-body+hydrodynamical simulations, demonstrating the effectiveness of our method.
The action-based chemical mapping framework provides a computationally efficient and flexible tool for linking dynamical and chemical evolution in barred galaxies.
Its low computational cost enables the use of high-resolution $N$-body models and facilitates extensive parameter studies, making it a powerful approach for interpreting the chemo-dynamical structure of the BPX bulges.

\section*{Data availability}
The data and code used in this study are available at \url{https://zenodo.org/records/20846005} and \url{https://github.com/tetsuroasano/action_chemistry_mapping}, respectively.
Full snapshots of the MW $N$-body models of \citet{2019MNRAS.482.1983F} are available at \url{http://galaxies.astron.s.u-tokyo.ac.jp/}.

\begin{acknowledgements}
VPD thanks the University of Tokyo and the Department of Astronomy for support and hospitality when this project was initiated. 
Simulations of \citet{2019MNRAS.482.1983F} were performed using GPU clusters, HA-PACS at Tsukuba University, Piz Daint at the Swiss National Supercomputing Centre, and Little Green Machine II.
TA acknowledges the support of of the JSPS Overseas Research Fellowship and the Project PID2024-160244NB-I00 funded by MICIU/AEI /10.13039/501100011033 and by FEDER, UE, and the Institute of Cosmos Sciences University of Barcelona (ICCUB, Unidad de Excelencia `Mar\'{\i}a de Maeztu’) through grant CEX2019-000918-M.

This work used \texttt{Agama} \citep{2019MNRAS.482.1525V} and \texttt{pynbody} \citep{pynbody} for simulation data analysis.
\end{acknowledgements}

   \bibliographystyle{aa} %
   \bibliography{extracted}

\begin{appendix}
\nolinenumbers
\raggedbottom
\section{Kinematic fractionation}\label{sec:kinematic_fractionation}
\begin{figure*}
	\begin{center}
		\includegraphics[width=0.45\hsize]{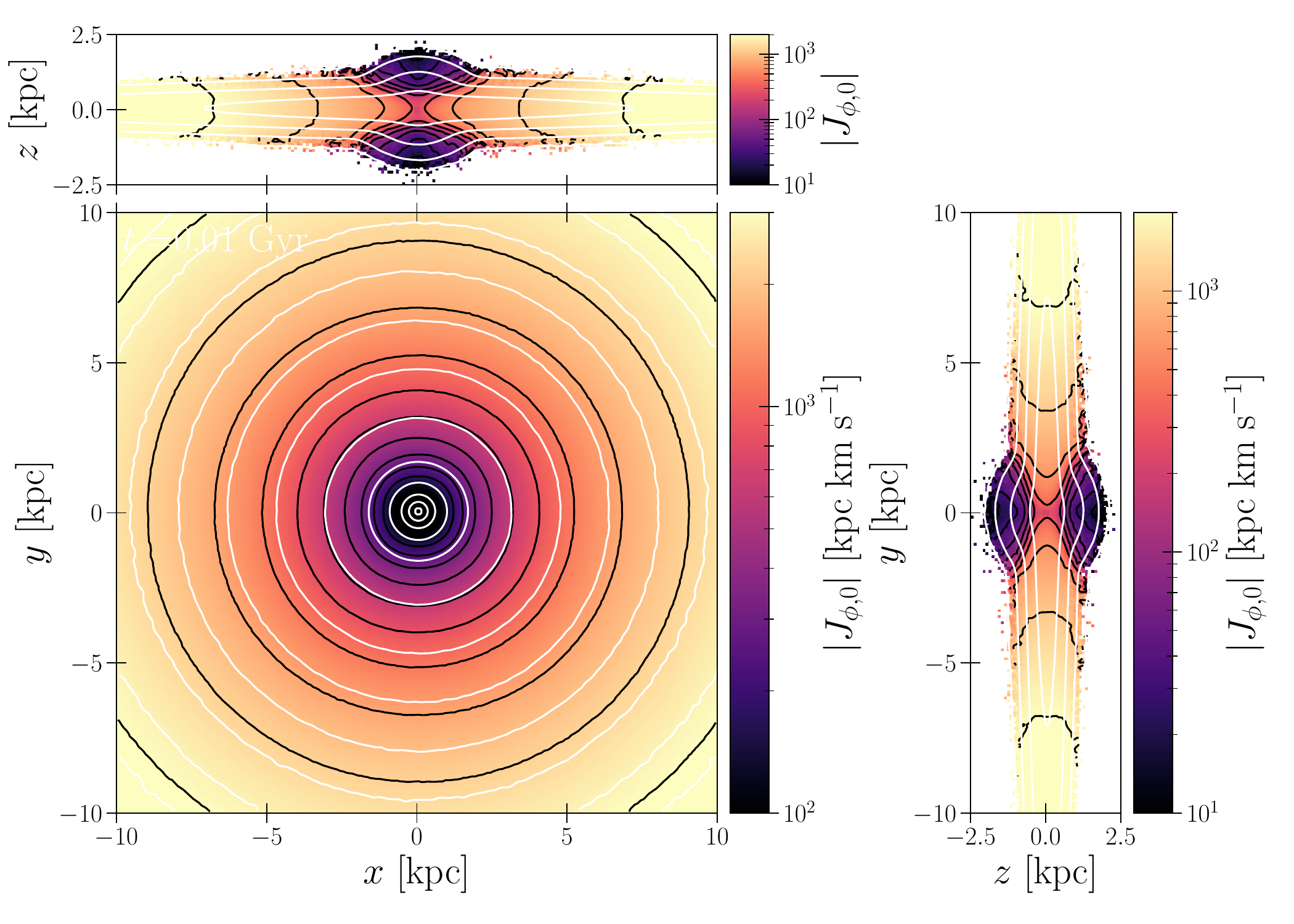}
		\includegraphics[width=0.45\hsize]{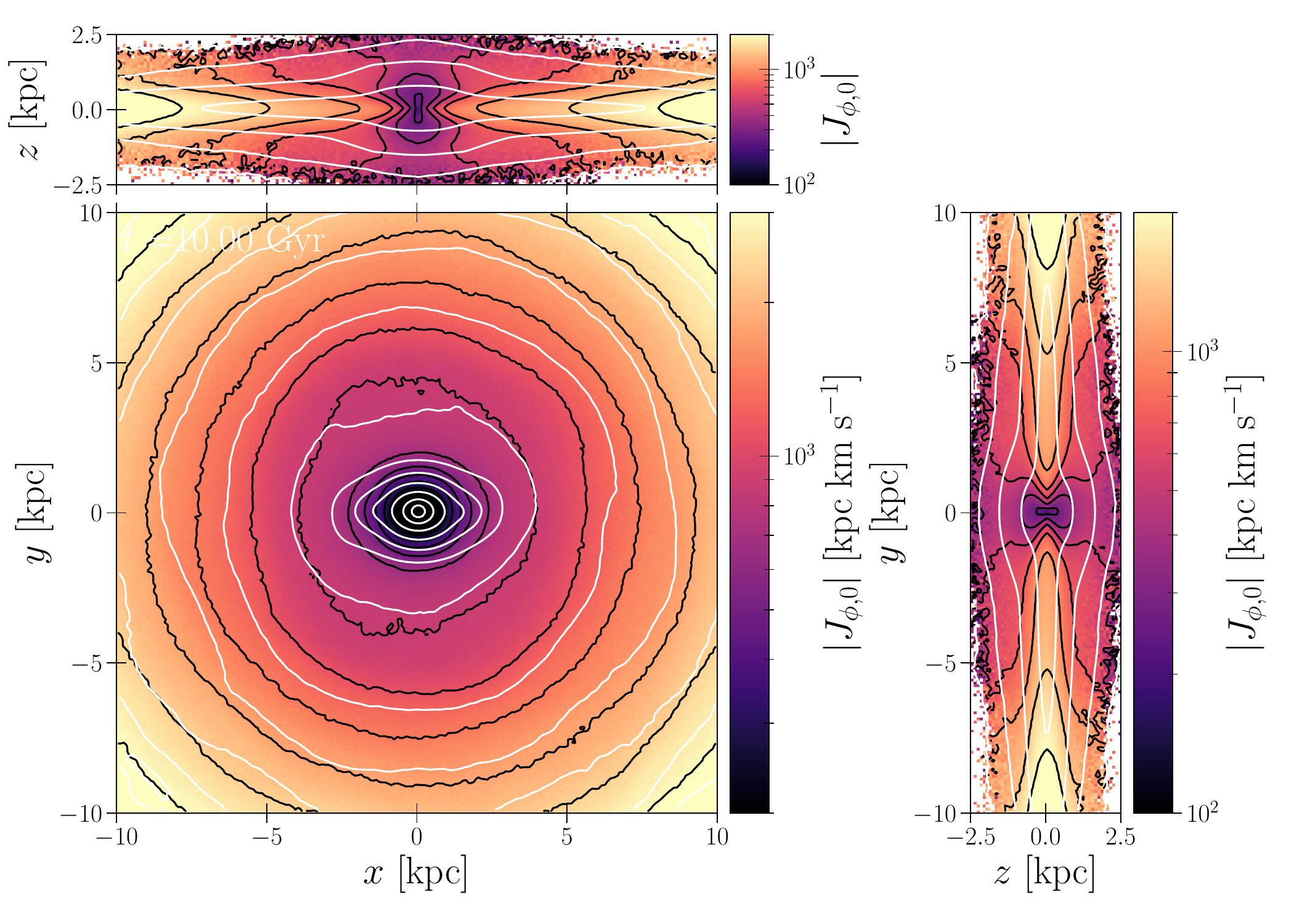}
		\includegraphics[width=0.45\hsize]{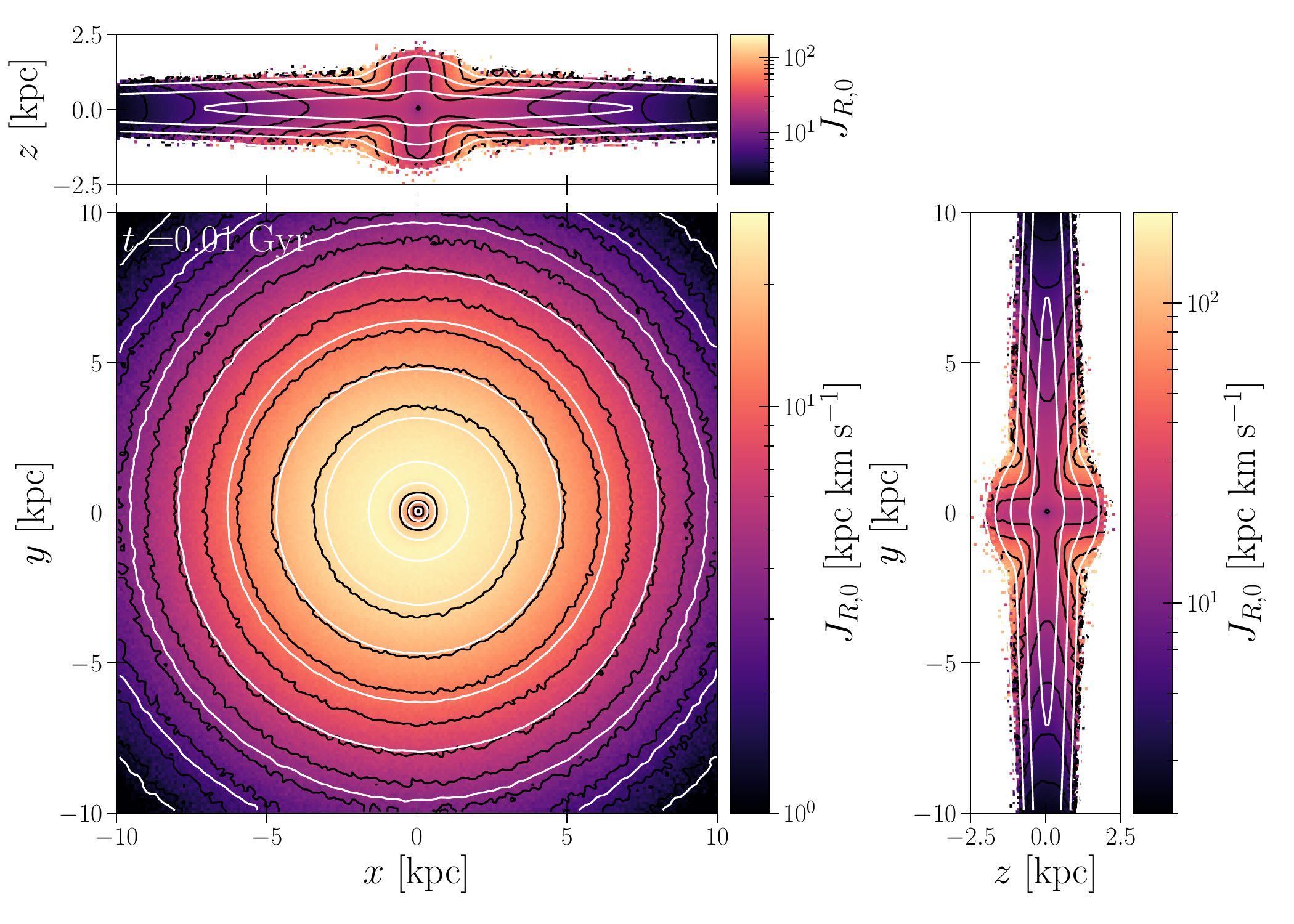}
		\includegraphics[width=0.45\hsize]{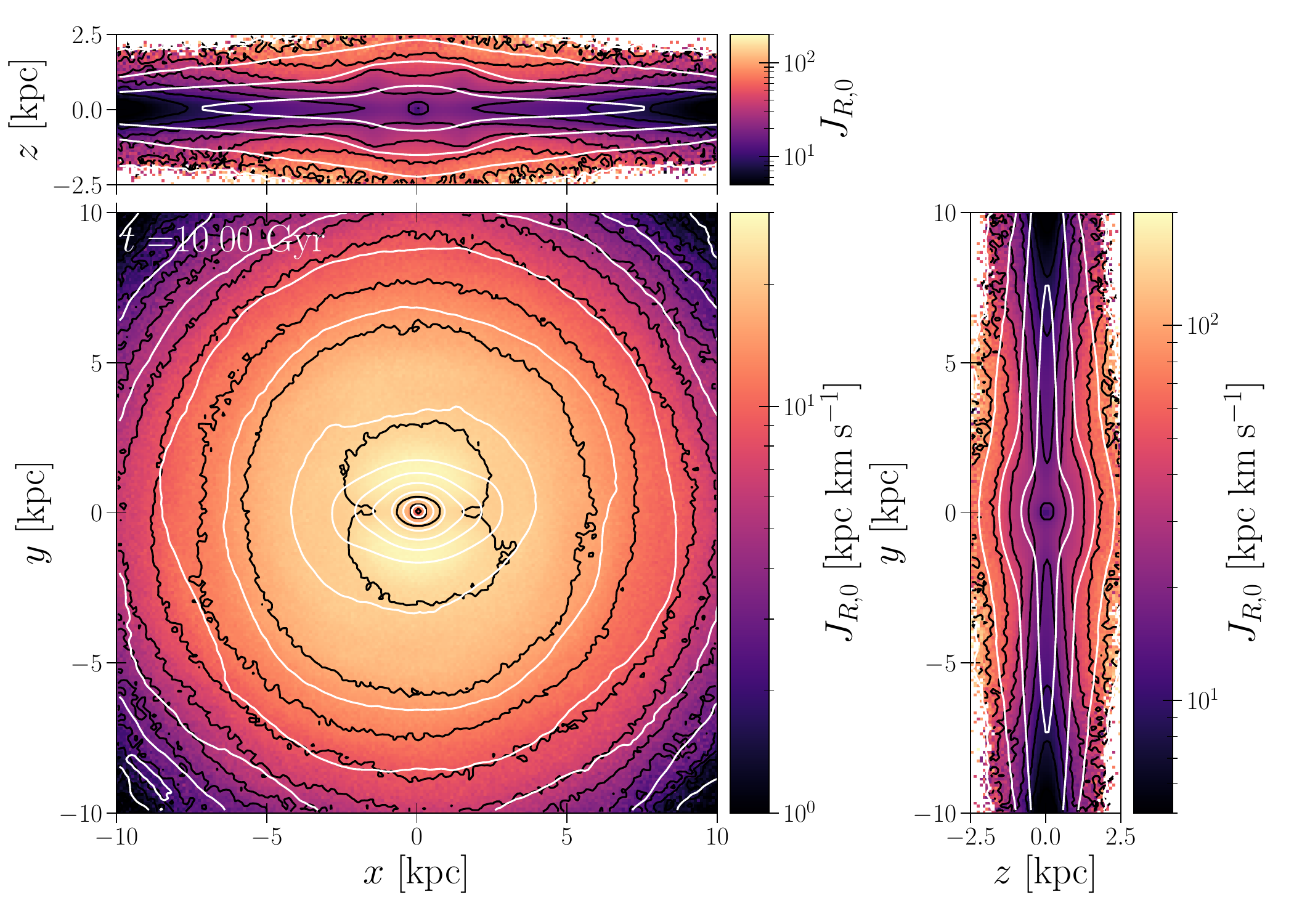}
		\includegraphics[width=0.45\hsize]{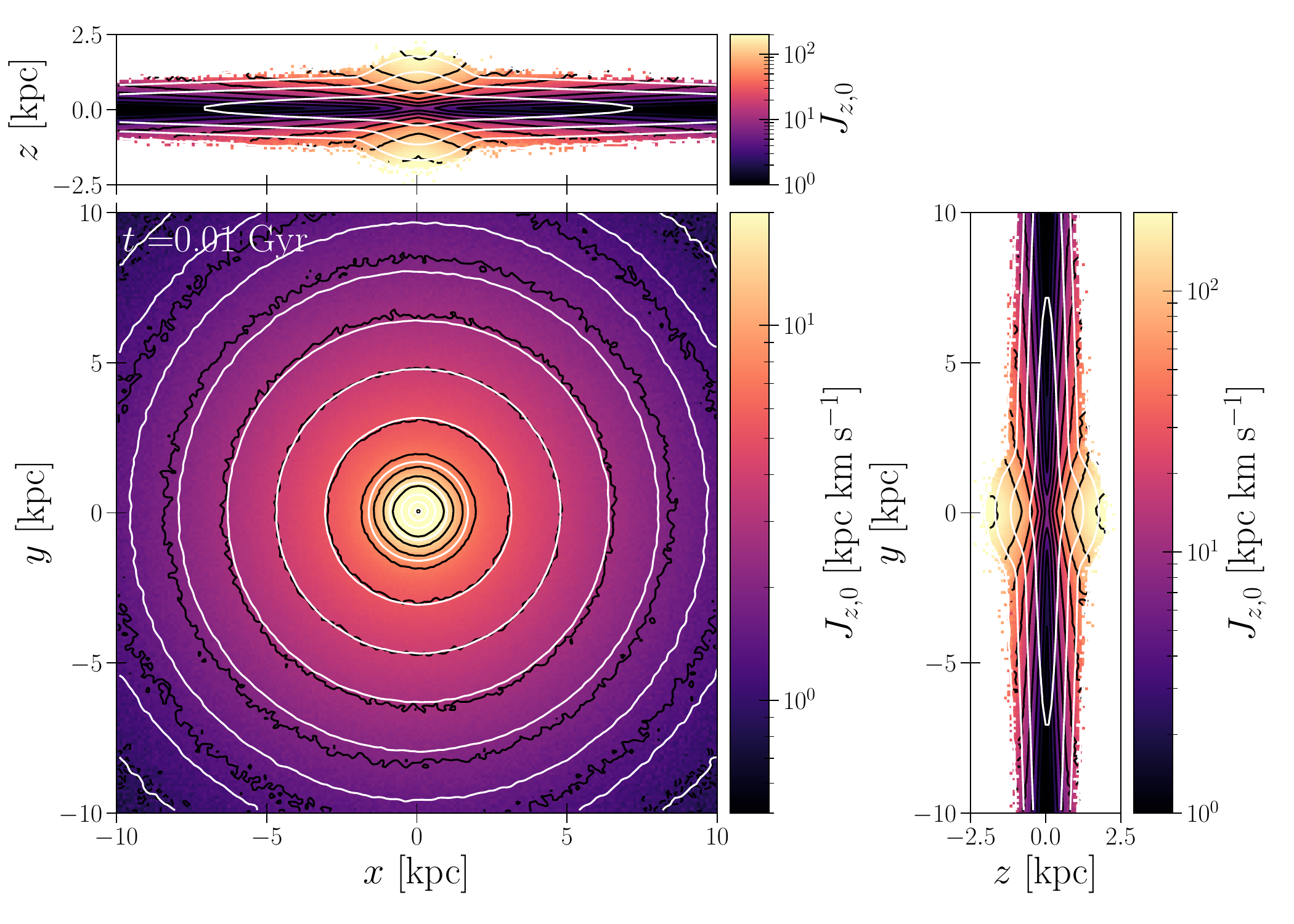}
		\includegraphics[width=0.45\hsize]{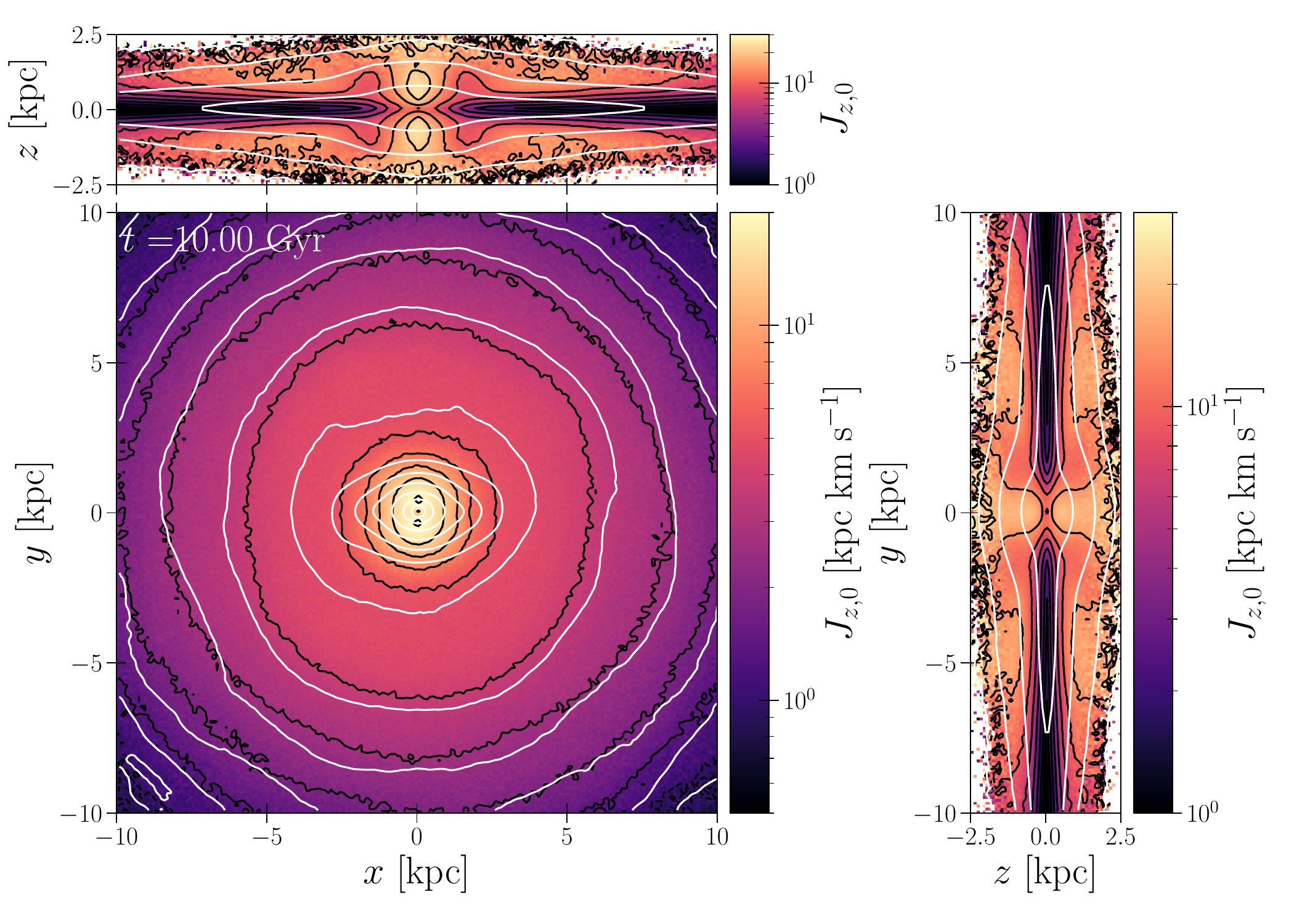}
		\caption{Face-on and edge-on  maps of MWa colour-coded by mean azimuthal action $|J_{\phi,0}|$ (\textit{first row}), mean radial action $J_{R,0}$ (\textit{second row}) and mean vertical action $J_{z,0}$ (\textit{third row}). \textit{Left column:} The maps at $t=0$. \textit{Right column:} The maps in the final snapshot ($t=10$~Gyr). White and black lines represent the contours for the surface density in logarithmic scale and the mean actions, respectively.}\label{fig:face_on_edge_on_action_MWa}
	\end{center}
\end{figure*}

The panels on the left side of Fig~\ref{fig:face_on_edge_on_action_MWa} show the face-on and edge-on maps colour-coded by the mean actions in the initial snapshot ($t=0$) of MWa. The bar has been aligned along the $x$-axis throughout the analysis. Following \citet{2020MNRAS.498.3334D}, we refer to these actions as $J_{R,0}$, $J_{\phi,0}$, and $J_{z,0}$.
The panels on the right side show maps of the same initial actions in the final snapshot ($t=10$~Gyr), but are colour-coded by the initial actions rather than the final actions.

The first row of Fig.~\ref{fig:face_on_edge_on_action_MWa}  is colour-coded by $|J_{\phi, 0}|$.
In the first snapshot, $|J_{\phi, 0}|$ increases with $R$.
Within $R\lesssim2$~kpc, where the classical bulge dominates, $|J_{\phi, 0}|$ decreases vertically.
Beyond $R\gtrsim2$~kpc, the dependence of $|J_{\phi, 0}|$  on $|z|$ becomes weaker.
In the face-on maps of the final snapshot, $|J_{\phi, 0}|$ is slightly elongated along the bar's major axis ($x$-axis) in the bar region.
In the $x$-$z$ and $y$-$z$ planes, the contours bend towards the galactic centre, and the vertical gradient becomes steeper than in the first snapshot outside of the bulge region.
At $R\lesssim 2$~kpc, $|J_{\phi, 0}|$ shows a weaker dependence on $|z|$.

The second row of Fig.~\ref{fig:face_on_edge_on_action_MWa} is colour-coded by $J_{R,0}$.
At $t=0$, $J_{R,0}$ decreases radially and increases vertically at $R\gtrsim2$~kpc. In the central part ($R\lesssim 2$~kpc), $J_{R,0}$ decreases slightly towards the galactic centre.
The contours are vertical in this region when viewed edge-on.
In the face-on map of the final snapshot, the contour is elongated along the bar's minor axis ($y$-axis).
In the edge-on map, $J_{R,0}$ increases with $|z|$.
The contour is pinched vertically in the side-on view ($x$-$z$ plane).

The third row of Fig.~\ref{fig:face_on_edge_on_action_MWa} is colour-coded by $J_{z,0}$. In the first snapshot, $J_{z,0}$ decreases radially and increases vertically. In the final snapshot, the contour is elongated along the bar's minor axis in the face-on map. The contour is strongly pinched when viewed side-on. Outside the bulge region, $J_{z,0}$ increases with $|z|$ as it did in the initial snapshot. At $R\lesssim 1$~kpc, the vertical gradient is erased.

\begin{figure*}
	\begin{center}
		\includegraphics[width=0.24\hsize]{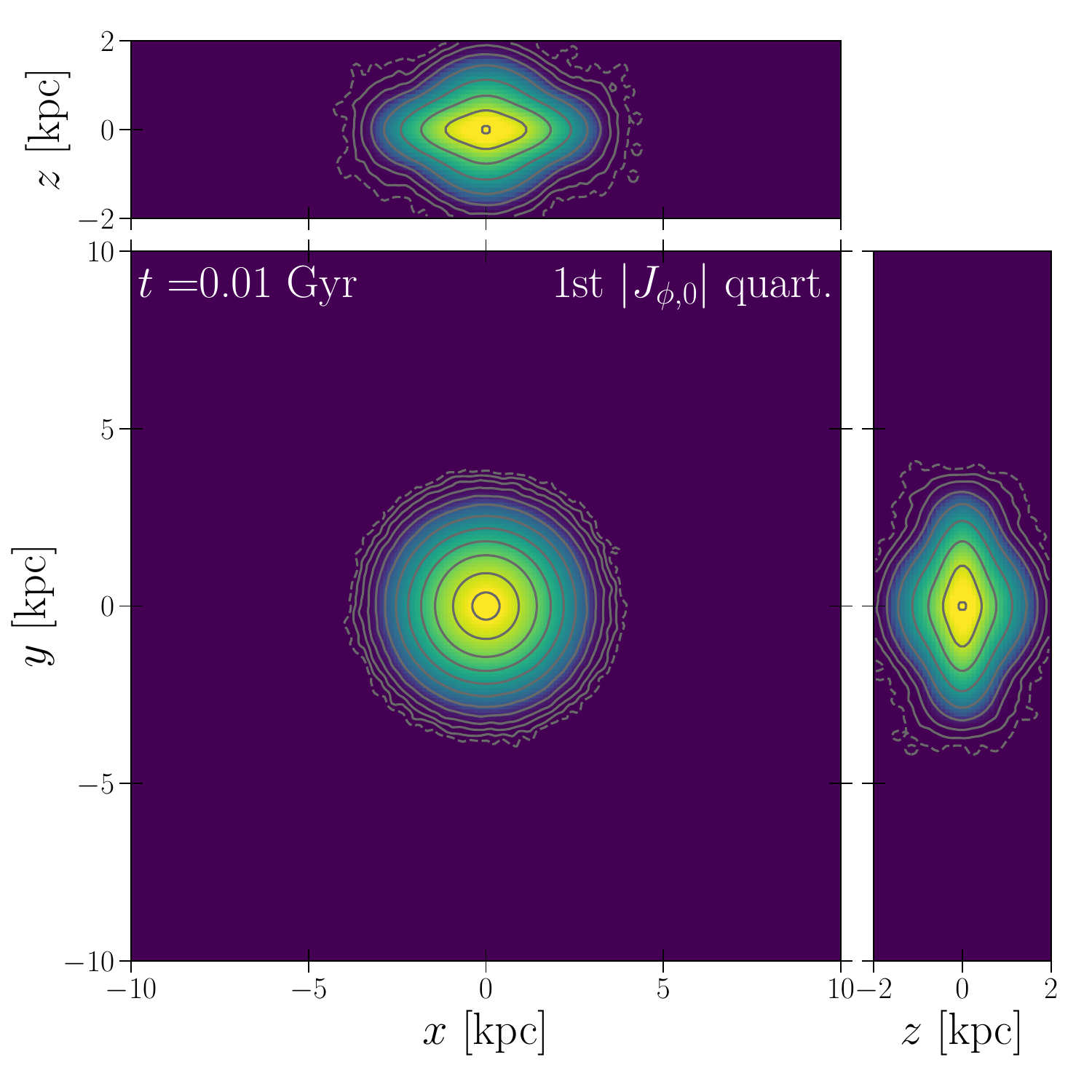}
		\includegraphics[width=0.24\hsize]{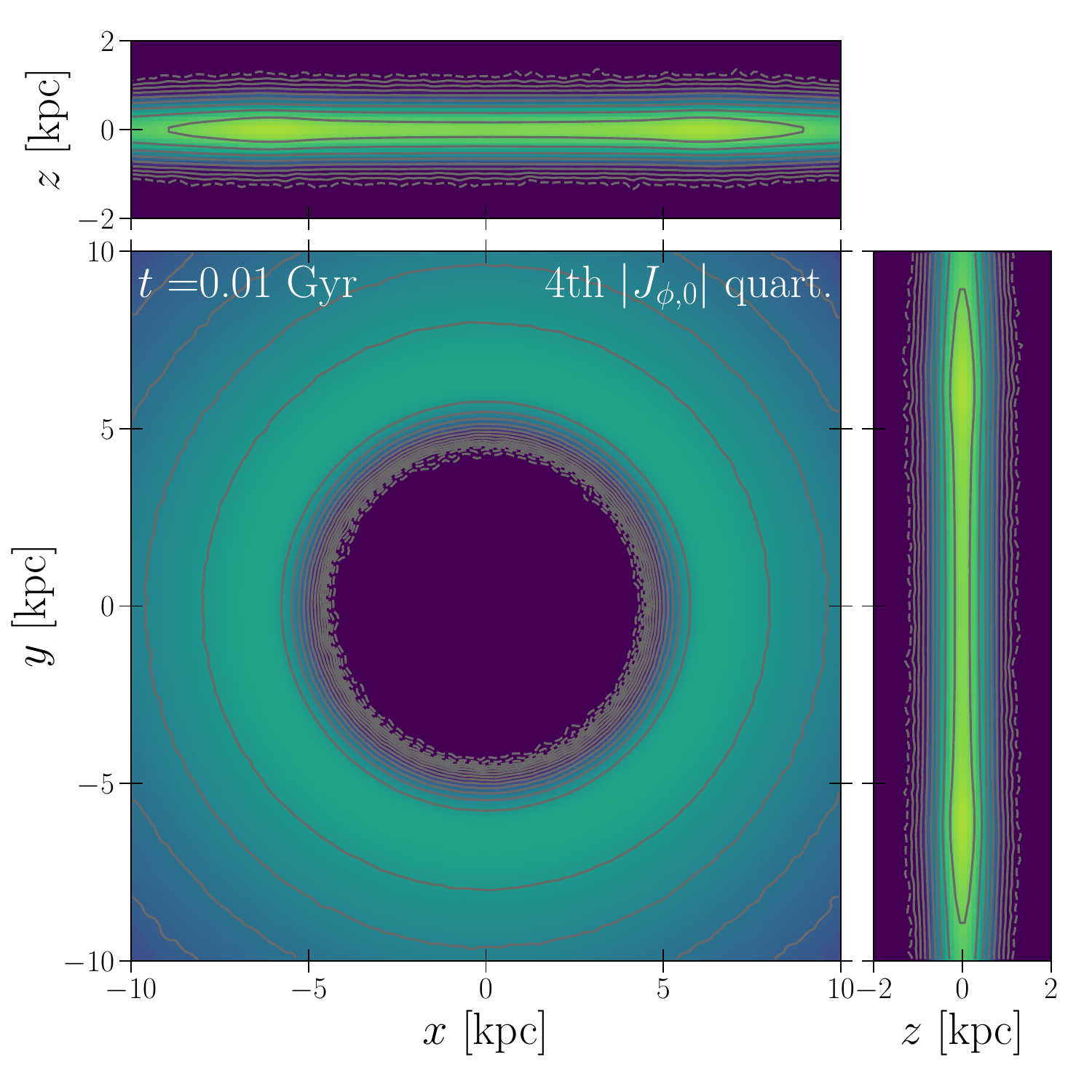}
		\includegraphics[width=0.24\hsize]{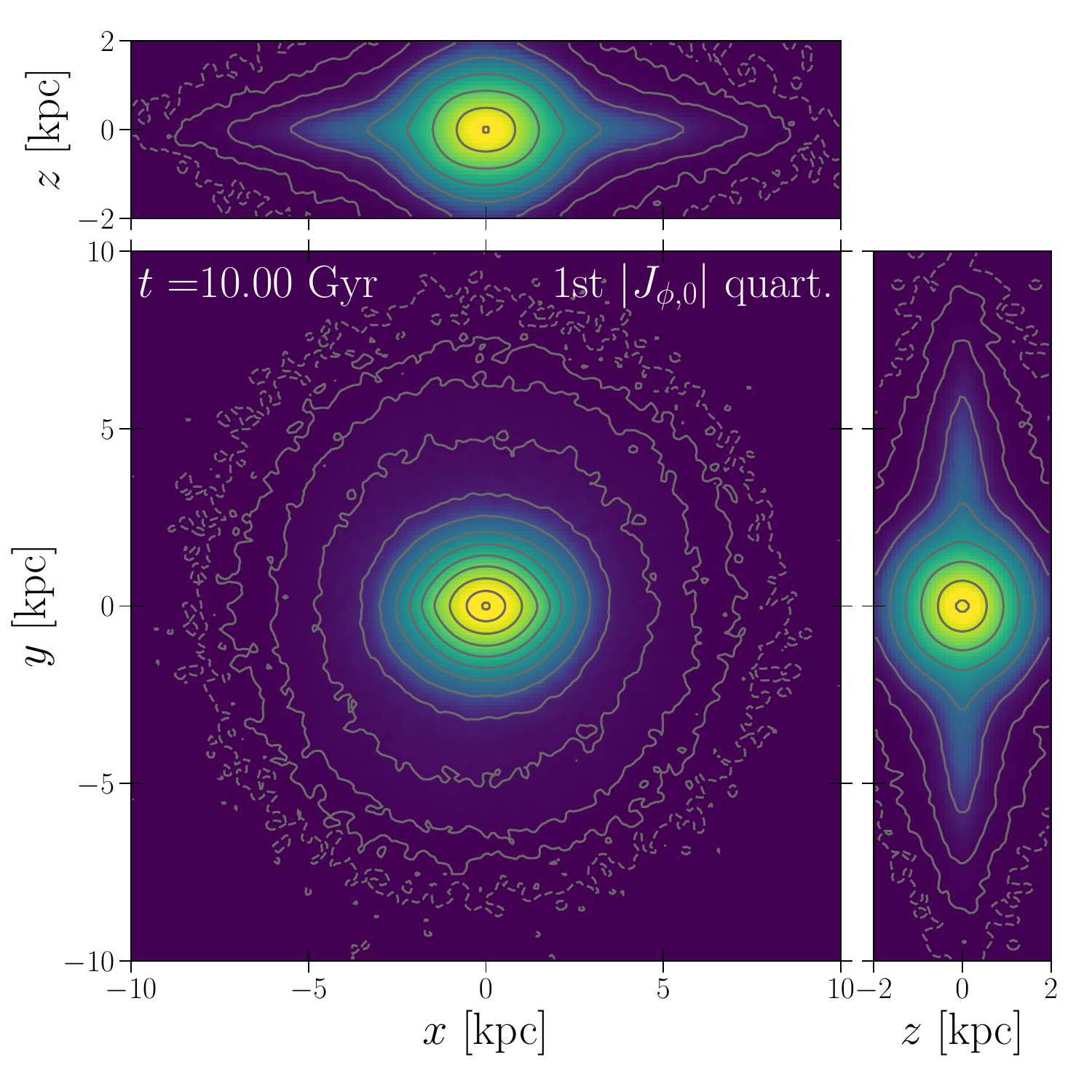}
		\includegraphics[width=0.24\hsize]{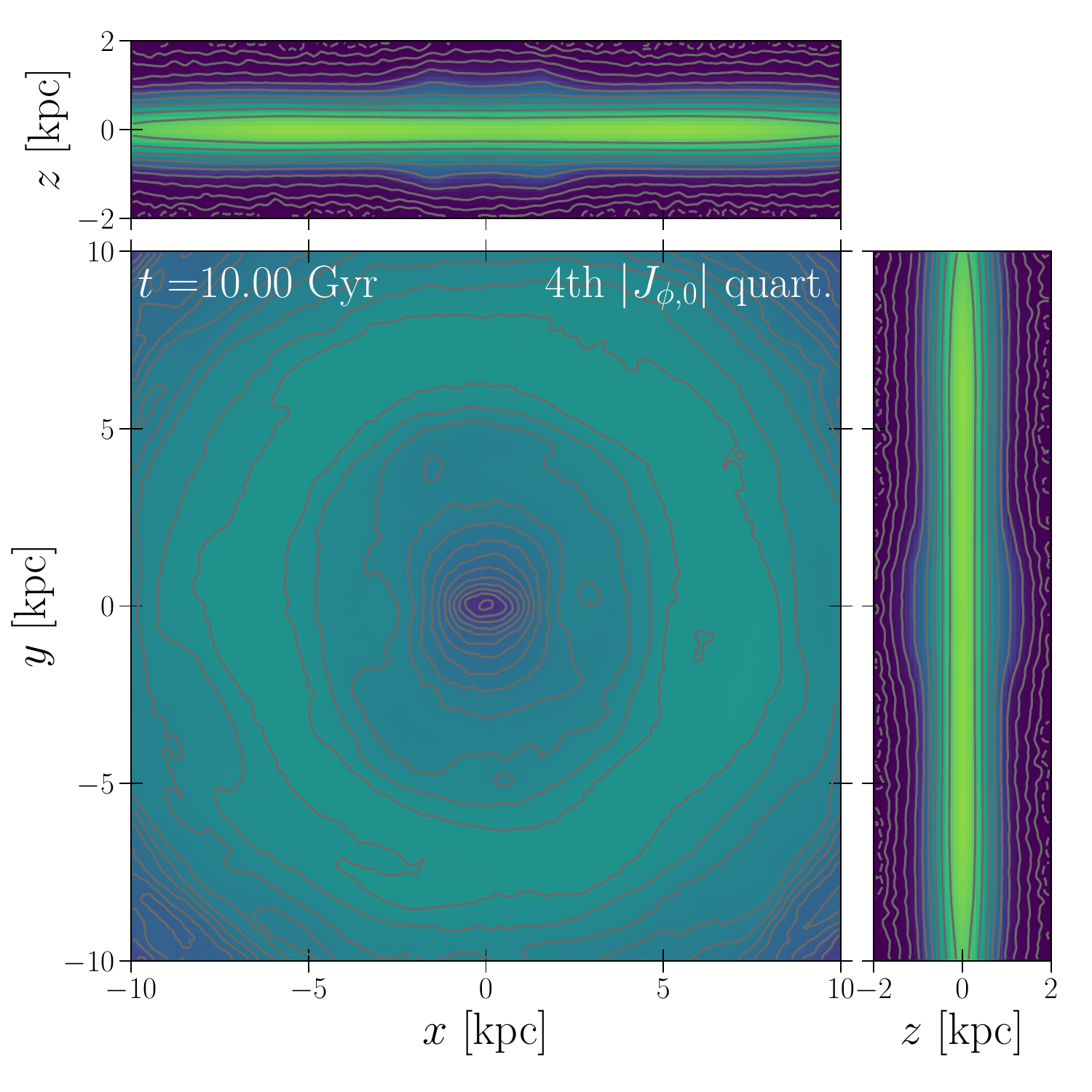}
		\includegraphics[width=0.24\hsize]{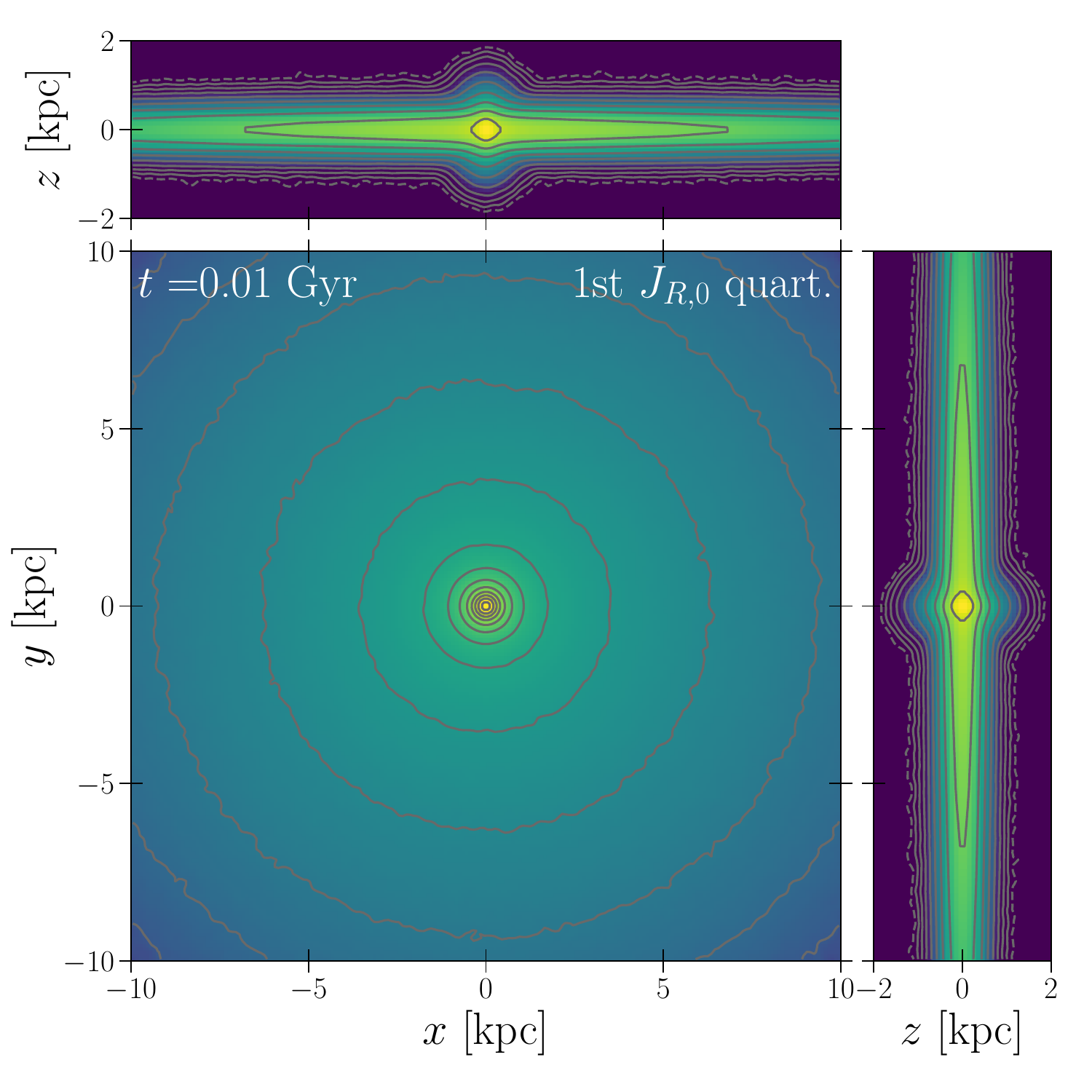}
		\includegraphics[width=0.24\hsize]{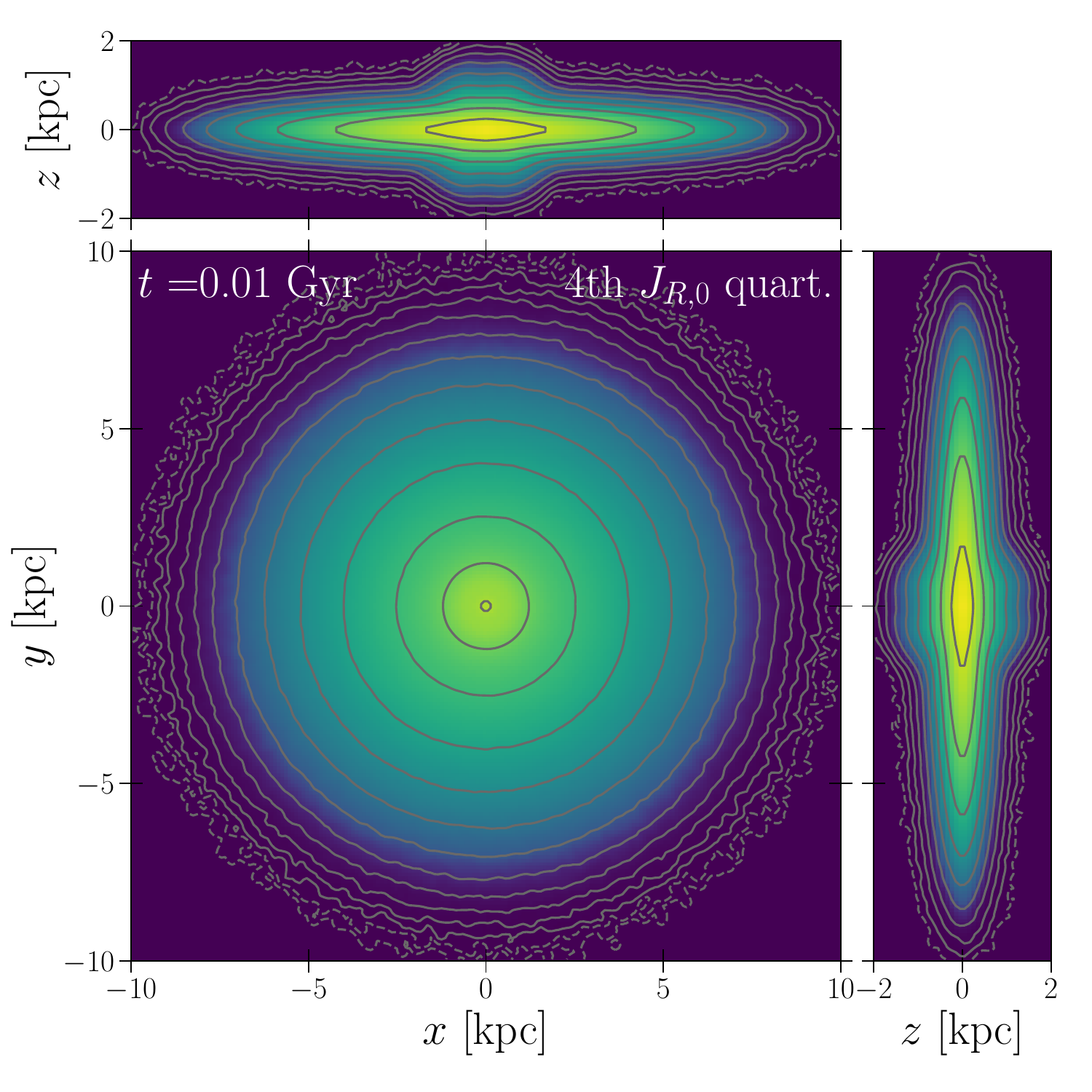}
		\includegraphics[width=0.24\hsize]{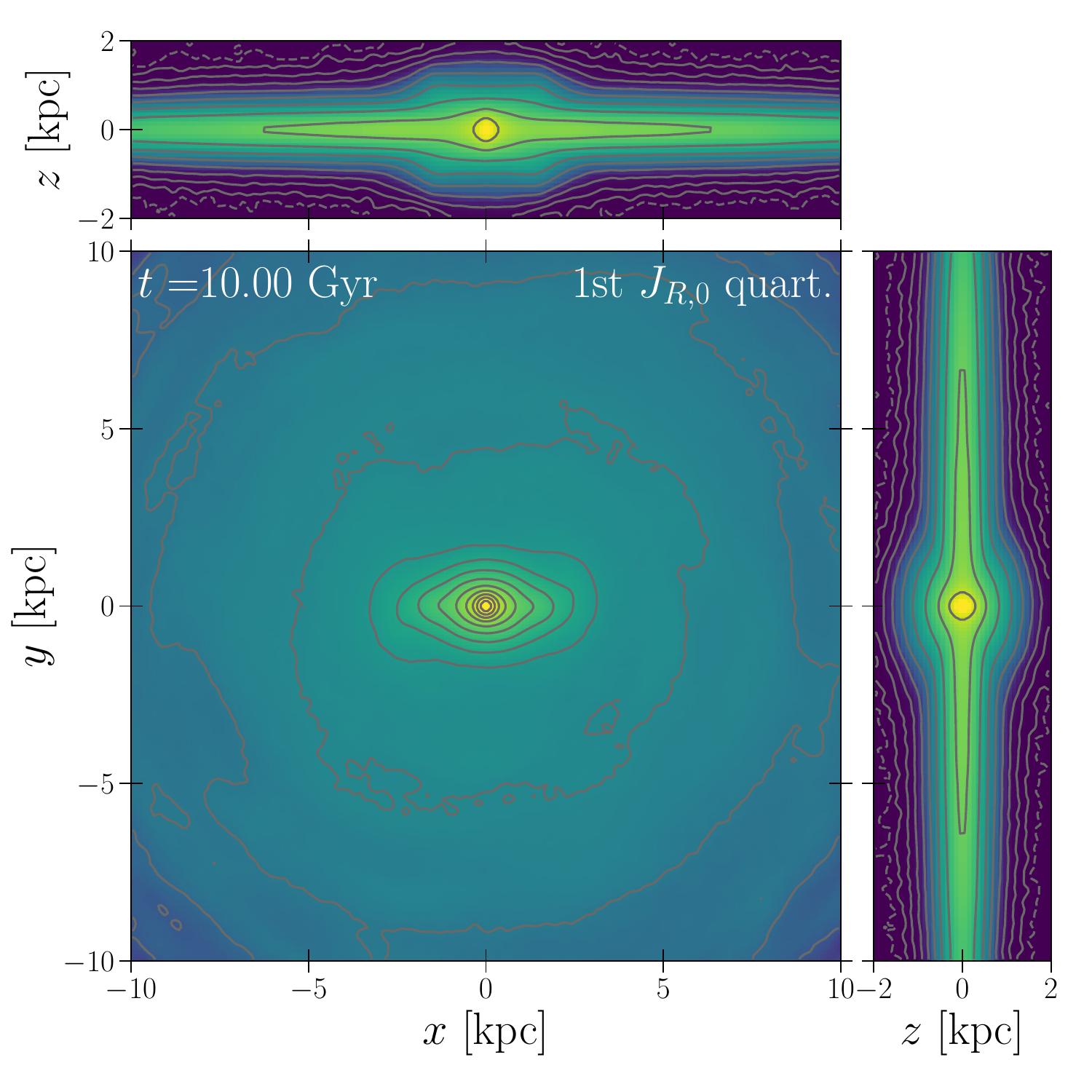}
		\includegraphics[width=0.24\hsize]{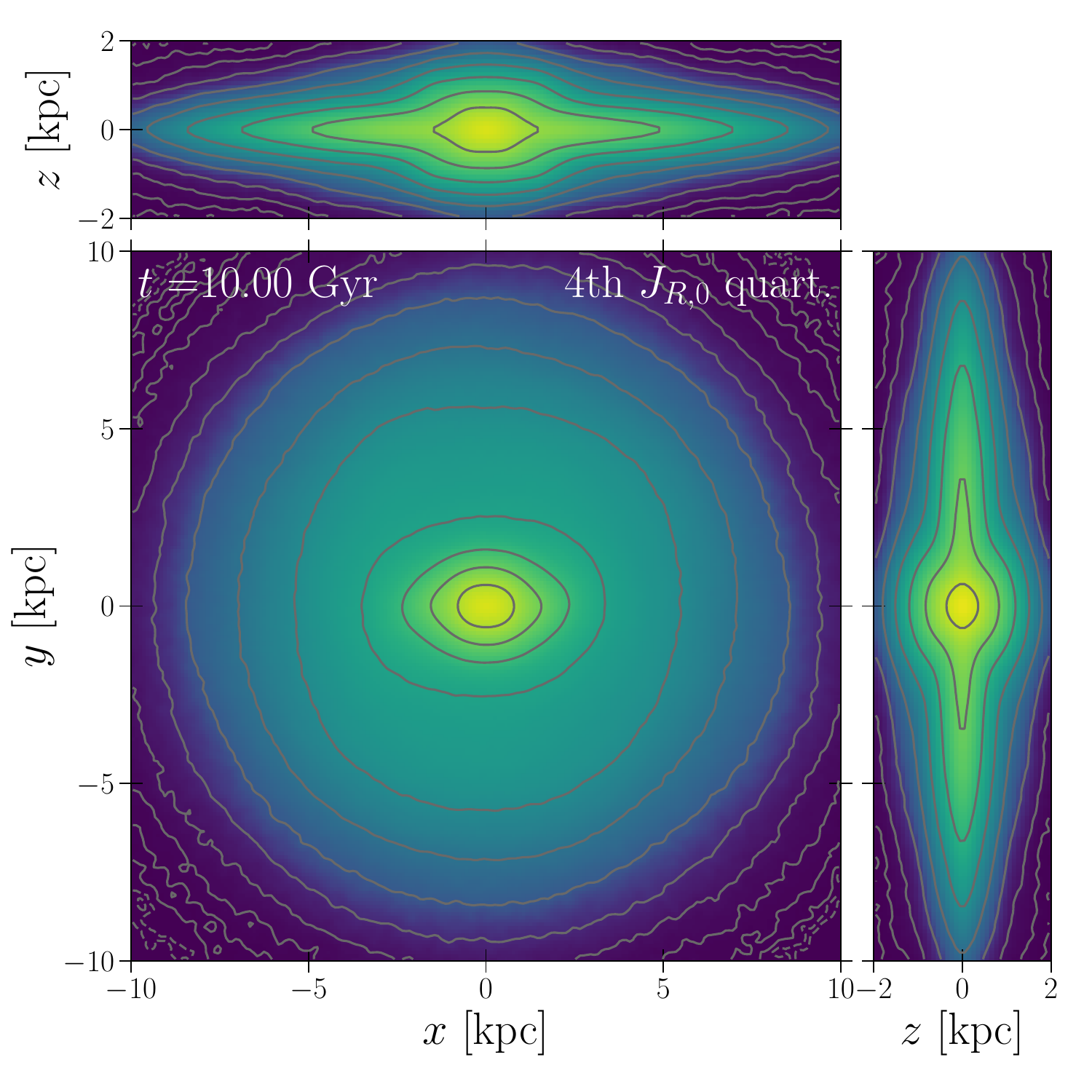}
		\includegraphics[width=0.24\hsize]{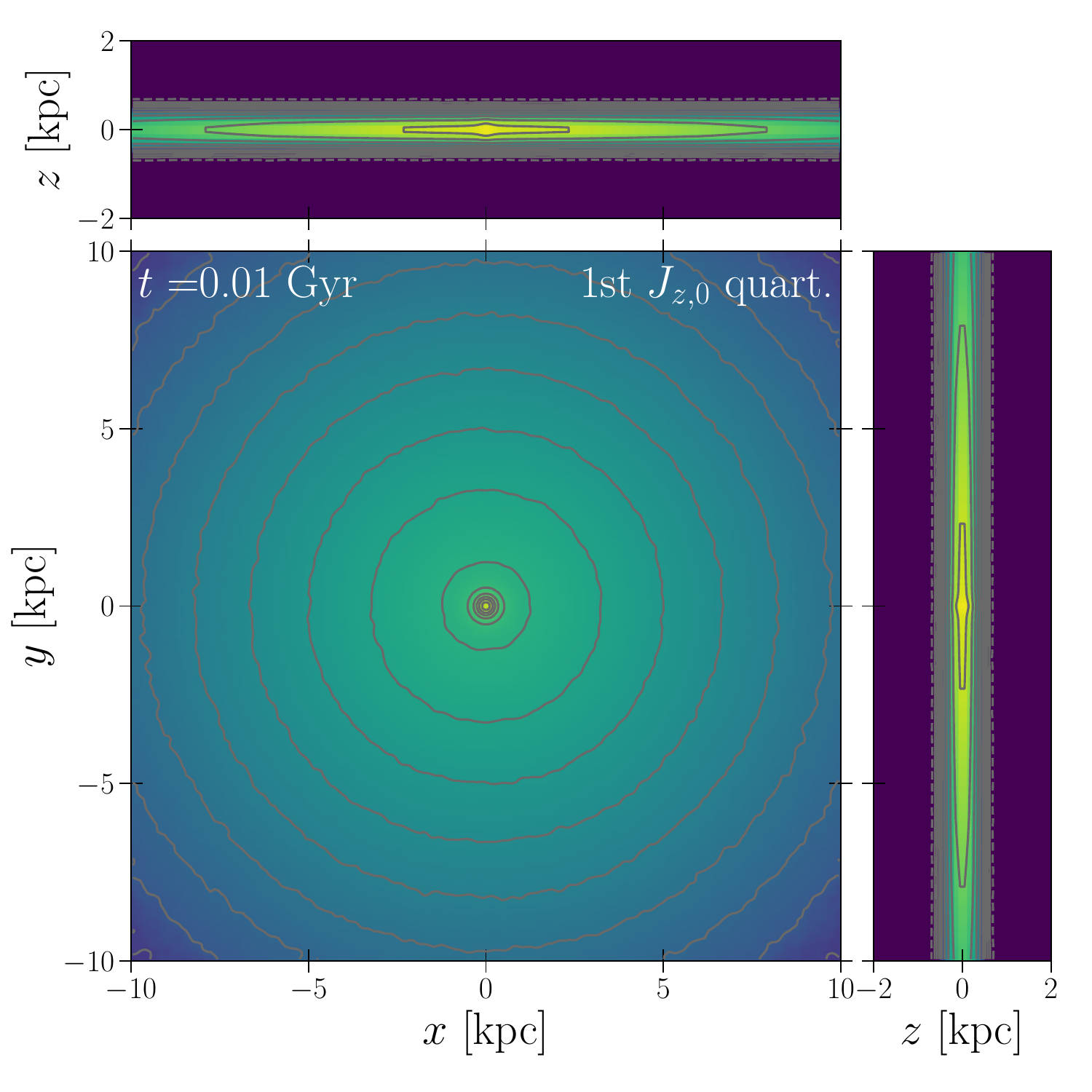}
		\includegraphics[width=0.24\hsize]{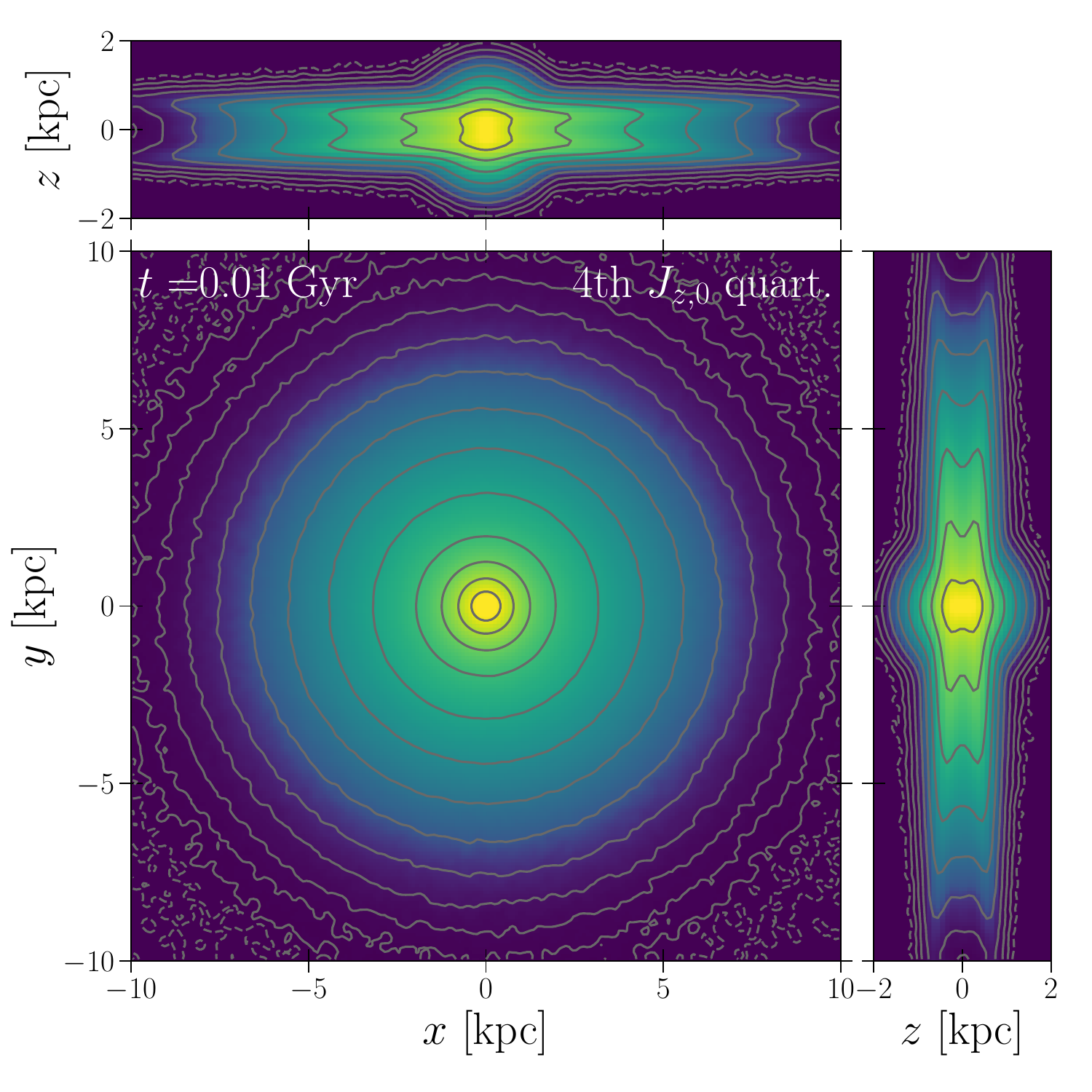}
		\includegraphics[width=0.24\hsize]{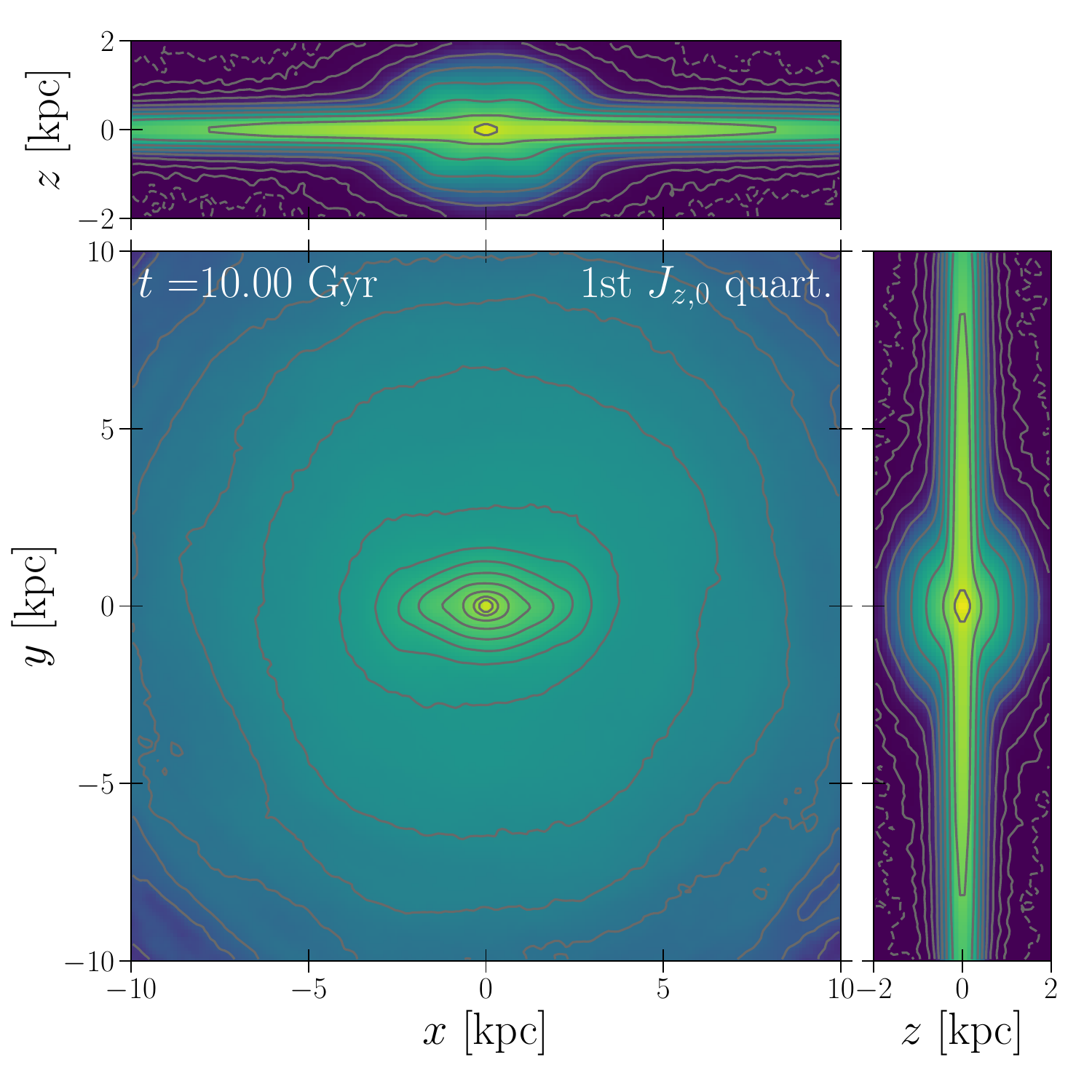}
		\includegraphics[width=0.24\hsize]{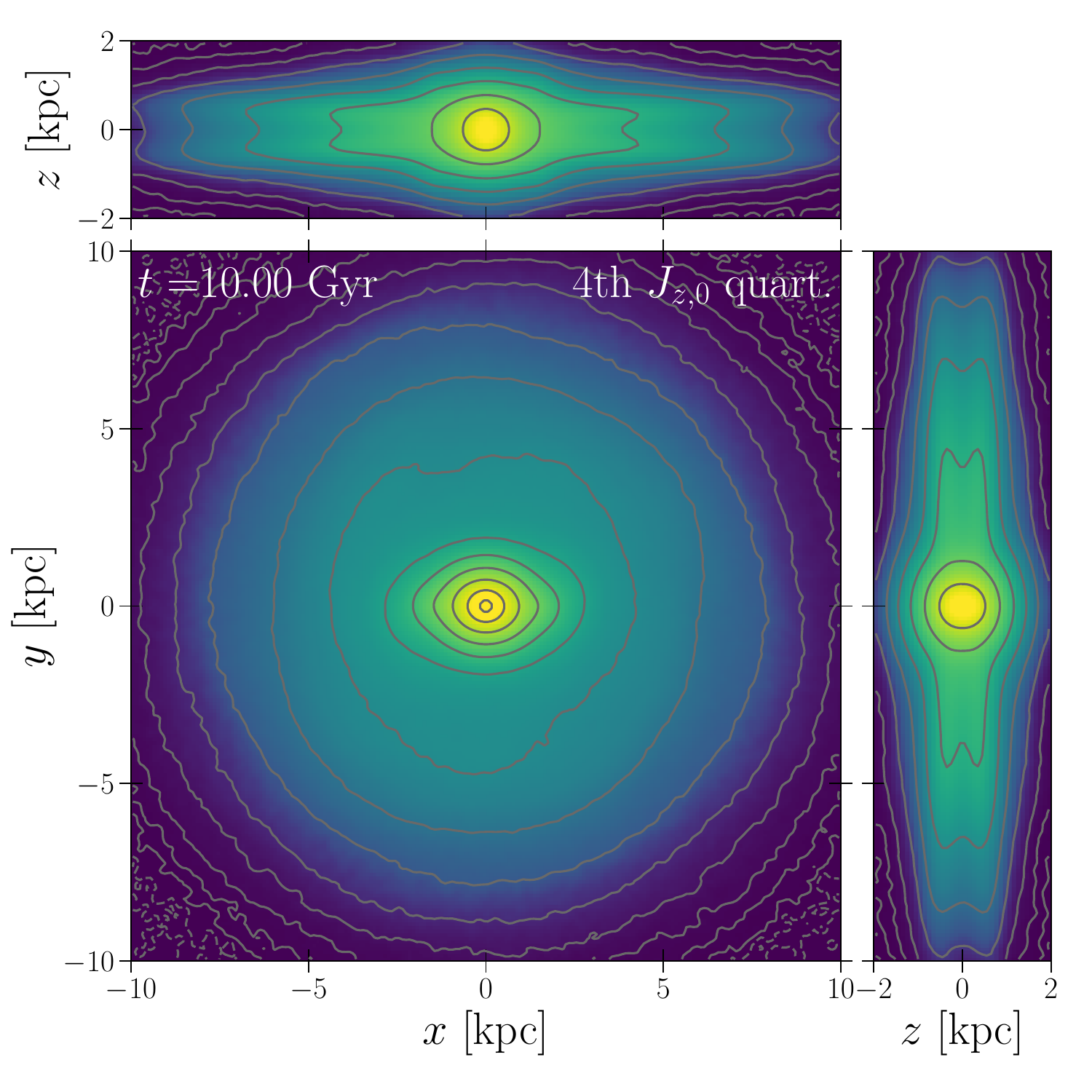}
		\caption{Density maps of MWa  separated into the quartiles of $|J_{\phi,0}|$ (\textit{top row}), $J_{R,0}$ (\textit{middle row}) and $J_{z,0}$ (\textit{bottom row}).
			\textit{First column: } First quartiles in the first snapshot.
			\textit{Second column: } Fourth quartiles in the first snapshot.
			\textit{Third column: } First quartiles in the final snapshot.
			\textit{Fourth column: } Fourth quartiles in the final snapshot.
		}\label{fig:face_on_quartile_MWa}
	\end{center}
\end{figure*}

Fig.~\ref{fig:face_on_quartile_MWa} shows the density maps of MWa separated into the quartiles of the initial actions. The first and the second columns show the first and the fourth quartiles in the first snapshot, respectively. The third and the fourth columns show the first and the fourth quartiles in the final snapshot, respectively.
The top row shows the evolution of the $|J_{\phi,0}|$ quartiles. In the first snapshot, particles with low (high) $|J_{\phi, 0}|$ distribute in the inner (outer) galaxy as naturally expected. In the final snapshot, the distribution of the stars of the first quartile is slightly elongated along the bar's major axis in the face-on map and exhibits spheroidal shape in edge-on maps. The distribution of the fourth quartile stars is weakly pinched at $|z|\gtrsim1$~kpc in the $x$-$z$ plane.
The middle and the bottom rows show the evolution of the $J_{R,0}$ and $J_{z,0}$ quartiles, respectively. They show similar evolution. Initially, the  distribution of the particles with low (high) $J_{R,0}$ and $J_{z,0}$ are radially extended (concentrated) and vertically thin (thick).
In the final snapshot, the bar is stronger in the first quartiles than in the fourth quartiles. In the side-on views, the particles in the first quartiles form a peanut-bulge. On the other hand, in the fourth quartiles the distribution is boxy/spheroidal rather than a peanut shape.

To summarise the above, the  peanut shape of the bulge is more prominent in the populations with higher $J_{\phi,0}$, lower $J_{R,0}$ and lower $J_{z,0}$ than in the populations with lower $J_{\phi,0}$, higher $J_{R,0}$ and higher $J_{z,0}$.
\citet{2017MNRAS.469.1587D} refer to this separation of stellar populations by the bar as kinematic fractionation.
Similar plots for model 2 are shown in Figures~4, 7 and 8 of \citet{2020MNRAS.498.3334D}.
We see a similar trend in model 2 although the bar in model 2 is larger than that in MWa and strongly buckled.

\begin{figure*}
    \centering
    \includegraphics[width=0.9\linewidth]{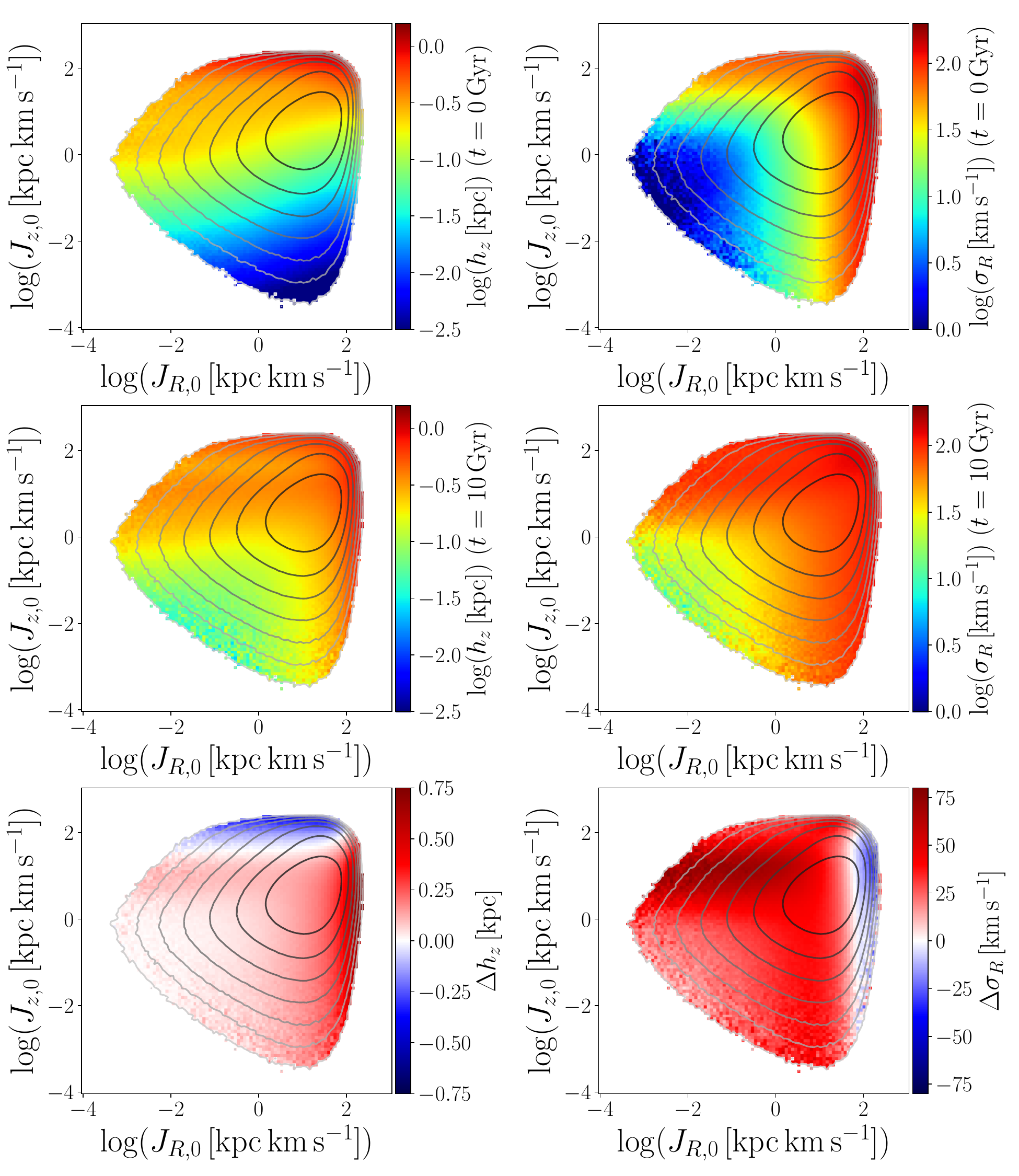}
    \caption{Evolution of the vertical thickness $h_z$ (\textit{left panels}) and the radial velocity dispersion $\sigma_R$ (\textit{right panels}) in the plane of initial radial action $J_{R,0}$ and vertical action $J_{z,0}$ for the MWa model. The top and middle rows show the initial ($t=0$) and final ($t=10$~Gyr) states, respectively, while the bottom row shows the change between the two snapshots ($\Delta h_z$ and $\Delta \sigma_R$). Grey contours represent the number density in logarithmic scale. }
    \label{fig:Jr_Jz_MWa}
\end{figure*}

Furthermore, Fig.~\ref{fig:Jr_Jz_MWa} shows the evolution of the vertical thickness and the radial velocity dispersion as a function of the initial radial and vertical actions in MWa, following the analysis of \citet{2020MNRAS.498.3334D}.
This further confirms that the kinematic fractionation process is operating in this model.
The left panels display the distribution of particles in the plane of initial radial action $J_{R,0}$ and vertical action $J_{z,0}$, colour-coded by the standard deviation of the vertical position in each bin (vertical thickness), $h_z$.
Panels in the top and middle rows are colour-coded by the initial and final values, respectively, while the bottom panels show their difference.
In the initial state, $h_z$ increases with $J_{z,0}$, as naturally expected, while it also shows a weaker dependence on $J_{R,0}$.
In the final state, the gradient of $h_z$ becomes steeper in the $J_{R,0}$ direction, while the gradient in the $J_{z,0}$ direction becomes shallower.
The difference, $\Delta h_z$, increases with $J_{R,0}$, and particularly, populations with $\log (J_{R,0}\; [\mathrm{kpc\, km\,s^{-1}}]) \gtrsim 2$ show a significant increase in $h_z$ of $\sim 0.5\,\mathrm{kpc}$.
In contrast, $\Delta h_z$ for the highest-$J_{z,0}$ populations with $\log (J_{z,0}\; [\mathrm{kpc\, km\,s^{-1}}]) \gtrsim 1.5$ is negative.
These results indicate that $J_{R,0}$ contributes more strongly to the vertical thickening than $J_{z,0}$, and populations with higher $J_{R,0}$ become thicker than those with lower $J_{R,0}$.
The same feature is observed in model 2 as well (see Fig.~12 of \citealt{2020MNRAS.498.3334D}).
The right panels show the same plots but colour-coded by the radial velocity dispersion $\sigma_R$.
$\sigma_R$ increases with both $J_{\mathrm{R},0}$ and $J_{z,0}$ in the initial and final states, although the gradient becomes shallower.
As evident in the bottom panel, $\sigma_{\mathrm{R}}$ increases over most of the $J_{\mathrm{R},0}$--$J_{z,0}$ plane but decreases at $\log (J_{\mathrm{R},0}\;[\mathrm{kpc\, km \, s^{-1}}]) \gtrsim 1.5$.
This highest-$J_{R,0}$ population is heated vertically but cooled radially after the bar formation.

\section{Supplementary figures}\label{sec:supplementary_figures}
\begin{figure}
	\begin{center}
		\includegraphics[width=\hsize]{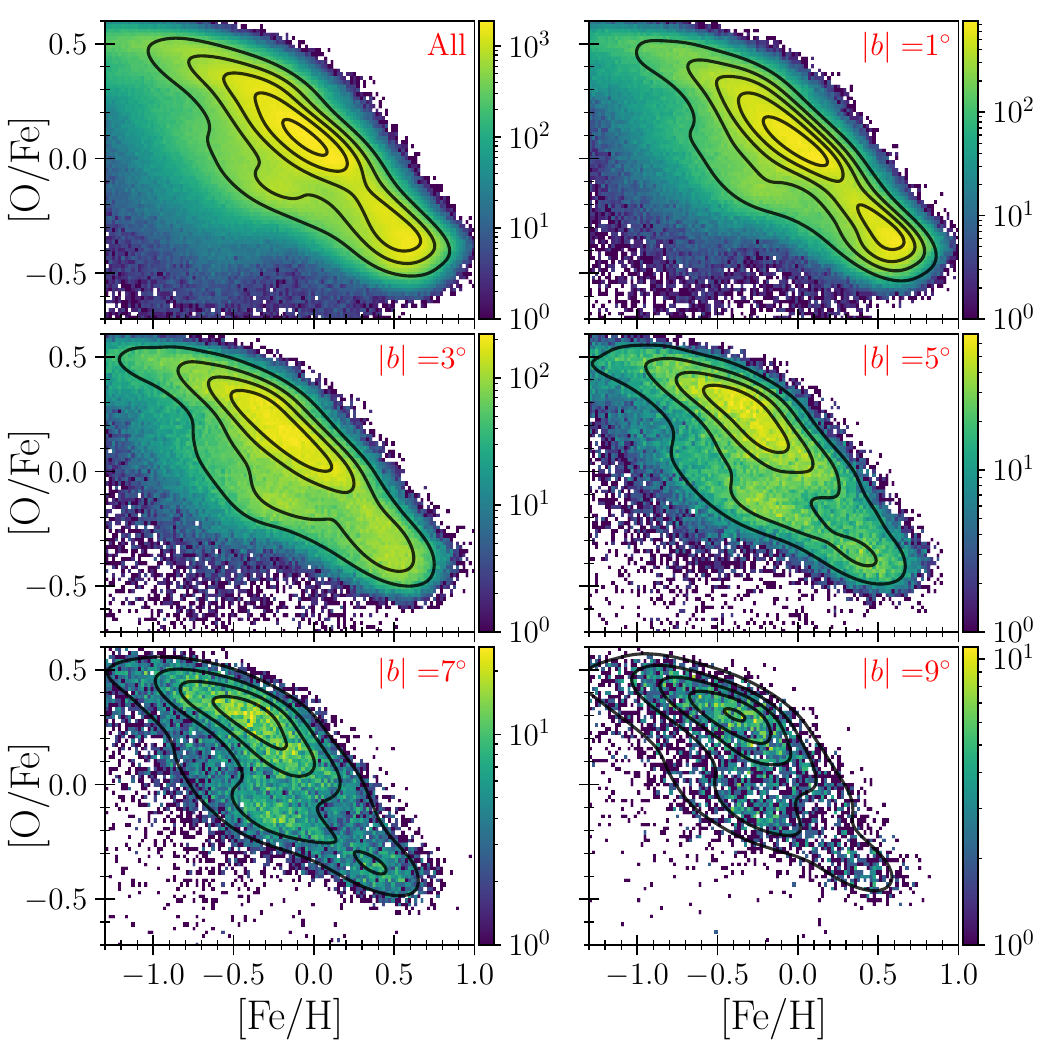}
		\caption{Same as Fig.~\ref{fig:feh_ofe_MWa_run733} but for (model 2, clumpy).}\label{fig:feh_ofe_run741_run733}
	\end{center}
\end{figure}

\begin{figure}
	\begin{center}
		\includegraphics[width=0.85\hsize]{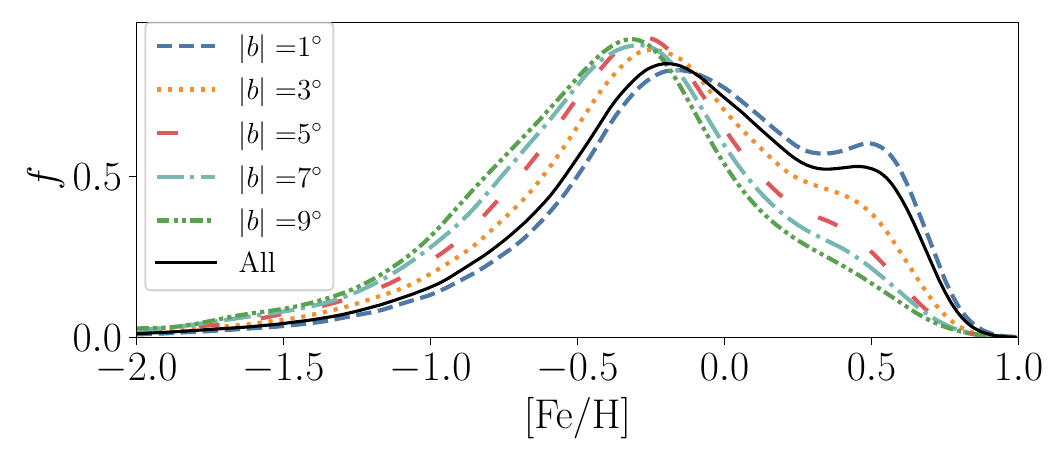}
		\includegraphics[width=0.85\hsize]{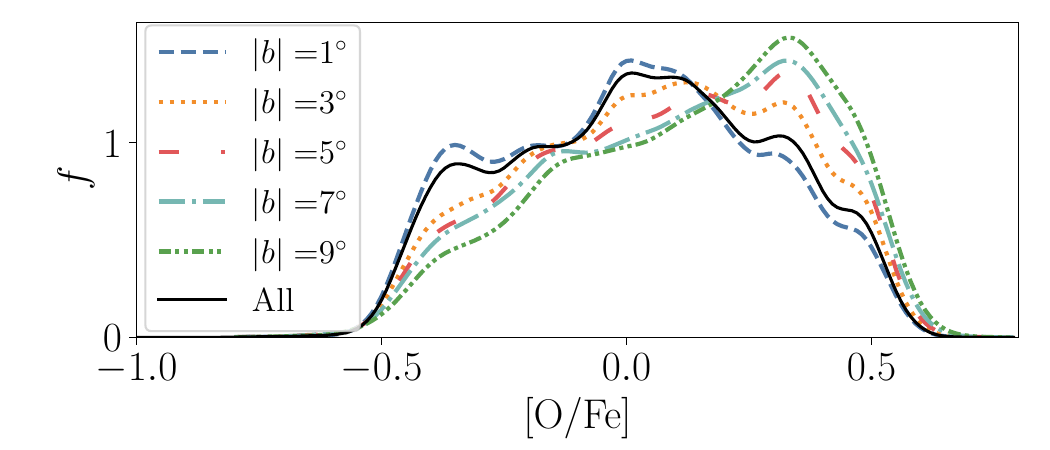}
		\caption{Same as Fig.~\ref{fig:feh_ofe_1Dhist_MWa_run733} but for (model 2, clumpy).}\label{fig:feh_hist_b_kde_run741_run733}
	\end{center}
\end{figure}

\begin{figure}
	\begin{center}
		\includegraphics[width=\hsize]{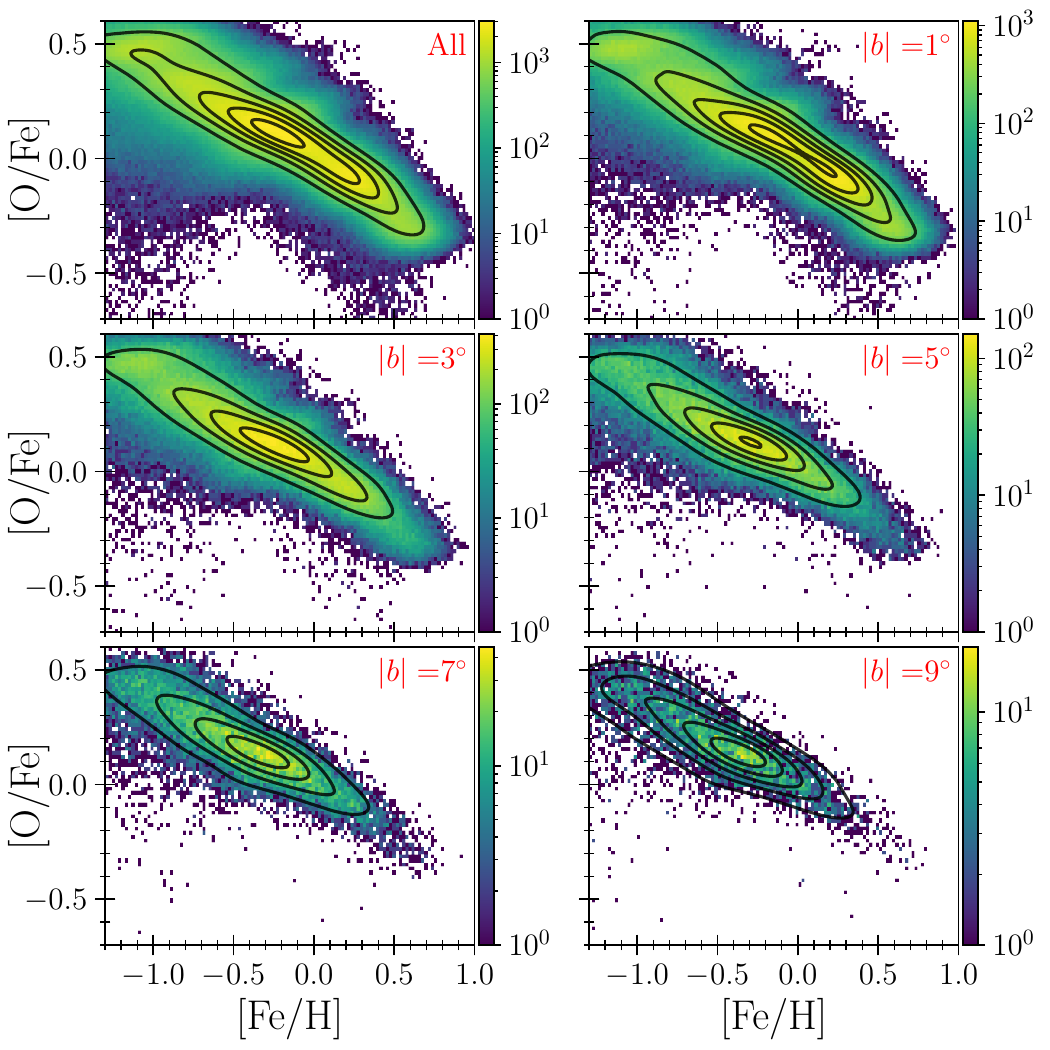}
		\caption{Same as Fig.~\ref{fig:feh_ofe_MWa_run733} but for (model 2, M1\_c\_b).}\label{fig:feh_ofe_run741_run739}
	\end{center}
\end{figure}

\begin{figure}
	\begin{center}
		\includegraphics[width=0.85\hsize]{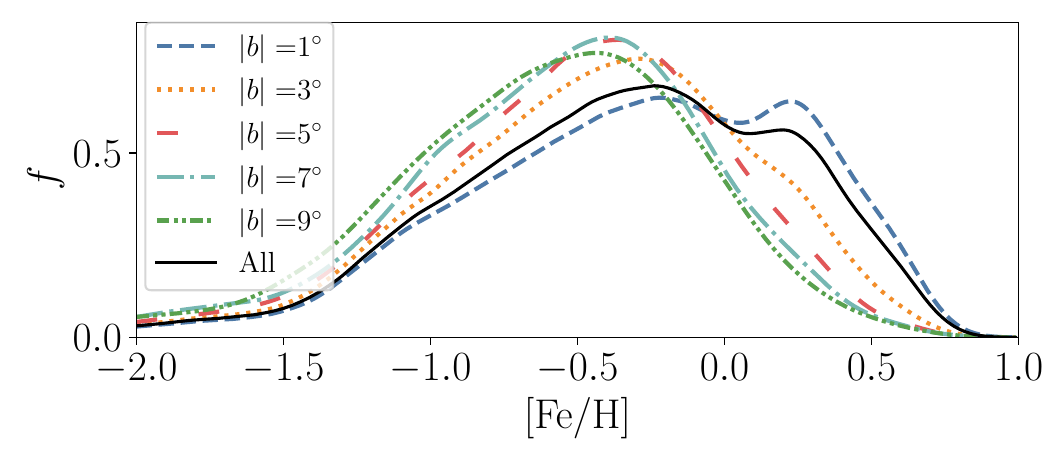}
		\includegraphics[width=0.85\hsize]{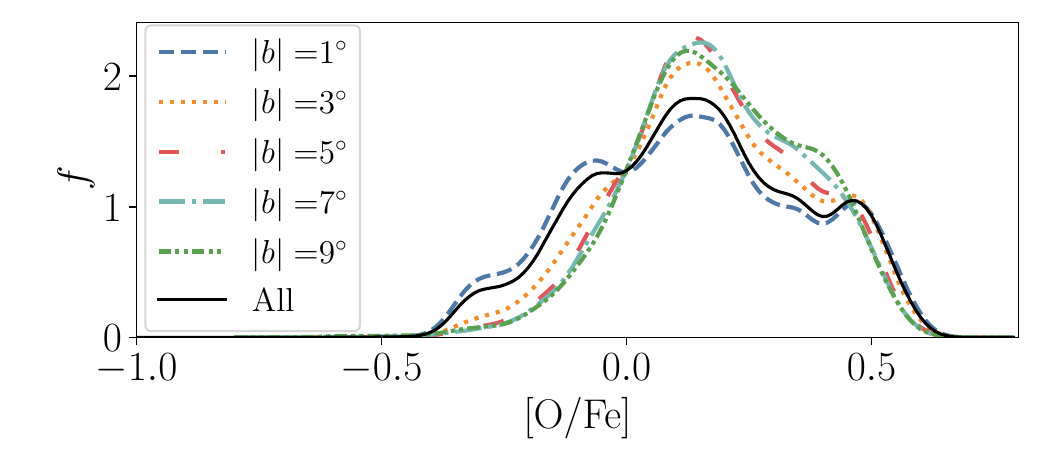}
		\caption{Same as Fig.~\ref{fig:feh_ofe_1Dhist_MWa_run733} but for (model 2, M1\_c\_b).}\label{fig:feh_hist_b_kde_run741_run739}
	\end{center}
\end{figure}

\begin{figure*}
	\begin{center}
		\includegraphics[width=\hsize]{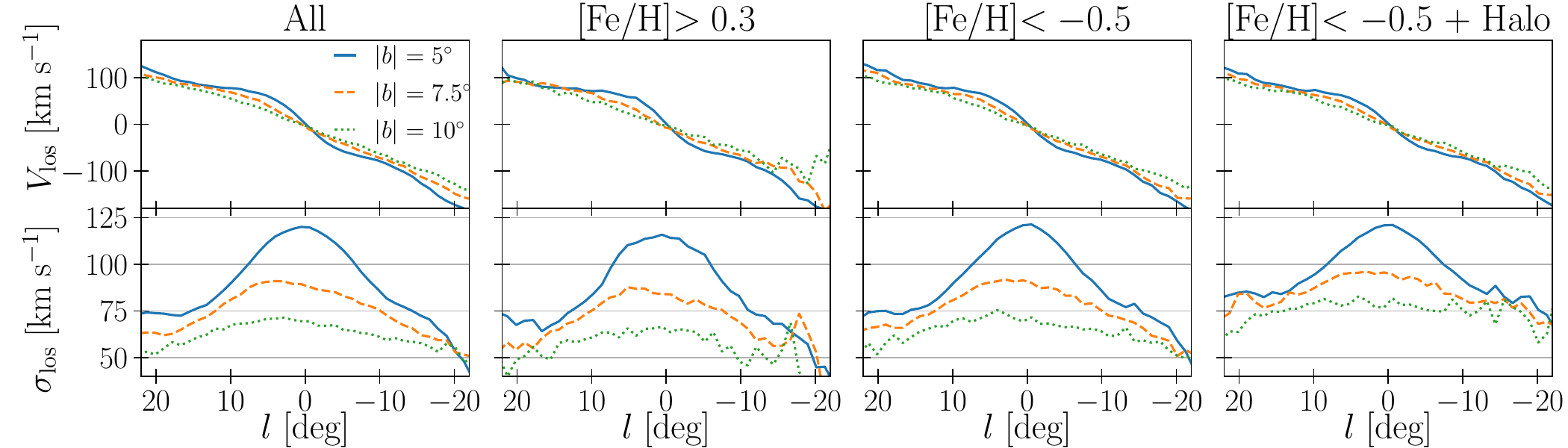}
		\caption{Same as Fig.~\ref{fig:bulge_kinematics_MWa_run733} but for (model 2, clumpy).}\label{fig:bulge_kinematics_run741_run733}
	\end{center}
\end{figure*}

\begin{figure}
	\begin{center}
		\includegraphics[width=0.95\hsize]{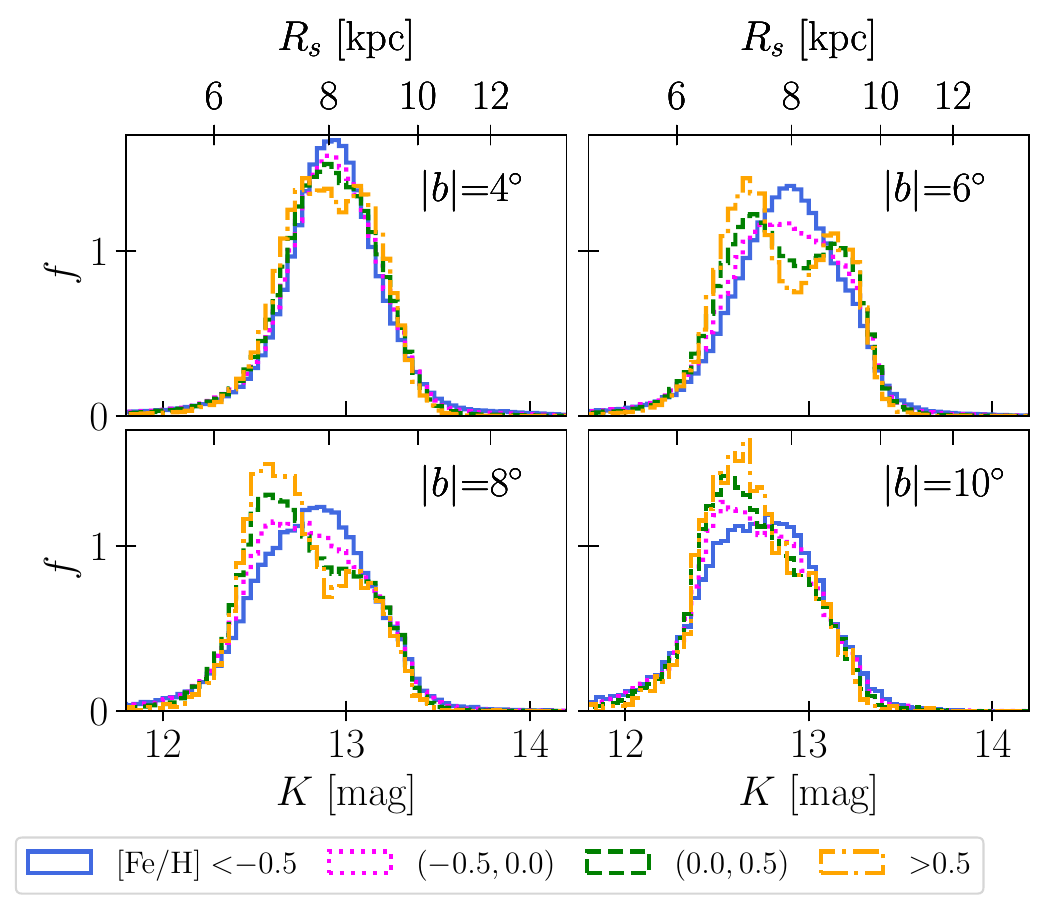}
		\caption{Same as Fig.~\ref{fig:Kmag_MWa_run733} without Gaussian convolution.}\label{fig:Kmag_MWa_run733_unconvolved}
	\end{center}
\end{figure}

In this section, we present supplementary figures for model 2 and additional details for MWa.
Figures~\ref{fig:feh_ofe_run741_run733} and \ref{fig:feh_hist_b_kde_run741_run733} show the [Fe/H]--[O/Fe] distributions and their 1D histograms for (model 2, clumpy).
As discussed in Section~\ref{sec:results}, model 2 exhibits a clear latitude-dependent bimodality, similar to MWa.
Figures~\ref{fig:feh_ofe_run741_run739} and \ref{fig:feh_hist_b_kde_run741_run739} show the results for (model 2, M1\_c\_b), which follows a single chemical track.
Figure~\ref{fig:bulge_kinematics_run741_run733} displays the bulge kinematics for (model 2, clumpy).
In this model, the metallicity dependence of the velocity dispersion is more evident than in MWa, although it still remains weaker than observed in the MW due to the lack of a stellar halo component.
Figure~\ref{fig:Kmag_MWa_run733_unconvolved} provides the mock $K$-band magnitude distribution of red clump stars in (MWa, clumpy) without Gaussian convolution.
This reveals the intrinsic bimodality in the distance distribution of the metal-rich population, which is smoothed out by the larger observational errors adopted in Fig.~\ref{fig:Kmag_MWa_run733}.


\end{appendix}
\end{document}